\documentclass[12pt]{article}
\usepackage{jheppub}

\pdfoutput=1

\usepackage{amsmath,bbm,array,amsfonts,graphicx,wrapfig,lscape,float,mathtools,multirow,longtable}
\usepackage[table]{xcolor}
\usepackage[export]{adjustbox}

\newcommand{\be}{\begin{equation}}
\newcommand{\ee}{\end{equation}}
\newcommand{\beq}{\begin{equation}}
\newcommand{\beql}[1]{\begin{equation}\label{#1}}
\newcommand{\eeq}{\end{equation}}
\newcommand{\ba}{\begin{array}}
\newcommand{\ea}{\end{array}}
\newcommand{\bea}{\begin{eqnarray}}
\newcommand{\beal}[1]{\begin{eqnarray}\label{#1}}
\newcommand{\eea}{\end{eqnarray}}
\newcommand{\ben}{\begin{enumerate}}
\newcommand{\een}{\end{enumerate}}
\newcommand{\bean}{\begin{eqnarray*}}
\newcommand{\eean}{\end{eqnarray*}}
\newcommand{\eref}[1]{(\ref{#1})}
\newcommand{\sref}[1]{\S\ref{#1}}

\newcommand{\fref}[1]{Figure \ref{#1}}
\newcommand{\btab}[1]{\begin{tabular}{#1}}
\newcommand{\etab}{\end{tabular}}

\def \Black {\pdfsetcolor{0 0 0 1}}

\newcommand{\comment}[1]{}

\newcommand{\qed}{\nobreak \ifvmode \relax \else
      \ifdim\lastskip<1.5em \hskip-\lastskip
      \hskip1.5em plus0em minus0.5em \fi \nobreak
      \vrule height0.75em width0.5em depth0.25em\fi}

\definecolor{darkspringgreen}{rgb}{0.09, 0.45, 0.27}
\definecolor{forestgreen}{rgb}{0.13, 0.55, 0.13}

\newcommand{\code}[1]{\texttt{#1}} 
\usepackage{algpseudocode} 
\usepackage{algorithm} 

\usepackage{array}
\newcolumntype{C}[1]{>{\centering\let\newline\\\arraybackslash\hspace{0pt}}m{#1}}

\newcommand{\mb}[1]{\mathbb{#1}} 

\title{A Comprehensive Survey of Brane Tilings}

\author[a,b]{Sebasti\'an Franco,} 
\author[c,d,e]{Yang-Hui He,}
\author[d]{Chuang Sun,}
\author[e]{Yan Xiao}

\affiliation[a]{
Physics Department, The City College of the CUNY \\
160 Convent Avenue, New York, NY 10031, USA}

\affiliation[b]{The Graduate School and University Center, The City University of New York  \\
365 Fifth Avenue, New York NY 10016, USA}

\affiliation[c]{
School of Physics, NanKai University, Tianjin, 300071, P.R.~China
}

\affiliation[d]{
Rudolf Peierls Centre for Theoretical Physics, University of Oxford\\ 
1 Keble Road, Oxford OX1 3NP, UK
}

\affiliation[e]{
Department of Mathematics, City, University of London, EC1V 0HB, UK
}

\emailAdd{sfranco@ccny.cuny.edu}
\emailAdd{hey@maths.ox.ac.uk}
\emailAdd{chuang.sun@physics.ox.ac.uk}
\emailAdd{Yan.Xiao@city.ac.uk}

\preprint{
\begin{flushright}
CCNY-HEP-17-02 
\end{flushright}
}

\abstract{An infinite class of $4d$ $\mathcal{N}=1$ gauge theories can be engineered on the worldvolume of D3-branes probing toric Calabi-Yau 3-folds. This kind of setup has multiple applications, ranging from the gauge/gravity correspondence to local model building in string phenomenology. Brane tilings fully encode the gauge theories on the D3-branes and have substantially simplified their connection to the probed geometries. The purpose of this paper is to push the boundaries of computation and to produce as comprehensive a database of brane tilings as possible. We develop efficient implementations of brane tiling tools particularly suited for this search. We present the first complete classification of toric Calabi-Yau 3-folds with toric diagrams up to area 8 and the corresponding brane tilings. This classification is of interest to both physicists and mathematicians alike.
}

\begin{document}

\maketitle

\section{Introduction}

A powerful approach for engineering $4d$ $\mathcal{N}=1$ gauge theories in string theory consists of realizing them on the worldvolume of D3-branes probing singular Calabi-Yau (CY) 3-folds. The case in which the CY 3-fold is toric is extremely rich, yet particularly tractable. 

More than a decade has passed since the first systematic treatment of the question ``what is the gauge theory given an arbitrary toric CY$_3$?'' \cite{Feng:2000mi}. A first approach for addressing this problem was the {\it Inverse Algorithm}, which generates the quiver and superpotential for a given toric singularity via partial resolution of an appropriate $\mb{C}^3/(\mb{Z}_N \times \mb{Z}_M$) orbifold. In practice, a chief bottleneck of this method was the exponential running time necessary for finding dual cones needed for partial resolution. Later, the connection between toric geometry and gauge theory was tremendously simplified with the advent of {\it brane tilings} \cite{Hanany:2005ve,Franco:2005rj,Franco:2005sm}, which have become the standard tools in this field. Brane tilings are Type IIB configurations of branes related to D3-branes at toric singularities by T-duality. Throughout this paper, we will equivalently refer to brane tilings as dimer models.

Configurations of D3-branes probing toric CY 3-folds have found a myriad of applications. In physics, they include: the understanding that toric duality is Seiberg duality \cite{Feng:2001bn,Beasley:2001zp}, one of the most fertile grounds for testing the AdS/CFT correspondence \cite{Benvenuti:2004dy,Franco:2005sm,Butti:2005sw, Benvenuti:2005ja}, connections to mirror symmetry and tropical geometry \cite{Feng:2005gw}, local string phenomenology \cite{Krippendorf:2010hj,Dolan:2011qu}, and bipartite field theories \cite{Franco:2012mm,Xie:2012mr,Franco:2012wv,Franco:2013pg,Heckman:2012jh,Franco:2013ana,Franco:2014nca}.

In parallel, in mathematics, the dialogue between gauge theory and the geometry and combinatorics of toric CY 3-folds also engendered numerous developments, including: new directions in Calabi-Yau algebras and quiver representations \cite{2006math.....12139G,rafCY,2007arXiv0704.0649D,2007arXiv0710.1898I,2009arXiv0901.4662B,2009arXiv0904.0676D,baur-king-marsh-16}, non-commutative crepant resolutions of toric singularities \cite{rafNCCS,balazs,MR,beil,BCQ}, connections with Grothendieck's dessins d'enfants and certain isogenies of elliptic curves \cite{Jejjala:2010vb,Hanany:2011ra,Hanany:2011bs,Vidunas:2016xun} and a geometric perspective on cluster algebras \cite{Eager:2011ns,MSW,LM,GK}.

The purpose of this paper is to push the boundaries of computation and to produce as comprehensive a database of brane tilings as possible. We will develop efficient implementations of dimer model tools particularly suited for this search and develop a catalogue of explicit brane tilings for a large class of toric geometries. We will also generate new computational tools, in the form of {\it Mathematica} modules, which we will make publicly available \cite{DimerSystem}. We expect a wide range of researchers will find this novel toolkit useful.

Until now, a large database of explicit brane tilings was lacking and we envision many applications for such a catalogue in both physics and mathematics. In our case, we plan to use these theories in the near future as starting points for a systematic and large scale investigation of phenomenological local models in string theory, following \cite{Krippendorf:2010hj, Dolan:2011qu}.

The organization of this paper is as follows. Section \sref{section_brane_tiling_technology} reviews brane tilings and outlines how to construct new ones by means of partial resolution. Section \sref{section_existing_classifications} summarizes the existing classifications of brane tilings. Section \sref{section_geometries} classifies all independent toric diagrams up to area 8. Section \sref{section_results} presents brane tilings for all toric CY 3-folds with toric diagrams of area 6, 7 and 8.\footnote{All the brane tilings for lower toric diagram areas can be found in \cite{Franco:2005rj,Franco:2005sm}. The few missing cases can be immediately determined from gauge theory information presented in \cite{Feng:2004uq}.} We collect our conclusions and directions for future research in section \sref{section_conclusions}. Appendix \sref{appendix_modules} explains the {\it Mathematica} modules we created for manipulating brane tilings.

\section{Brane Tiling Technology}

\label{section_brane_tiling_technology}

In this section we present a lighting review of brane tiling technology. In order to set up the stage for our computations, we also review the basics of the connection between brane tilings and geometry and the implementation of partial resolution in terms of them. We refer the interested reader to \cite{Franco:2005rj,Franco:2005sm,Kennaway:2007tq,Yamazaki:2008bt} and references therein for further details.

\subsection{D3-Branes Probing Toric CY 3-Folds and Brane Tilings}

The $4d$ $\mathcal{N}=1$ gauge theories living on the worldvolume of D3-branes probing affine toric CY 3-folds are described by bipartite graphs on $T^2$ called {\it brane tilings} \cite{Hanany:2005ve,Franco:2005rj,Franco:2005sm}. In fact a brane tiling is a physical brane configuration, related to the D3-branes at a toric singularity by T-duality, consisting of an NS5-brane wrapping a holomorphic surface from which D5-branes are suspended. The geometry of a non-compact toric CY 3-fold is captured by a {\it toric diagram}, which is convex lattice polygon.\footnote{An affine toric variety of complex dimension $n$ is usually described by a convex polyhedral cone in $\mathbb{R}^n$ but the Calabi-Yau condition imposes the extra condition that the endpoints of the vector generators of the cone are co-hyperplanar. Thus for 3-folds, the toric diagram can be taken to be a convex lattice polygon in $2d$.} The probed CY$_3$ arises as the vacuum moduli space of the gauge theory on the D3-branes, which is defined by the vanishing of $D$- and $F$-terms.

A brane tiling encodes a $4d$ $\mathcal{N}=1$ quiver gauge theory as follows:
\begin{enumerate}
\item Every face (say labeled by $i$) corresponds to a $U(N_i)$ gauge group factor in a product gauge group structure.
\item Every edge between faces $i$ and $j$ corresponds to a bifundamental chiral field $X_{ij}$ of $U(N_i) \times U(N_j)$. If $i$ is equal to $j$, then $X_{ii}$ is an adjoint field of $U(N_i)$. The orientation of fields is a convention, e.g. clockwise and counterclockwise around black and white nodes of the tiling, respectively.
\item Every node corresponds to a monomial term in the superpotential, obtained by multiplying all the edges adjacent to the node. Like the orientation of chiral fields, the sign of the monomial is controlled by the color of the node.
\end{enumerate}

In order to illustrate these ideas, below we present an explicit example that corresponds to the complex cone over $F_0$.\footnote{In fact there are two toric phases, i.e. two theories described by brane tilings, for this geometry. They are related by Seiberg duality \cite{Feng:2001bn}.} The red dashed lines indicate the boundary of the unit cell.

\begin{table}[!htbp] 
\centering 
\begin{tabular}{c|c|c}
Toric Diagram & Brane Tiling & Gauge Theory \\
\hline 
\adjustimage{height=3cm,valign=m}{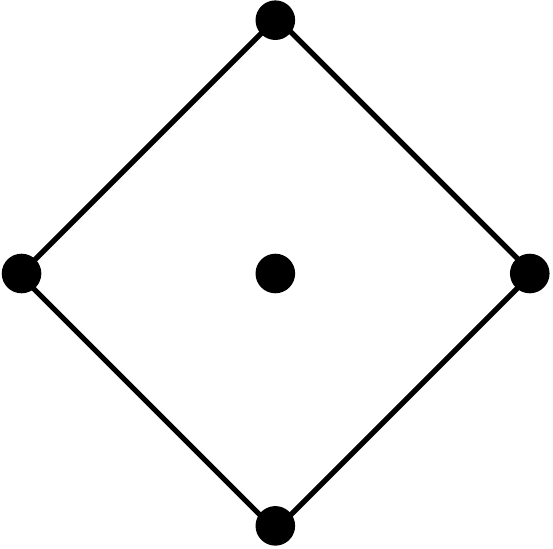} & 
\adjustimage{height=5cm,valign=m}{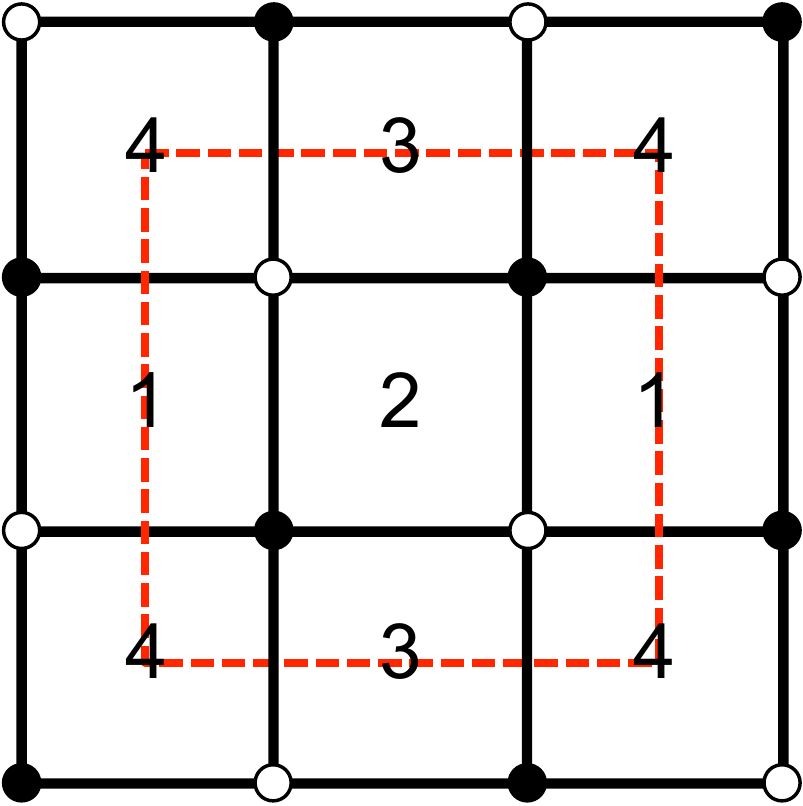} &
\adjustimage{height=6cm,valign=m}{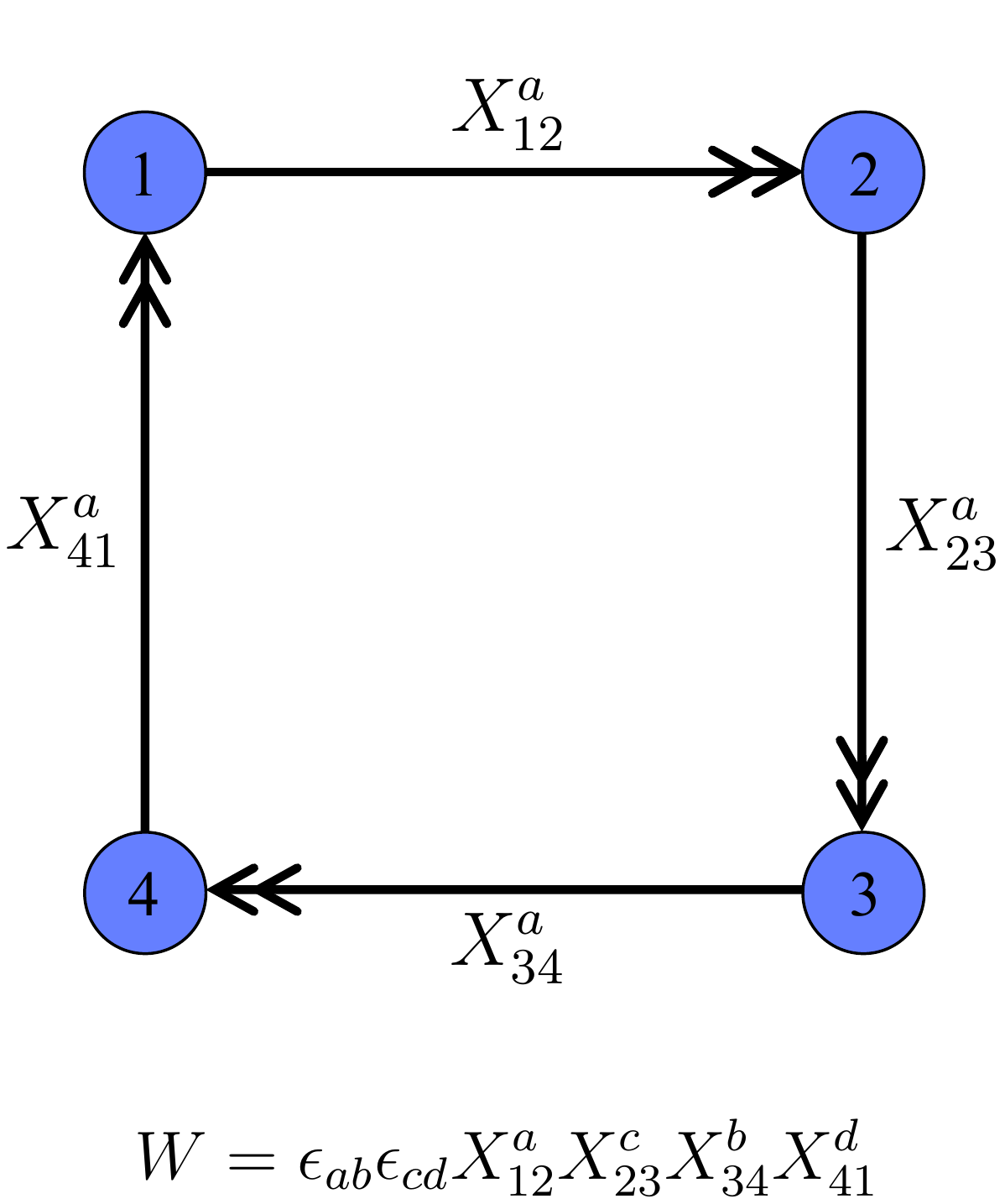} 
\end{tabular}
\end{table}

\subsection{Geometry and Perfect Matchings}

{\it Perfect matchings} are combinatorial objects that play a central role in the study of bipartite graphs. A perfect matching $p$ is defined as a collection of edges in the brane tiling such that every node is the endpoint of exactly one edge in $p$.  

Perfect matching substantially simplify the connection between brane tilings and geometry. Let us consider the following map between chiral fields in the quiver $X_\alpha$, equivalently edges in the brane tiling, and perfect matchings $p_\mu$
\beq
X_\alpha = \prod_{\mu=1}^c p_\mu^{P_{\alpha\mu}}\, ,
\label{X_pm_map}
\eeq
where $c$ is the total number of perfect matchings. The $P$-matrix summarizes the edge content of perfect matchings and is defined as follows
\beq
P_{\alpha\mu}=\left\{ \begin{array}{ccccc} 1 & \rm{ if } & X_\alpha  & \in & p_\mu \\
0 & \rm{ if } & X_\alpha  & \notin & p_\mu
\end{array}\right.
\label{Xi_to_pmu}
\eeq
A remarkable feature of the map in (\ref{X_pm_map}) is that when chiral fields are expressed in terms of perfect matching variables in this way, all $F$-terms automatically vanish. Perfect matchings are thus in one-to-one correspondence with fields in the GLSM description of the toric CY 3-fold, namely points in its toric diagram \cite{Franco:2005rj}. 

Perfect matchings and the toric diagram can be efficiently determined using the {\it Kasteleyn matrix} $K$. We define $K$ as the adjacency matrix of the graph in which rows are indexed by black nodes and columns are indexed by white nodes, i.e. for every edge $X_\alpha$ in the bipartite graph between nodes ${\bf b}_\mu$ and ${\bf w}_\nu$, we introduce a contribution $X_\alpha$ to the $K_{\mu \nu}$ entry. In addition, whenever an edge crosses the boundary of the unit cell in the $x$ and/or $y$ directions, we multiply the contribution by $x^{\pm 1}$ and $y^{\pm1}$ weights, respectively. The exponents are positive or negative depending on whether the crossing occurs in the positive or negative direction, which is determined by conventionally orienting edges from white to black nodes. 

\begin{figure}[ht]
	\centering
	\includegraphics[width=5cm]{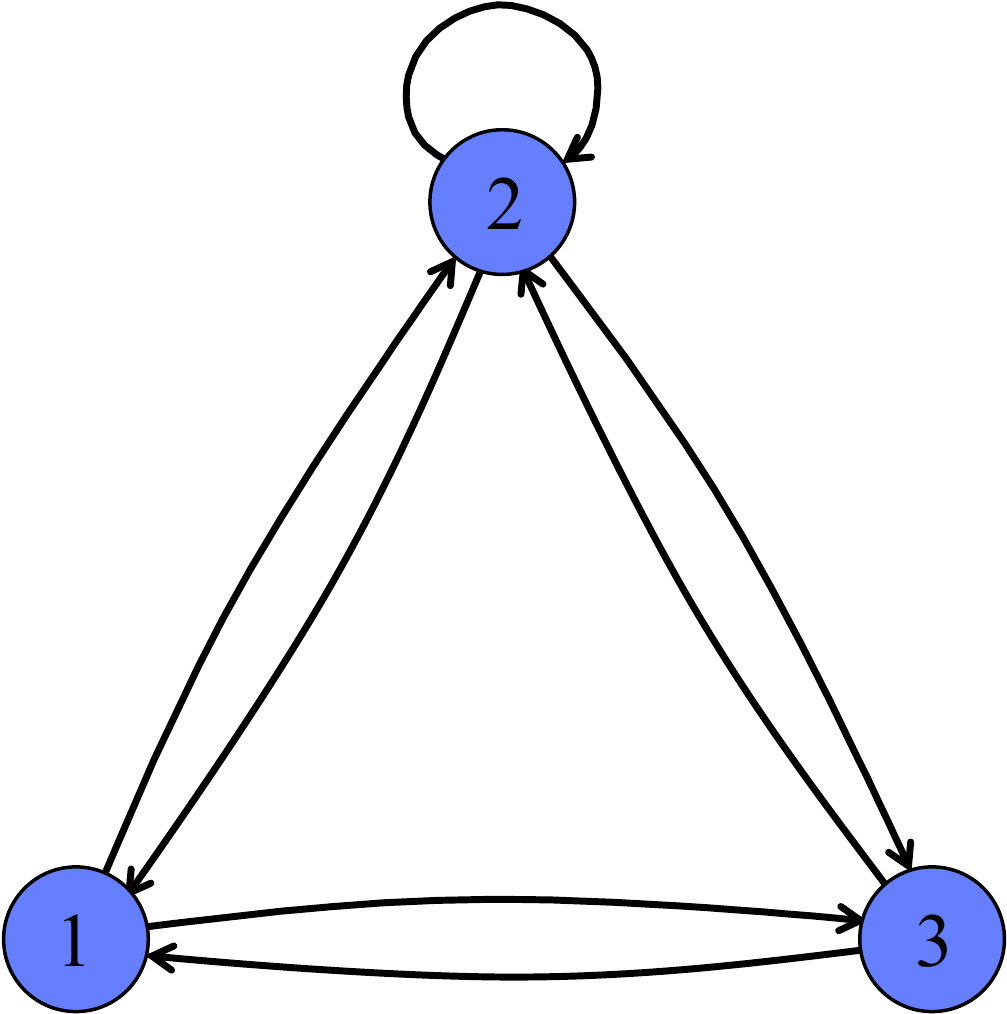}
\caption{Quiver diagram for SPP. Nodes represent gauge groups. The arrow from $i \rightarrow j$ corresponds to the chiral field $X_{ij}$.}
	\label{quiver_SPP}
\end{figure}

Let us consider a concrete example. \fref{quiver_SPP} shows the quiver diagram for the suspended pinch point (SPP). The corresponding superpotential is 
\beq
W = X_{12} X_{21} X_{22}-X_{22} X_{23} X_{32}+X_{13} X_{23} X_{31} X_{32}-X_{12} X_{13} X_{21} X_{31} \, .
\label{W_SPP}
\eeq

\begin{figure}[ht]
	\centering
	\includegraphics[width=6cm]{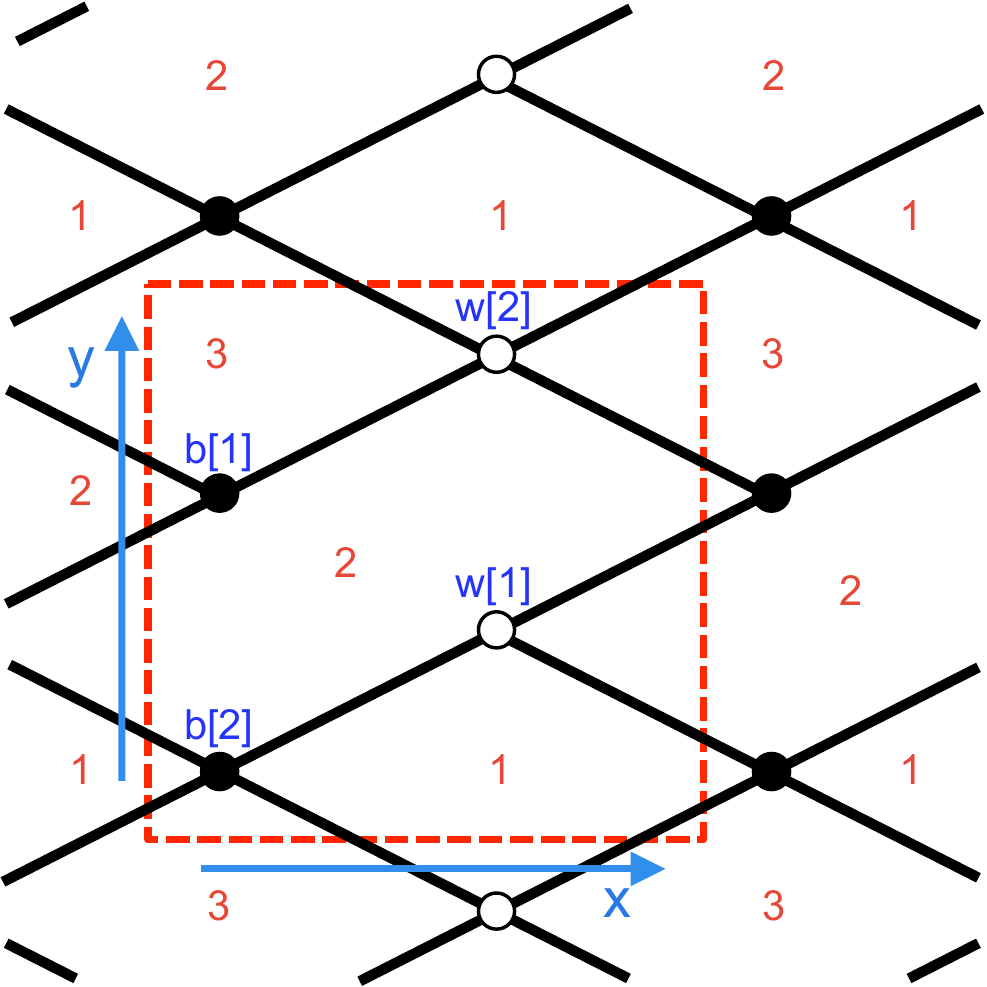}
\caption{Brane tiling for SPP.}
	\label{dimer_SPP}
\end{figure}

All this information is encoded in the brane tiling shown in \fref{dimer_SPP}. 

\begin{figure}[ht]
	\centering
	\includegraphics[width=4.5cm]{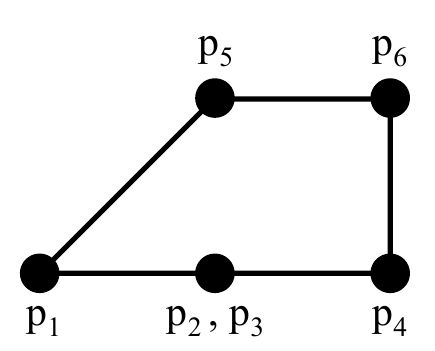}
\caption{Toric diagram for SPP. We indicate the perfect matching associated to each point.}
	\label{toric_SPP}
\end{figure}

The superpotential has four terms, which are represented in the brane tiling by two white and two black nodes. We have labeled the nodes in blue to facilitate the construction of the Kasteleyn matrix, which is given by
\beq
K=\left(
\begin{array}{c|c|c}
 & w[1] & w[2] \\ \hline
b[1] \ & X_{22} \, x & X_{23} + X_{32} \, x \\ \hline 
b[2] \ & \  X_{12} + X_{21} \, x \ & \ X_{31} \, y + X_{13} \, x y \
\end{array}
\right) \, .
\eeq
The determinant of the Kasteleyn matrix generates the perfect matchings. In this case, we get
\beq
\det K = -X_{12} X_{23} - (X_{21} X_{23}+X_{12} X_{32}) \, x - X_{21} X_{32} \, x^2 + X_{22} X_{31} \, xy + X_{13} X_{22} \, x^2 y  \,.
\eeq
Every monomial in this expression corresponds to a perfect matching. Furthermore, the powers of $x$ and $y$ indicate their position in the toric diagram, as shown in \fref{toric_SPP}. The perfect matching can be summarized in the $P$-matrix as follows
\beq
P= \left(
\begin{array}{c|cccccc}
& p_1 & p_2 & p_3 & p_4 & p_5 & p_6 \\ \hline
X_{22} & 0 & 0 & 0 & 0 & 1 & 1 \\
X_{12} & 1 & 0 & 1 & 0 & 0 & 0 \\
X_{21} & 0 & 1 & 0 & 1 & 0 & 0 \\
X_{23} & 1 & 1 & 0 & 0 & 0 & 0 \\
X_{32} & 0 & 0 & 1 & 1 & 0 & 0 \\
X_{31} & 0 & 0 & 0 & 0 & 1 & 0 \\
X_{13} & 0 & 0 & 0 & 0 & 0 & 1
\end{array}
\right) \, .
\eeq

\subsection{Partial Resolution and Brane Tilings}

Brane tilings completely solved the problem of finding the gauge theory associated to a generic toric CY 3-fold and vice versa. There are well established procedures for going from brane tilings to geometry and in the opposite direction: the {\it fast forward} \cite{Franco:2005rj} and {\it fast inverse algorithms} \cite{Hanany:2005ss,Feng:2005gw}, respectively. One of the main goals to this paper is to develop a practical approach to determine the brane tiling associated to a general toric diagram. While the fast inverse algorithm provides an answer to this question, its automation remains challenging. We thus opt for an alternative approach, which admits a rather simple computer implementation.

Our strategy will be to perform partial resolution, which translates to higgsing in the gauge theory. In terms of brane tilings, it corresponds to removing the edges associated to the fields acquiring non-zero vacuum expectation values (vevs). We will exploit the map between perfect matchings and fields in the gauge theory to systematically identify the vevs that are turned on when certain points in the toric diagram are deleted. 

\begin{figure}[ht]
	\centering
	\includegraphics[width=10cm]{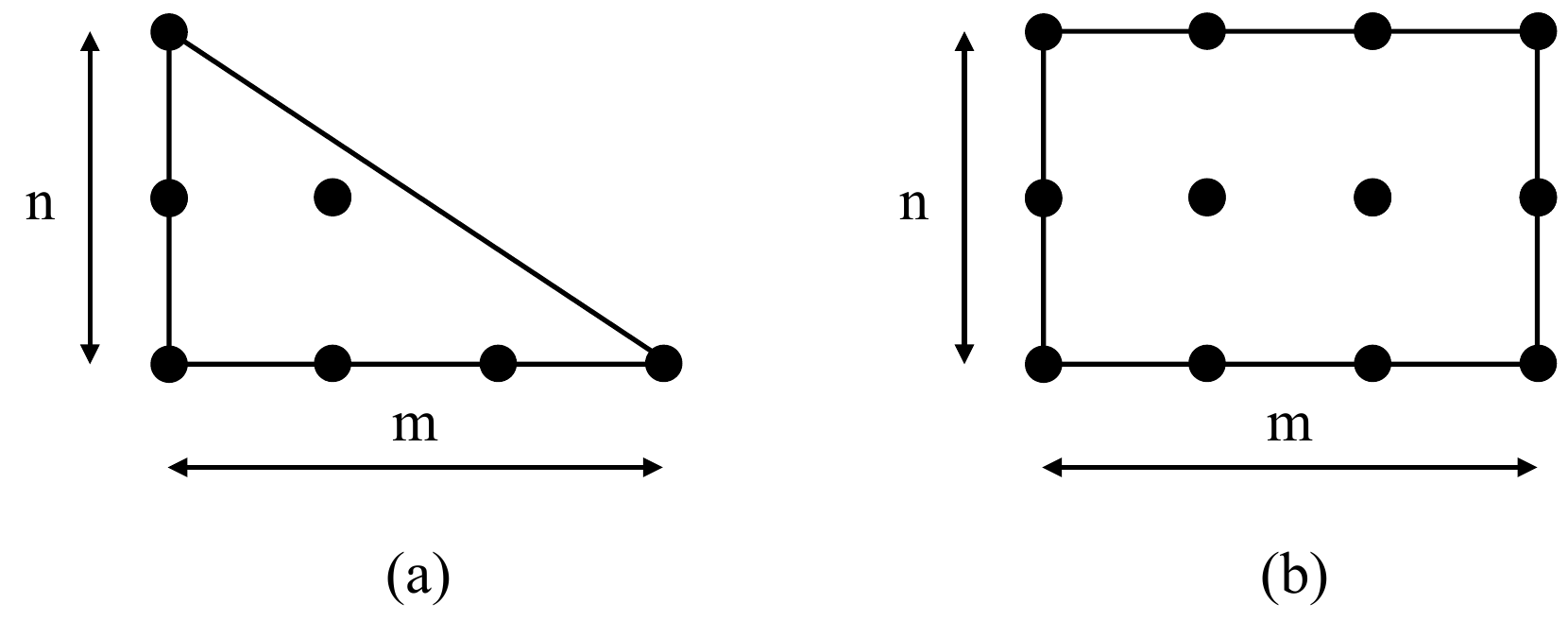}
\caption{Toric diagrams for: a) $\mathbb{C}^3/(\mathbb{Z}_m \times \mathbb{Z}_n)$ and b) $\mathcal{C}/(\mathbb{Z}_m \times \mathbb{Z}_n)$. We will use the second class of geometries as the starting points for partial resolution.}
	\label{orbifolds_C3_and_conifold}
\end{figure}

Any geometry for which the brane tiling is known can be used as the starting point for partial resolution. There are two canonical classes of initial theories that have been broadly used in the literature for this purpose. The first one is $\mathbb{C}^3/(\mathbb{Z}_m \times \mathbb{Z}_n)$ orbifolds, with the two generators of the orbifold group acting on $\mathbb{C}^3$ as: $(X,Y,Z)\mapsto (e^{i 2 \pi/N} X, e^{-i 2 \pi/N}Y, Z)$ and $(X,Y,Z)\mapsto (X, e^{i 2 \pi/M}Y, e^{-i 2 \pi/M} Z)$. The resulting toric diagram is shown in \fref{orbifolds_C3_and_conifold}.a, and the corresponding brane tiling is an hexagonal lattice with an $N\times M$ unit cell. The second standard class of starting points are $\mathbb{Z}_m \times \mathbb{Z}_n$ orbifolds of the conifold $\mathcal{C}$. Given the defining equation for the conifold $x y = u v$, the two generators of the orbifold group act as follows: $(x,y,u,v)\mapsto (e^{i 2 \pi/N} x, e^{-i 2 \pi/N}y, u,v)$ and $(x,y,u,v)\mapsto (x,y, e^{i 2 \pi/M}u, e^{-i 2 \pi/M} v)$. The toric diagram for this class of geometries is shown in \fref{orbifolds_C3_and_conifold}.b and the brane tiling is a square lattice with an $N\times M$ unit cell. We will adopt the orbifolds of the conifold as our initial theories. 

\begin{figure}[h]
\centering
	\includegraphics[width=6.7cm]{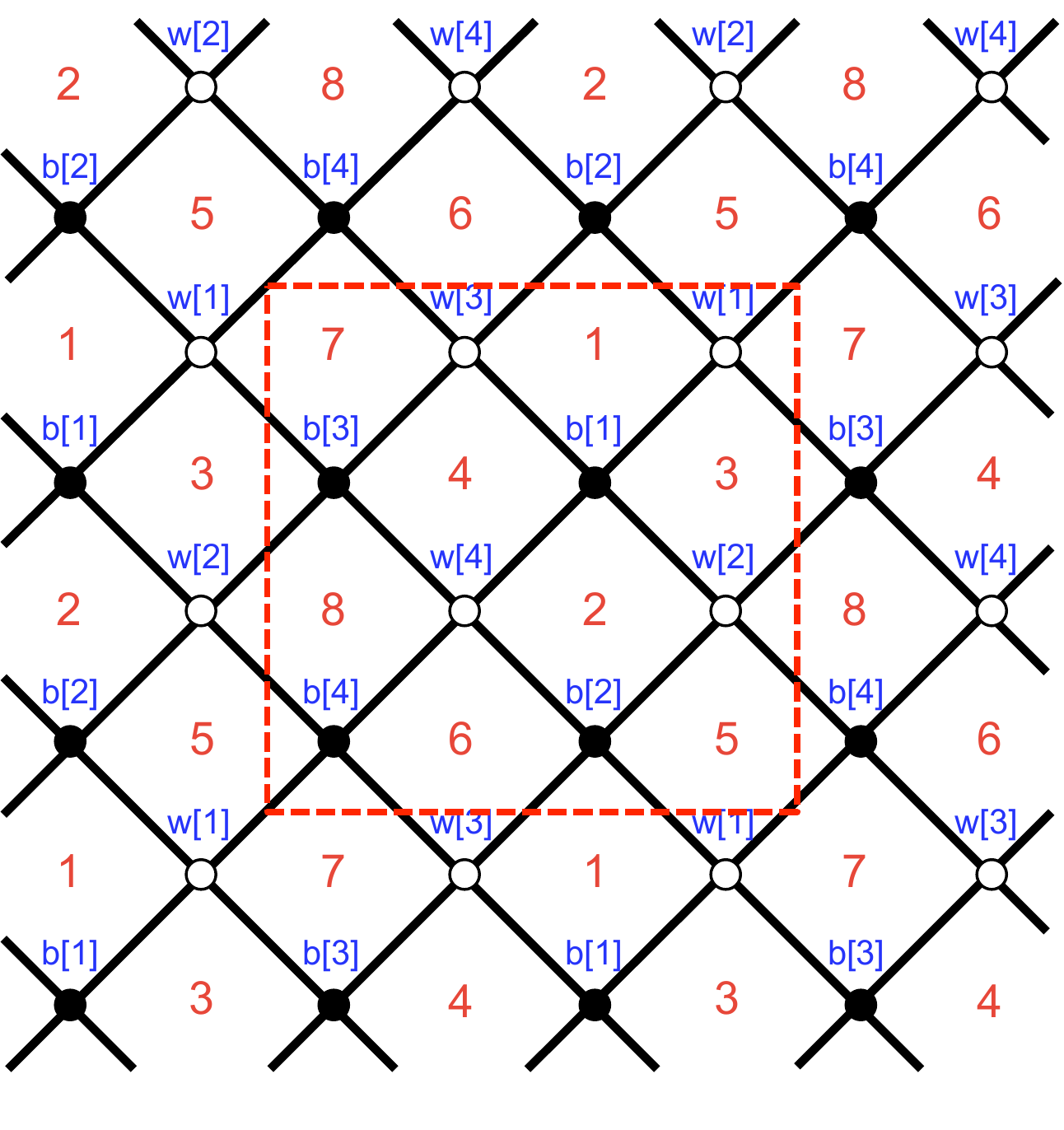}
\caption{Brane tiling for $\mathcal{C}/(\mathbb{Z}_2\times \mathbb{Z}_2)$.}
\label{dimer_conifold_Z2xZ2}
\end{figure}

We now illustrate the dimer implementation of partial resolution with an explicit example. Let us derive the brane tiling for the SPP from a $\mathcal{C}/(\mathbb{Z}_m \times \mathbb{Z}_n)$ orbifold. Considering the toric diagrams, it is clear that it would be sufficient to start from $\mathcal{C}/\mathbb{Z}_2$. However, in order to demonstrate the methods in a more involved partial resolution, let us use  $\mathcal{C}/(\mathbb{Z}_2\times \mathbb{Z}_2)$ as the initial theory. The brane tiling for $\mathcal{C}/(\mathbb{Z}_2\times \mathbb{Z}_2)$ is shown in \fref{dimer_conifold_Z2xZ2}.\footnote{There are other brane tilings for $\mathcal{C}/(\mathbb{Z}_2\times \mathbb{Z}_2)$, which correspond to additional toric phases obtained from this one by Seiberg duality.}

\begin{figure}[ht]
	\centering
	\includegraphics[width=8cm]{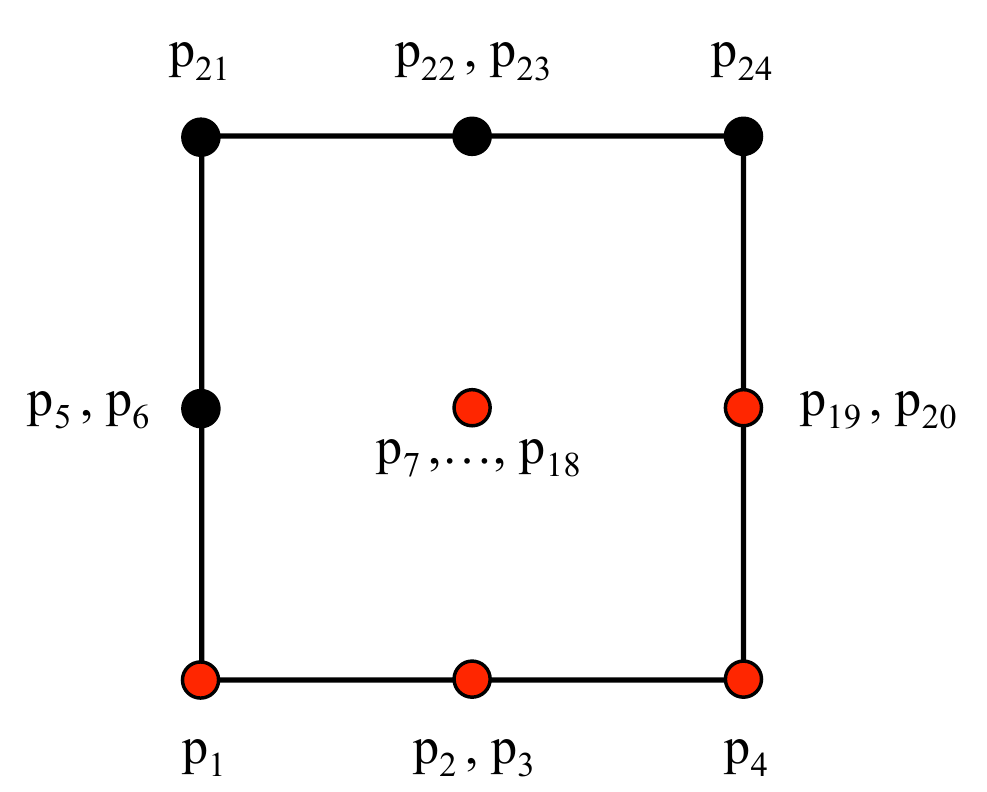}
\caption{Toric diagram for $\mathcal{C}/(\mathbb{Z}_2\times \mathbb{Z}_2)$ We indicate the perfect matching associated to each point and a possible embedding of the SPP toric diagram (in red).}
	\label{toric_Z2xZ2_SPP}
\end{figure}

The Kasteleyn matrix is given by
\begin{equation}
K=
\left(
\begin{array}{c|cccc}
 & w[1] & w[2] & w[3] & w[4] \\ \hline
b[1] \ & X_{13} & X_{32} & X_{41} & X_{24} \\ 
b[2] \ &  X_{51} \, y  & X_{25} & X_{16} \, y & X_{62} \\ 
b[3] \ &  X_{37} \, x & X_{83} \, x & X_{74} & X_{48} \\ 
b[4] \ &  X_{75} \, x y & X_{58} \, x & X_{67} \, y & X_{86} 
\end{array}
\right) \, .
\end{equation}
We obtain the perfect matchings by computing the determinant of the Kasteleyn matrix. They are summarized in the following $P$-matrix:
{\tiny \beq
\left(
\setlength\arraycolsep{3pt}
\begin{array}{c|c|cc|c|cc|cccccccccccc|cc|c|cc|c}
& (0,0) &  \multicolumn{2}{c|}{(1,0)} & (2,0) & \multicolumn{2}{c|}{(1,1)} & \multicolumn{12}{c|}{(2,1)} & \multicolumn{2}{c|}{(2,2)} & (0,2) &  \multicolumn{2}{c|}{(1,2)} & (2,2) \\ 
& p_1 & p_2 & p_3 & p_4 & p_5 & p_6 & p_7 & p_8 & p_9 & p_{10} & p_{11} & p_{12} & p_{13} & p_{14} & p_{15} & p_{16} & p_{17} & p_{18} & p_{19} & p_{20} & p_{21} & p_{22} & p_{23} & p_{24} \\ \hline
X_{13} & 1 & 1 & 0 & 0 & 1 & 0 & 1 & 0 & 0 & 0 & 0 & 0 & 0 & 0 & 1 & 0 & 1 & 0 & 0
   & 0 & 0 & 0 & 0 & 0 \\
\rowcolor{cyan!90!blue!60} X_{16} & 0 & 0 & 0 & 0 & 0 & 0 & 1 & 0 & 0 & 0 & 0 & 0 & 0 & 0 & 0 & 1 & 1 & 0 & 1
   & 0 & 0 & 1 & 0 & 1 \\
\rowcolor{cyan!90!blue!60} X_{24} & 0 & 0 & 0 & 0 & 0 & 0 & 0 & 0 & 1 & 0 & 1 & 0 & 1 & 0 & 0 & 0 & 0 & 0 & 1
   & 0 & 0 & 0 & 1 & 1 \\
X_{25} & 1 & 0 & 1 & 0 & 1 & 0 & 0 & 0 & 1 & 0 & 0 & 1 & 1 & 0 & 0 & 0 & 0 & 0 & 0
   & 0 & 0 & 0 & 0 & 0 \\
\rowcolor{cyan!90!blue!60} X_{32} & 0 & 0 & 0 & 0 & 0 & 1 & 0 & 0 & 0 & 1 & 0 & 0 & 0 & 1 & 0 & 1 & 0 & 0 & 0
   & 0 & 1 & 1 & 0 & 0 \\
X_{37} & 0 & 0 & 1 & 1 & 0 & 0 & 0 & 0 & 1 & 1 & 0 & 0 & 0 & 0 & 0 & 1 & 0 & 0 & 1
   & 0 & 0 & 0 & 0 & 0 \\
X_{41} & 0 & 0 & 1 & 1 & 0 & 0 & 0 & 1 & 0 & 0 & 0 & 1 & 0 & 0 & 0 & 0 & 0 & 1 & 0
   & 1 & 0 & 0 & 0 & 0 \\
\rowcolor{cyan!90!blue!60} X_{48} & 0 & 0 & 0 & 0 & 1 & 0 & 1 & 1 & 0 & 0 & 0 & 1 & 0 & 0 & 0 & 0 & 0 & 0 & 0
   & 0 & 1 & 1 & 0 & 0 \\
\rowcolor{cyan!90!blue!60} X_{51} & 0 & 0 & 0 & 0 & 0 & 1 & 0 & 1 & 0 & 0 & 1 & 0 & 0 & 0 & 0 & 0 & 0 & 1 & 0
   & 0 & 1 & 0 & 1 & 0 \\
X_{58} & 0 & 1 & 0 & 1 & 0 & 0 & 1 & 1 & 0 & 0 & 1 & 0 & 0 & 0 & 0 & 0 & 0 & 0 & 1
   & 0 & 0 & 0 & 0 & 0 \\
X_{62} & 0 & 1 & 0 & 1 & 0 & 0 & 0 & 0 & 0 & 1 & 0 & 0 & 0 & 1 & 1 & 0 & 0 & 0 & 0
   & 1 & 0 & 0 & 0 & 0 \\
X_{67} & 0 & 0 & 0 & 0 & 1 & 0 & 0 & 0 & 1 & 1 & 0 & 0 & 0 & 0 & 1 & 0 & 0 & 0 & 0
   & 0 & 1 & 0 & 1 & 0 \\
X_{74} & 1 & 1 & 0 & 0 & 0 & 1 & 0 & 0 & 0 & 0 & 1 & 0 & 1 & 1 & 0 & 0 & 0 & 0 & 0
   & 0 & 0 & 0 & 0 & 0 \\
X_{75} & 0 & 0 & 0 & 0 & 0 & 0 & 0 & 0 & 0 & 0 & 0 & 1 & 1 & 1 & 0 & 0 & 0 & 0 & 0
   & 1 & 0 & 1 & 0 & 1 \\
X_{83} & 0 & 0 & 0 & 0 & 0 & 0 & 0 & 0 & 0 & 0 & 0 & 0 & 0 & 0 & 1 & 0 & 1 & 1 & 0
   & 1 & 0 & 0 & 1 & 1 \\
X_{86} & 1 & 0 & 1 & 0 & 0 & 1 & 0 & 0 & 0 & 0 & 0 & 0 & 0 & 0 & 0 & 1 & 1 & 1 & 0
   & 0 & 0 & 0 & 0 & 0 \\
\end{array}
\right) 
\label{P_matrix_conifold_Z2xZ2}
\eeq}
where in the top row we have indicated the corresponding point in the toric diagram, which is shown in \fref{toric_Z2xZ2_SPP}. The significance of the rows that are highlighted in blue will be discussed soon. 

\begin{figure}[ht]
	\centering
	\includegraphics[width=4.4cm]{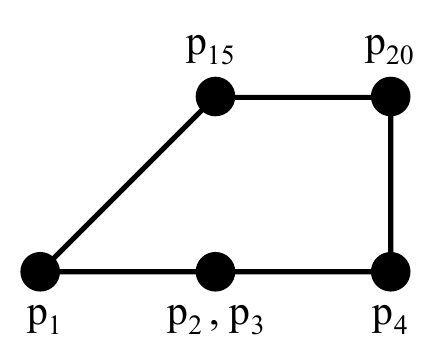}
\caption{Toric diagram for SPP obtained by partial resolution of $\mathcal{C}/(\mathbb{Z}_2\times \mathbb{Z}_2)$.}
	\label{toric_resolution_to_SPP}
\end{figure}

\fref{toric_Z2xZ2_SPP} shows a possible way of embedding the toric diagram of SPP, shown in red, into the one for $\mathcal{C}/(\mathbb{Z}_2\times \mathbb{Z}_2)$. According to \eref{Xi_to_pmu}, we should regard chiral fields as products of perfect matchings. The vev of a chiral field results from the product of the vevs of its perfect matching constituents. Then, a chiral field gets a vev and is removed from the brane tiling only when all the perfect matchings that contain it are deleted. Even after picking an embedding of the final toric diagram into the parent one there are, in general, multiple ways of achieving the desired partial resolution. For the embedding in \fref{toric_Z2xZ2_SPP}, one possibility is to turn on vevs for $\{ X_{16},X_{24},X_{32},X_{48},X_{51} \}$. The corresponding rows in the $P$-matrix are highlighted in blue in \eref{P_matrix_conifold_Z2xZ2}. It is straightforward to verify that this set of vevs achieves the desired resolution. Some perfect matchings can be removed from the surviving points in the toric diagram. For example, all but $p_{15}$ are deleted in the point that originally contains $p_7,\ldots,p_{18}$. Similarly, $p_{19}$ is removed while leaving $p_{20}$ for that point. \fref{toric_resolution_to_SPP} shows the final toric diagram.

Having established the vevs that implement the desired partial resolution to the SPP, the associated brane tiling is obtained by deleting the corresponding edges in \fref{dimer_conifold_Z2xZ2}. When doing so, a pair of 2-valent nodes is generated. Such nodes correspond to mass terms in the superpotential. Massive fields are easily integrated out in terms of brane tilings \cite{Franco:2005rj}. The final result is precisely the brane tiling in \fref{dimer_SPP}, which corresponds to the quiver in \fref{quiver_SPP} and the superpotential \eref{W_SPP}.

\subsection{Brane Tiling Consistency}

Not every bipartite graph on a 2-torus corresponds to a {\it consistent brane tiling} and hence defines a well-behaved $4d$ $\mathcal{N}=1$ gauge theory. In fact, higgsing of consistent brane tilings can lead to inconsistent ones. It thus becomes imperative to check the consistency of the brane tilings generated via partial resolution.

Inconsistent brane tilings may naively seem to correspond to toric CY$_3$'s, but fail more subtly. The problems of inconsistent tilings manifest at all levels: the gauge theory, the D-brane configuration and its algebraic description. By now, this subject has been studied in depth and is well understood. We refer the interested readers to \cite{Hanany:2005ss,Gulotta:2008ef,2009arXiv0901.4662B,ishii2010note,davison2011consistency,bocklandt2012consistency,Hanany:2015tgh} and references therein for detailed discussions.

Consistency can be determined using multiple diagnostics, all of which are closely related. They range from physical considerations regarding the positivity of $R$-charges to graph-theoretic tests based on intersection properties of zig-zag paths. The latter condition is closely related to the concept of {\it reducibility} of brane tilings. A brane tiling is reducible, or equivalently inconsistent, if the number of faces can be decreased by deleting edges while preserving the toric diagram. On the other hand, the number of gauge groups should be equal to the area of the toric diagram, measured in terms of elementary triangles. These two points lead to a simple criterion for consistency of brane tilings, which is particularly well-suited for partial resolution. A brane tiling is inconsistent whenever the number of faces is larger than the area of the toric diagram. When this occurs, the brane tiling can be cured and turned into a consistent one by removing certain edges, i.e. by turning on vevs, without modifying the toric diagram.

This clarifies how inconsistent brane tilings can arise when partial resolution is not properly implemented. Sometimes, given an initial toric diagram and its corresponding brane tiling, a target toric diagram may be obtained by turning on an incomplete collection of vevs.\footnote{This was not a possibility in the example discussed in the previous section.} To avoid inconsistent tilings we should make sure that the set of vevs not only gives rise to the desired toric diagram but that it is also {\it maximal}.

\newpage

\section{Existing Classifications}

\label{section_existing_classifications}

A plethora of explicit brane tilings have been constructed in the literature. Below we summarize the existing systematic classifications of {\it classes} of models. Several additional scattered examples exist.

\begin{itemize}
\item Del Pezzo surfaces \cite{Franco:2005rj}. The brane tilings for all toric phases for cones over toric del Pezzo surfaces $dP_n$, $n=0,\ldots 3$, have been classified. Even before the development of brane tilings, the corresponding gauge theories were determined in \cite{Beasley:1999uz,Feng:2000mi,Feng:2001xr,Feng:2001bn,Beasley:2001zp}.
\item Abelian orbifolds of $\mathbb{C}^3$ \cite{Hanany:2010cx,Davey:2010px,Hanany:2010ne,Davey:2011dd,Hanany:2011iw}. It is straightforward to construct the brane tilings for abelian orbifolds of arbitrary geometries by appropriately enlarging the unit cell. The geometric action of the orbifold group is encoded in the periodicity conditions. However, a systematic classification of the orbifold possibilities of geometries beyond $\mathbb{C}^3$ does not currently exist.
\item The $Y^{p,q}$ \cite{Benvenuti:2004dy} and $L^{a,b,c}$ \cite{Franco:2005sm,Butti:2005sw, Benvenuti:2005ja} infinite families. In fact the $Y^{p,q}$ theories are fully contained within the $L^{a,b,c}$ class. The toric diagrams for these geometries have four external edges. Explicit metrics for the $Y^{p,q}$ and $L^{a,b,c}$ Sasaki-Einstein manifolds were introduced in \cite{Gauntlett:2004zh,Gauntlett:2004yd,Cvetic:2005ft,Martelli:2005wy}. The construction of the gauge theories for these geometries had a substantial impact on the AdS$_5$/CFT$_4$ correspondence with $\mathcal{N}=1$ supersymmetry. It allowed refined tests of the correspondence for the infinite classes of dual pairs.
\item The $X^{p,q}$ family \cite{Hanany:2005hq}. The toric diagrams for these geometries have five external edges. While this classification was not performed in the language of brane tilings, it is straightforward to translate it.
\item Finally, \cite{Davey:2009bp} classified all brane tilings up to six superpotential terms. These theories are substantially simpler than the ones studied in this paper.
\end{itemize}

\section{The Geometries}

\label{section_geometries}

A primary goal of this paper is to construct brane tilings for all toric CY 3-folds with toric diagrams up to area 8. The relative simple cases of area 1 to 5 have been extensively studied and brane tilings are known for all of them. We will thus concentrate on areas 6 to 8.  As mentioned earlier, part of our motivation for focusing on these geometries has to do with applications to local string phenomenology along the lines of \cite{Krippendorf:2010hj, Dolan:2011qu}.

\newpage

The first step in our survey is thus to identify these geometries, i.e. their toric diagrams. To do so, we need to determine, for every area, all the $SL(2,\mathbb{Z})$ inequivalent convex polytopes in $\mathbb{Z}^2$. Interestingly, these toric diagrams have only been established for areas 6 and 7 by mathematicians \cite{arnold,LiuZong}, whose results we reproduce. We find that the number of independent toric diagrams with area 6, 7 and 8 are 13, 11 and 27, respectively. Below we present all independent toric diagrams up to area 8. For completeness, we include those for areas 1 to 5. For every toric diagram we provide an arbitrary triangulation, in order to make its area manifest.

\vspace{-.3cm}
\begin{center}
\small
\begin{longtable}{cccccccccccc}
\hline \hline
\multicolumn{4}{c|}{\includegraphics[width=0.8cm]{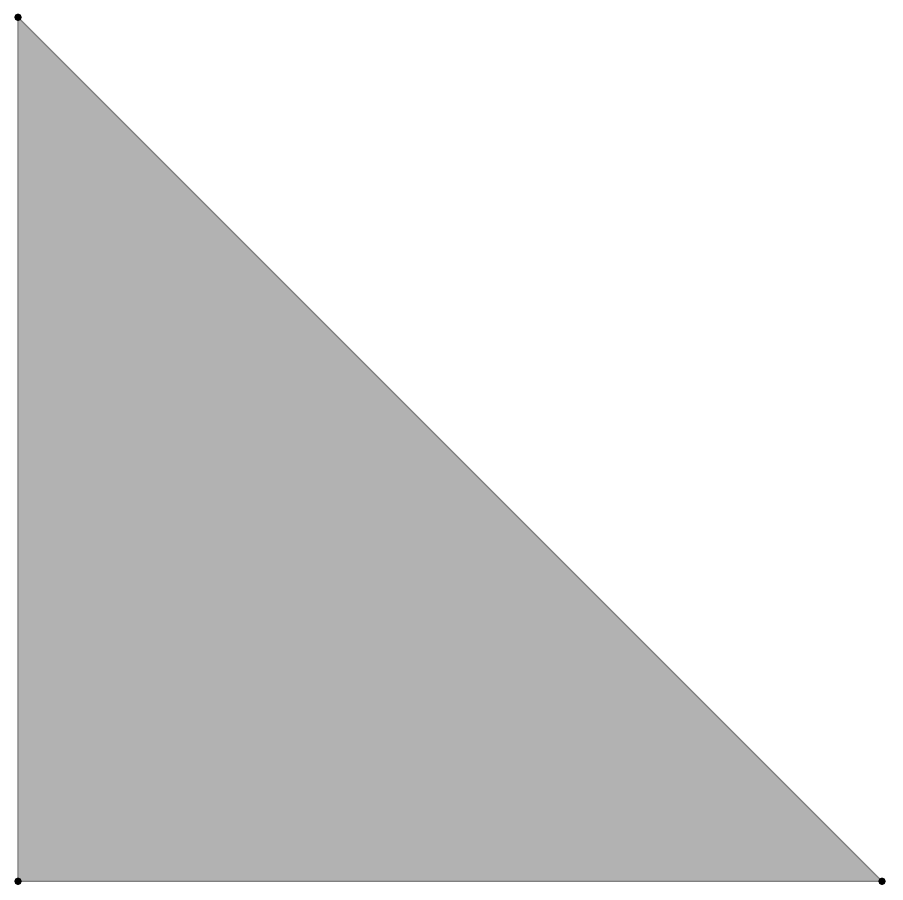}} &
\multicolumn{4}{c}{\includegraphics[width=1.6cm]{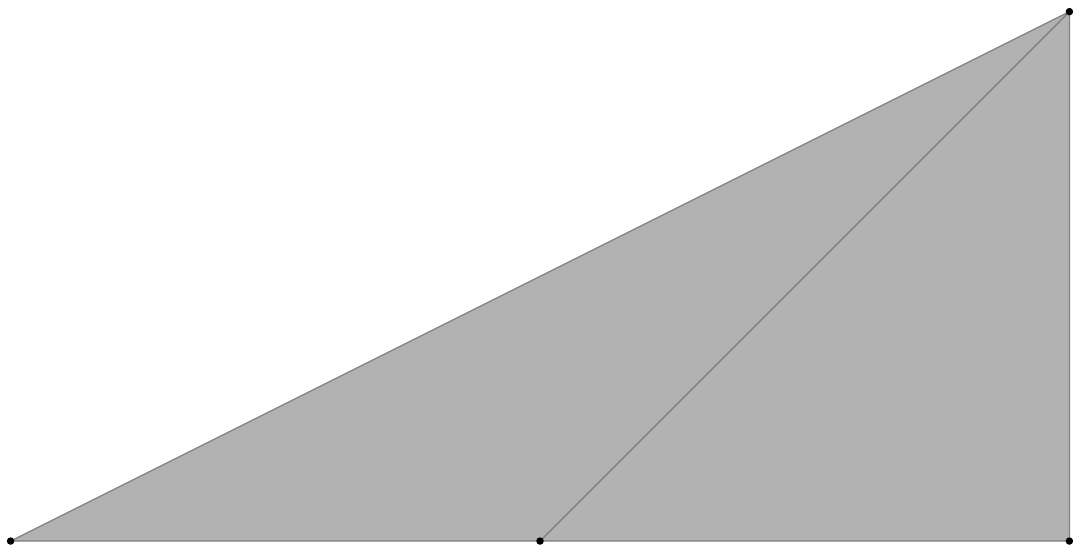}} &
\multicolumn{4}{c}{ \includegraphics[width=0.8cm]{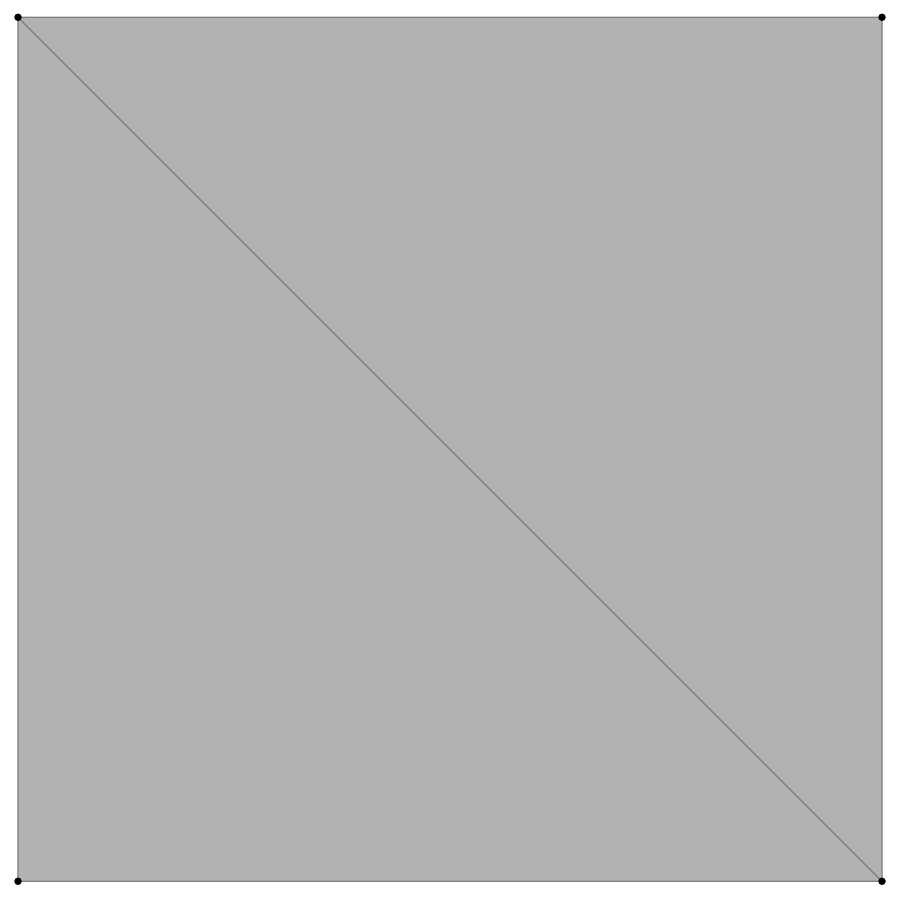}} \\
\multicolumn{4}{c|}{1} &
\multicolumn{4}{c}{1} &
\multicolumn{4}{c}{2} \\ \hline
\multicolumn{4}{c}{ \includegraphics[width=2.4cm]{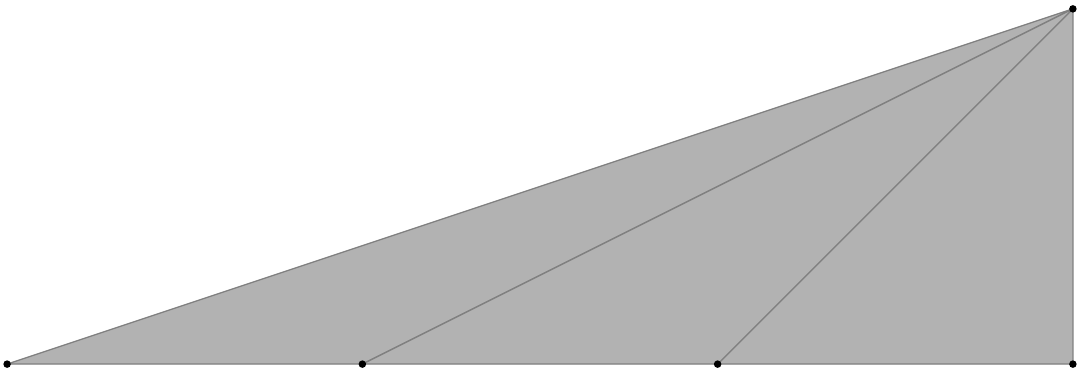}} &
\multicolumn{4}{c}{ \includegraphics[width=1.6cm]{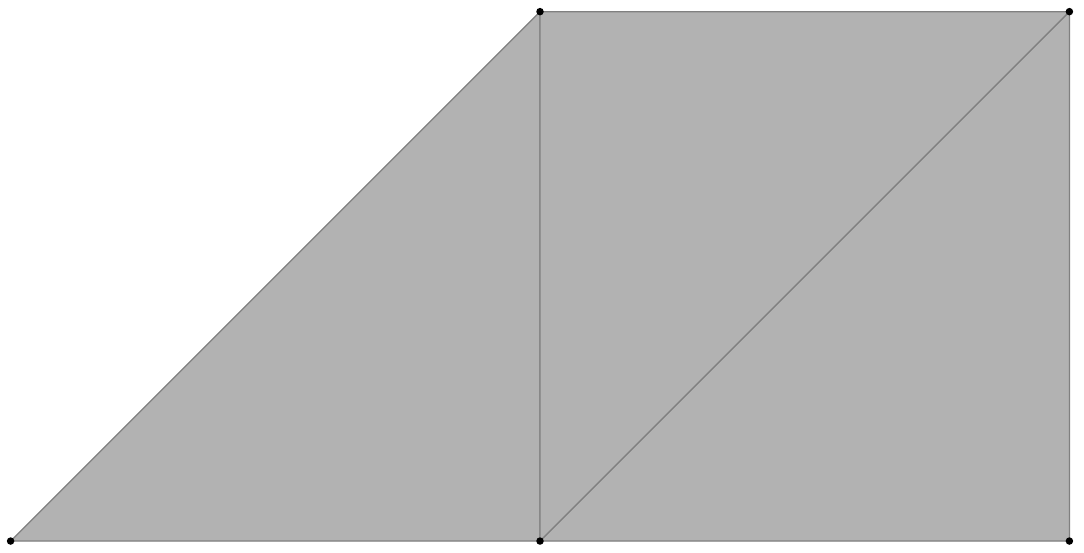}} &
\multicolumn{4}{c}{  \includegraphics[width=1.6cm]{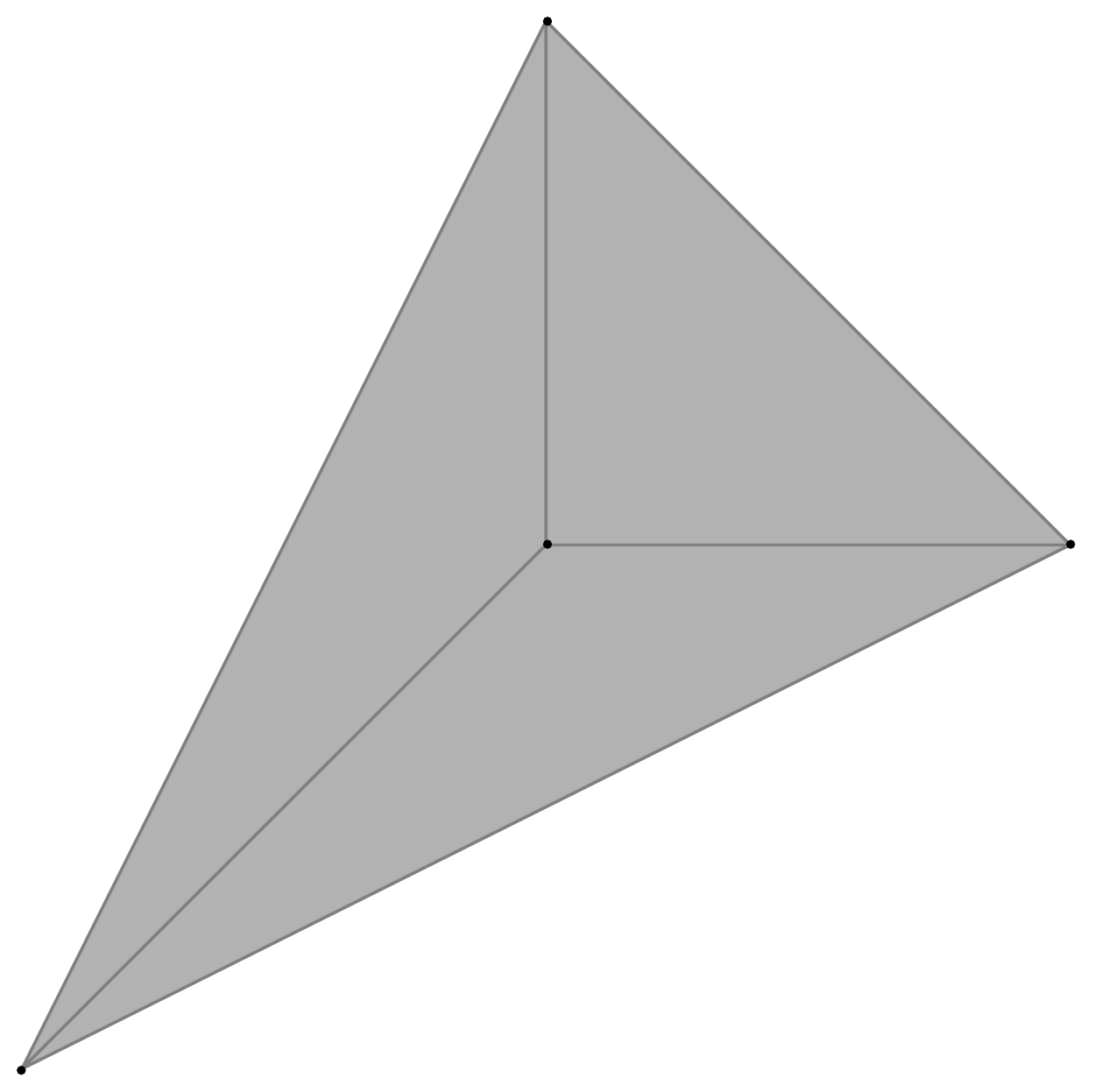}} \\
\multicolumn{4}{c}{1} &
\multicolumn{4}{c}{2} &
\multicolumn{4}{c}{3} \\ \hline \hline
\\ \caption{Toric diagrams of areas 1, 2 and 3.}
\label{table:convpoly123}
\end{longtable}
\normalsize
\end{center}

\vspace{-1.8cm}
\begin{center}
\small
\begin{longtable}{cccccccccccc}
\hline \hline 
\multicolumn{4}{c}{ \includegraphics[width=3.2cm]{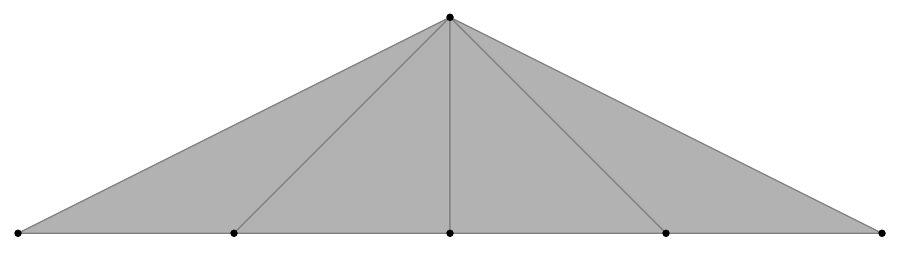}} &
\multicolumn{4}{c}{\includegraphics[width=2.4cm]{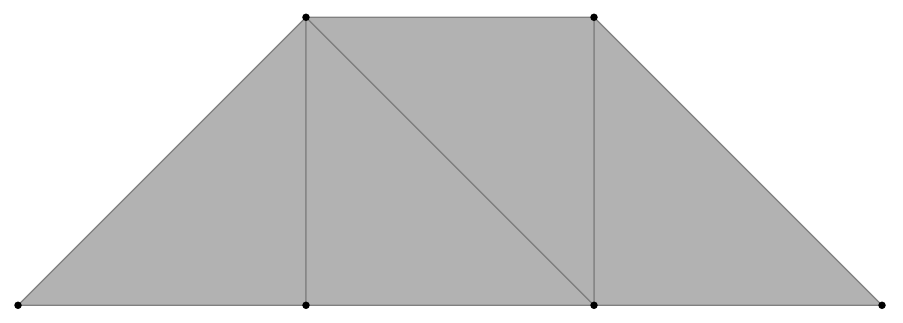}} &
\multicolumn{4}{c}{\includegraphics[width=1.6cm]{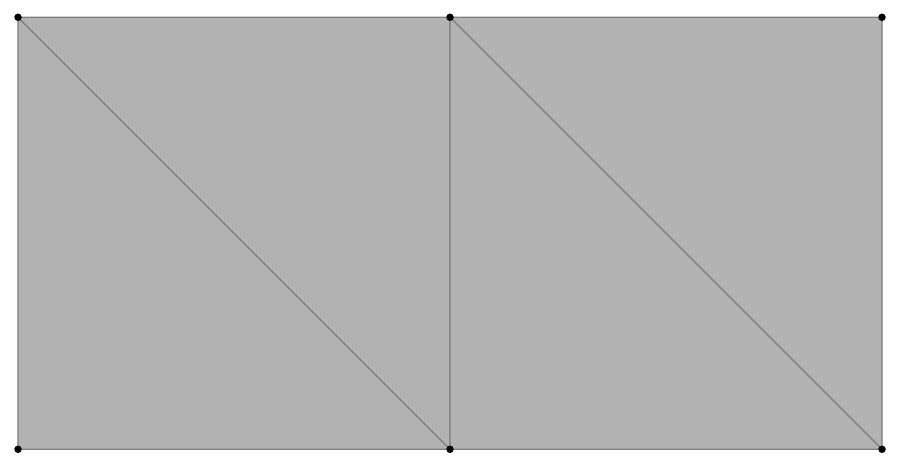}} \\
\multicolumn{4}{c}{1} &
\multicolumn{4}{c}{2} &
\multicolumn{4}{c}{3} \\ \hline 
\multicolumn{3}{c}{\includegraphics[width=1.6cm]{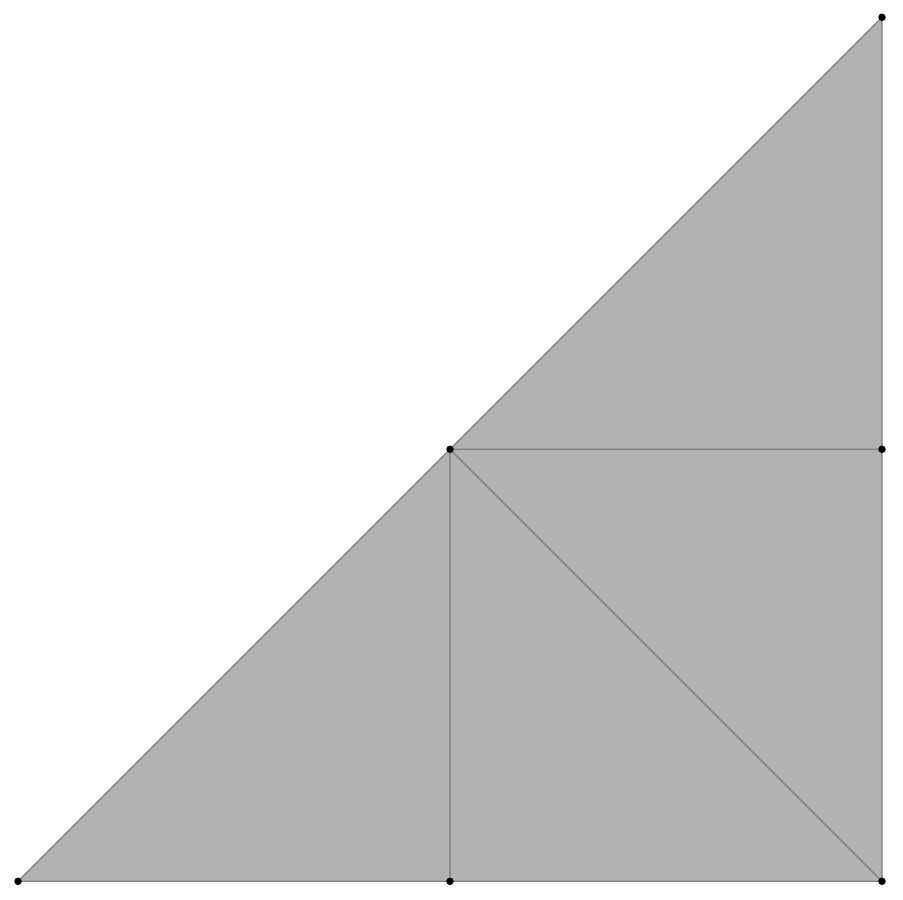}} &
\multicolumn{3}{c}{\includegraphics[width=1.6cm]{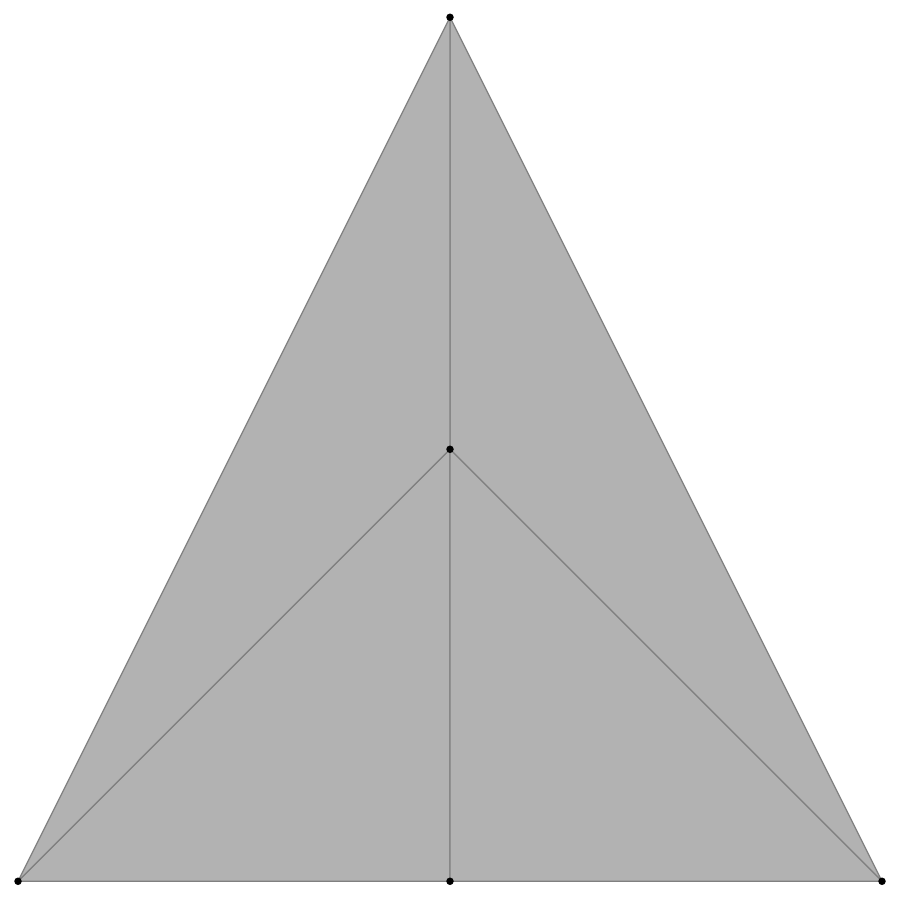}} &
\multicolumn{3}{c}{\includegraphics[width=1.6cm]{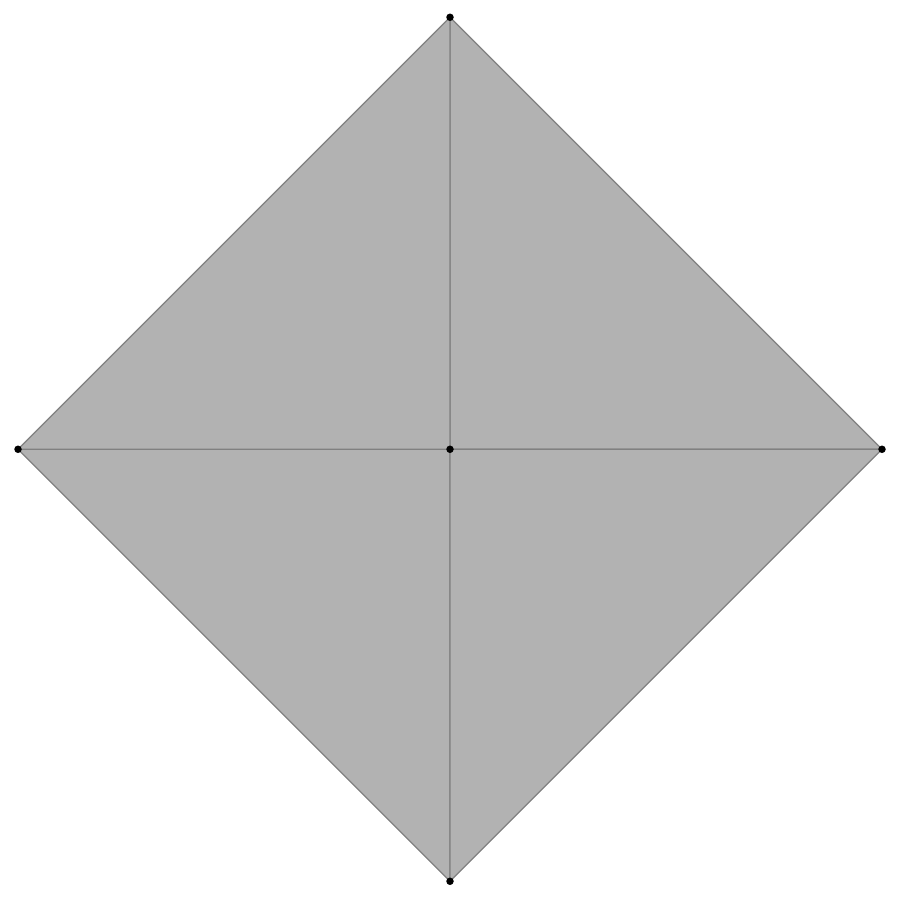}} &
\multicolumn{3}{c}{\includegraphics[width=1.6cm]{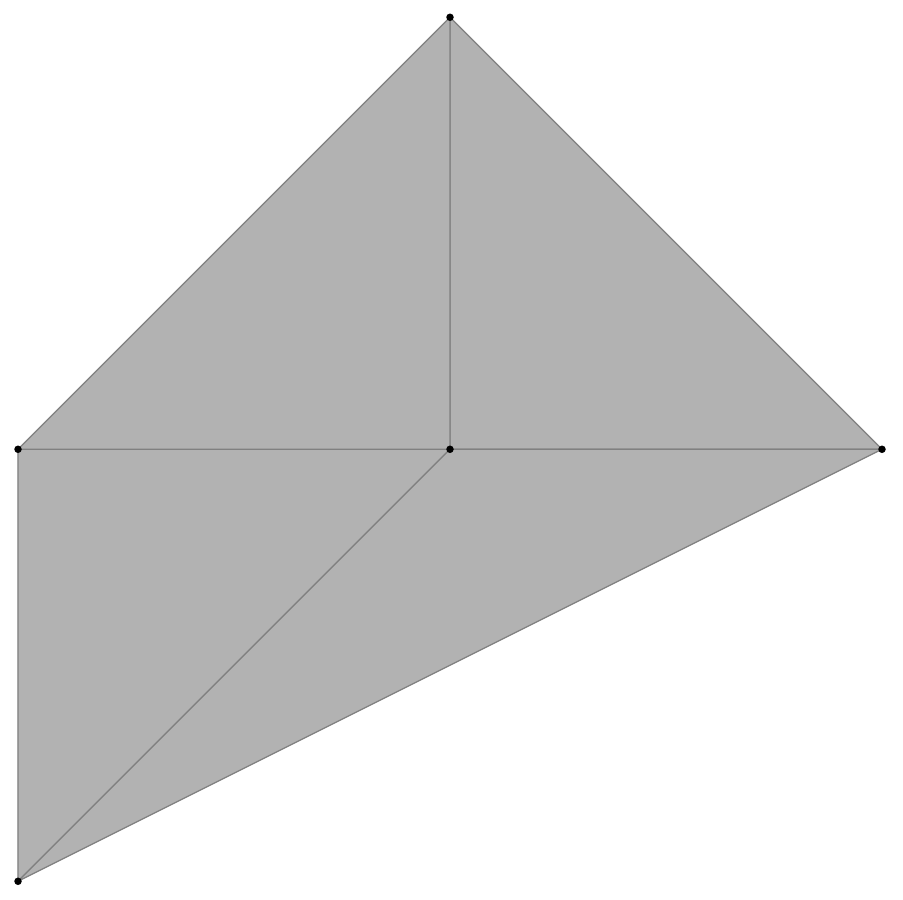}} \\
\multicolumn{3}{c}{4} &
\multicolumn{3}{c}{5} &
\multicolumn{3}{c}{6} &
\multicolumn{3}{c}{7} \\ \hline \hline
\\ \caption{Toric diagrams of area 4.}
\label{table:convpoly4}
\end{longtable}
\normalsize
\end{center}

\vspace{-1.8cm}
\begin{center}
\small
\begin{longtable}{cccccccccccc}
\hline \hline 
\multicolumn{6}{c}{ \includegraphics[width=4cm]{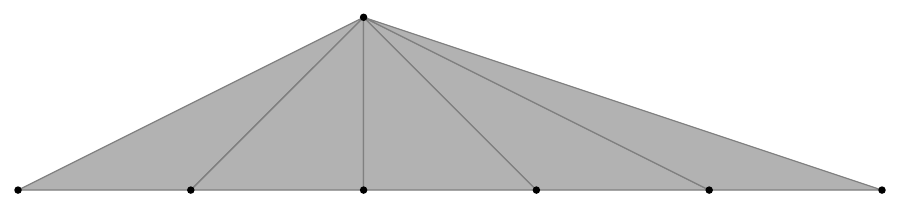}} &
\multicolumn{6}{c}{\includegraphics[width=3.2cm]{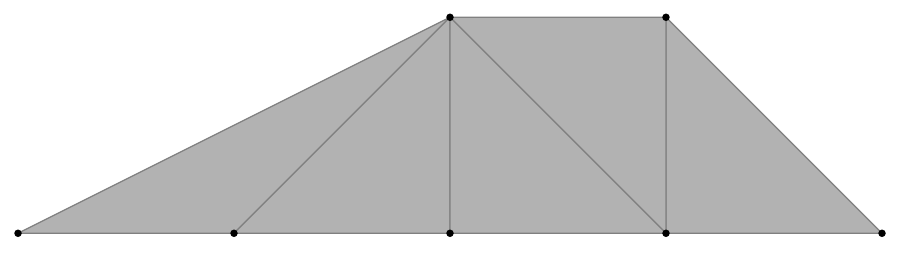}} \\
\multicolumn{6}{c}{1} &
\multicolumn{6}{c}{2} \\ \hline
\multicolumn{3}{c}{\includegraphics[width=2.4cm]{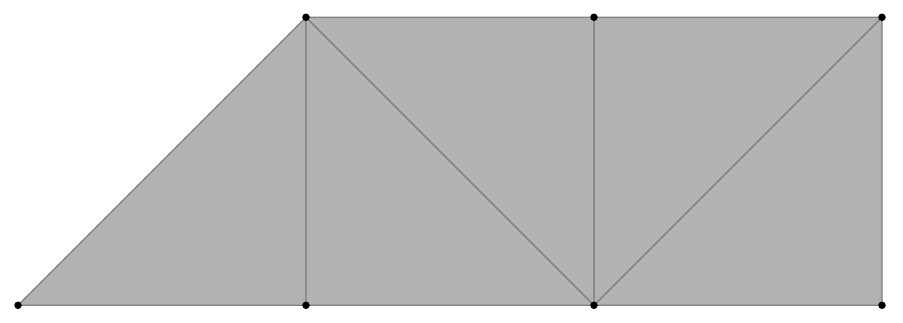}} &
\multicolumn{3}{c}{ \includegraphics[width=1.6cm]{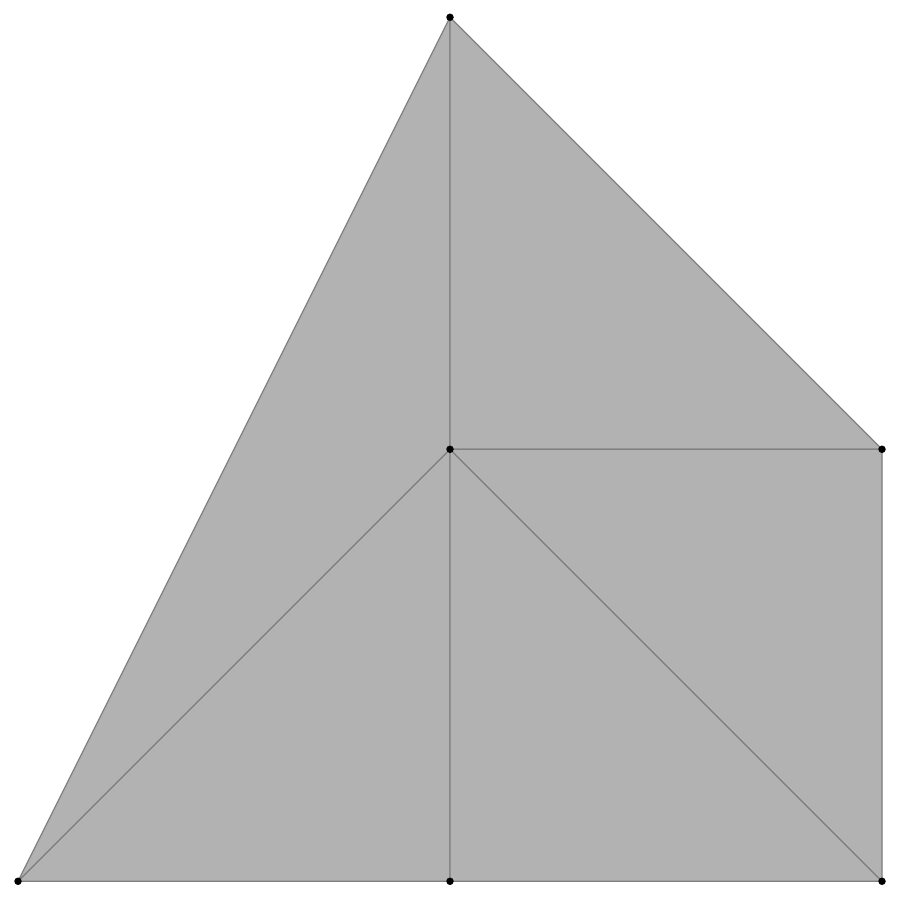}} &
\multicolumn{3}{c}{\includegraphics[width=1.6cm]{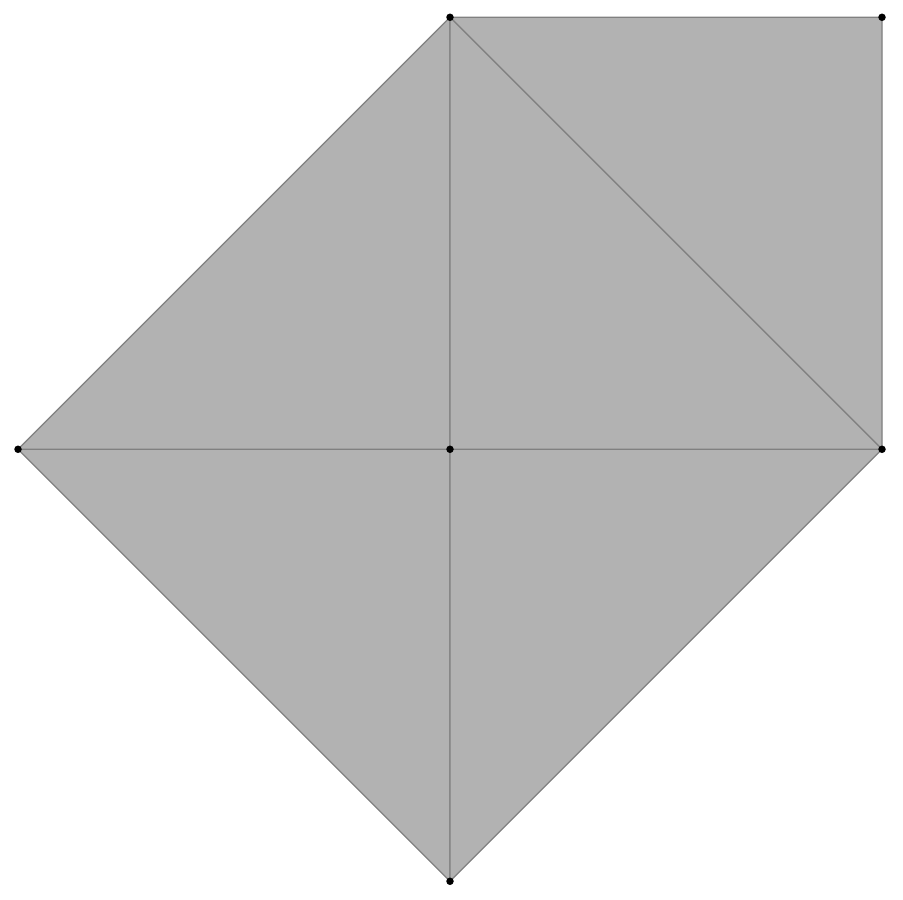}} &
\multicolumn{3}{c}{\includegraphics[width=2.4cm]{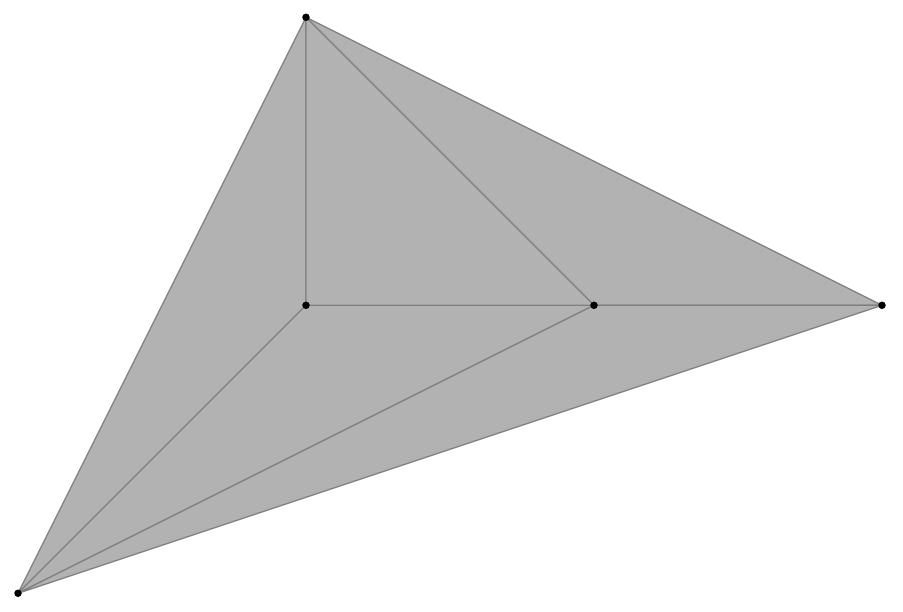}} \\ 
\multicolumn{3}{c}{3} &
\multicolumn{3}{c}{4} &
\multicolumn{3}{c}{5} &
\multicolumn{3}{c}{6}  \\ \hline \hline
\\ \caption{Toric diagrams of area 5.}
\label{table:convpoly5}
\end{longtable}
\normalsize
\end{center}

\vspace{-1.5cm}
\begin{center}
\small
\begin{longtable}{cccccccccccc}
\hline \hline 
\multicolumn{6}{c}{ \includegraphics[width=4.8cm]{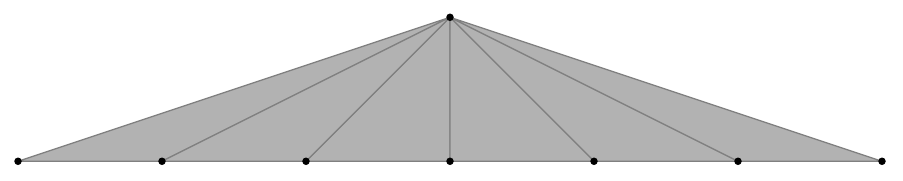}} &
\multicolumn{6}{c}{\includegraphics[width=4cm]{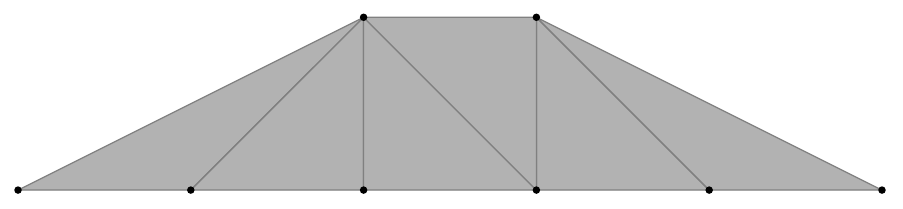}} \\
\multicolumn{6}{c}{1} &
\multicolumn{6}{c}{2} \\ \hline
\multicolumn{3}{c}{ \includegraphics[width=3.2cm]{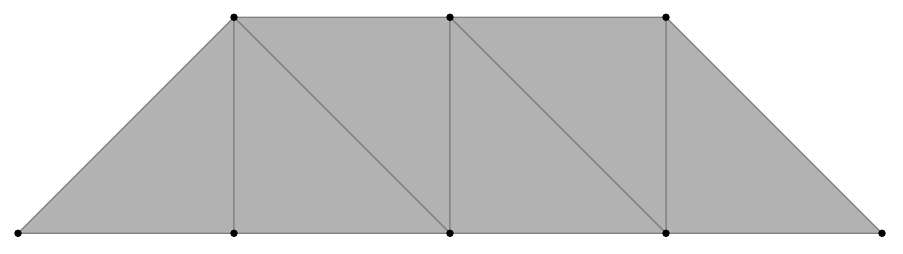}} &
\multicolumn{6}{c}{ \includegraphics[width=2.4cm]{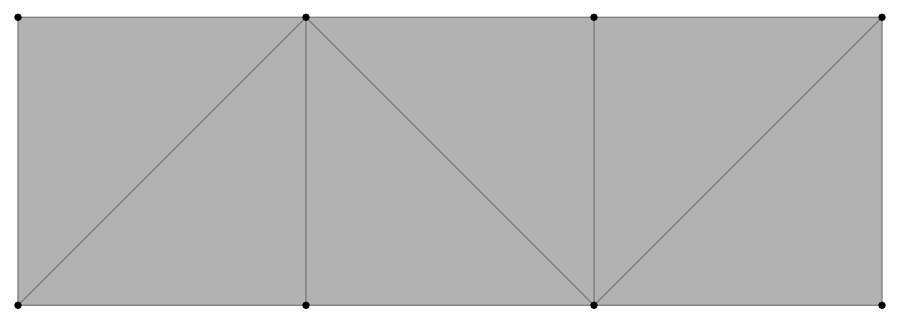}} &
\multicolumn{3}{c}{ \includegraphics[width=2.4cm]{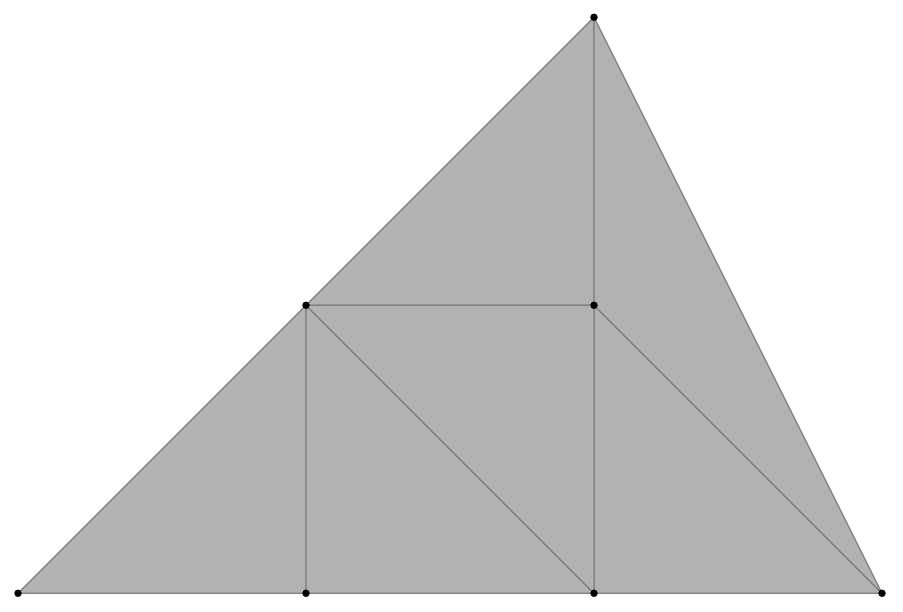}} \\
\multicolumn{3}{c}{3} &
\multicolumn{6}{c}{4} &
\multicolumn{3}{c}{5} \\ \hline
\multicolumn{3}{c}{  \includegraphics[width=1.6cm]{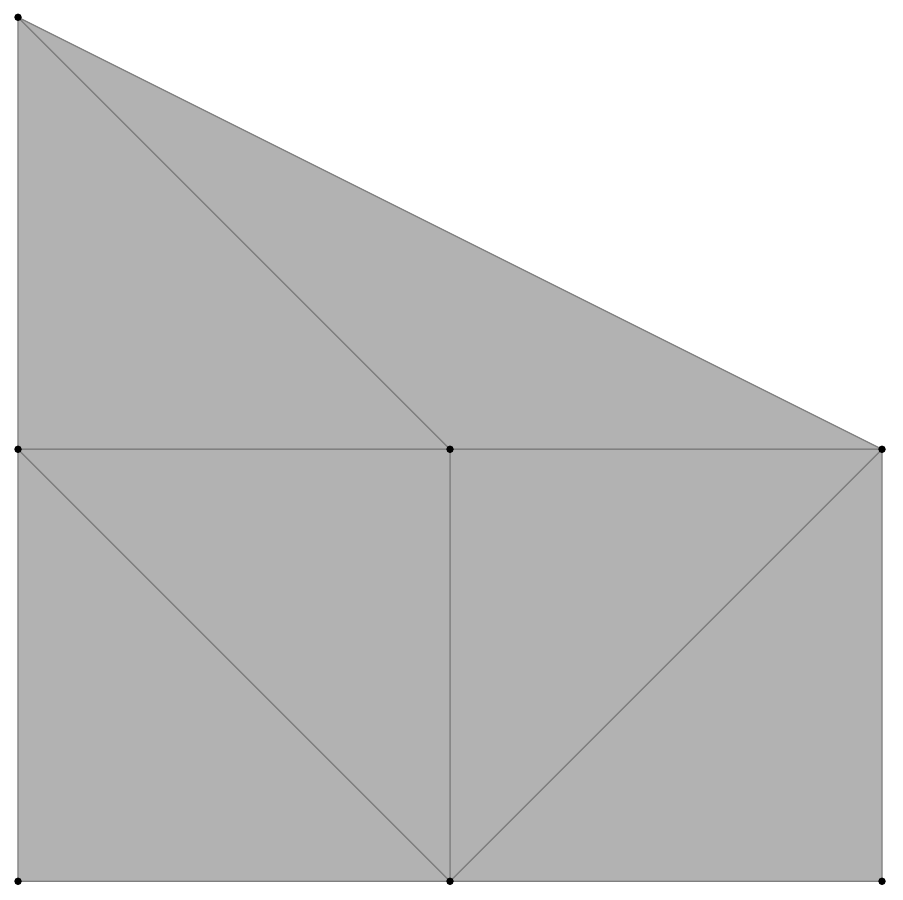}} &
\multicolumn{3}{c}{\includegraphics[width=1.6cm]{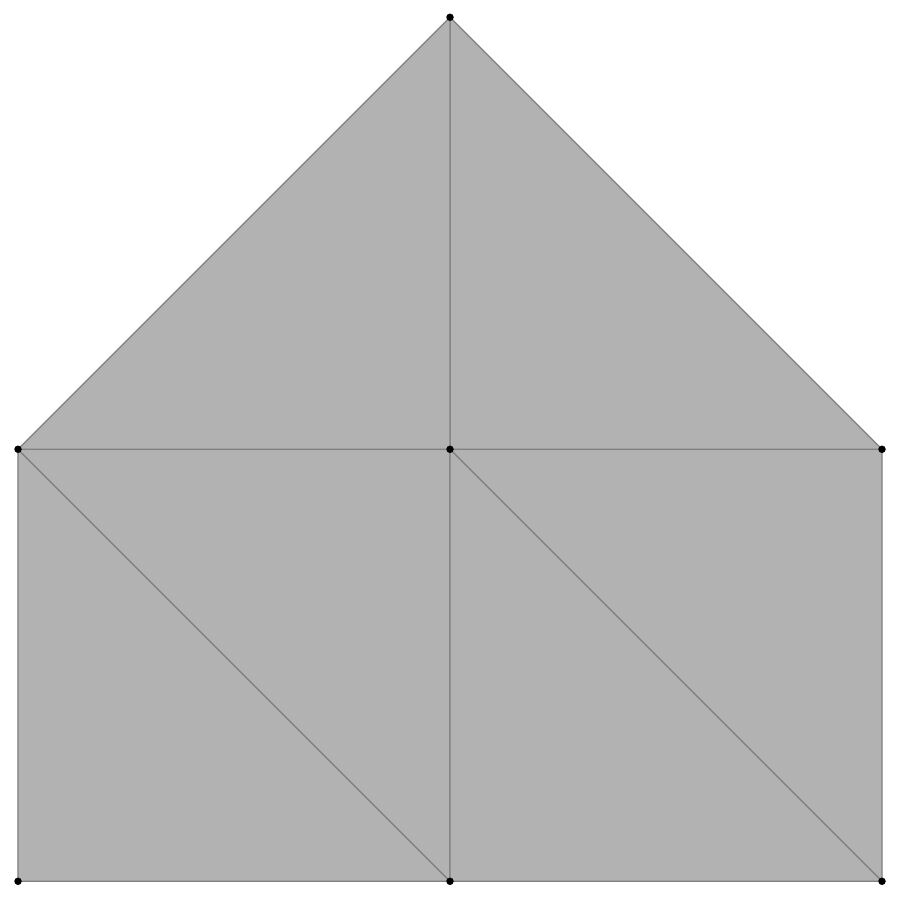}} &
\multicolumn{3}{c}{\includegraphics[width=1.6cm]{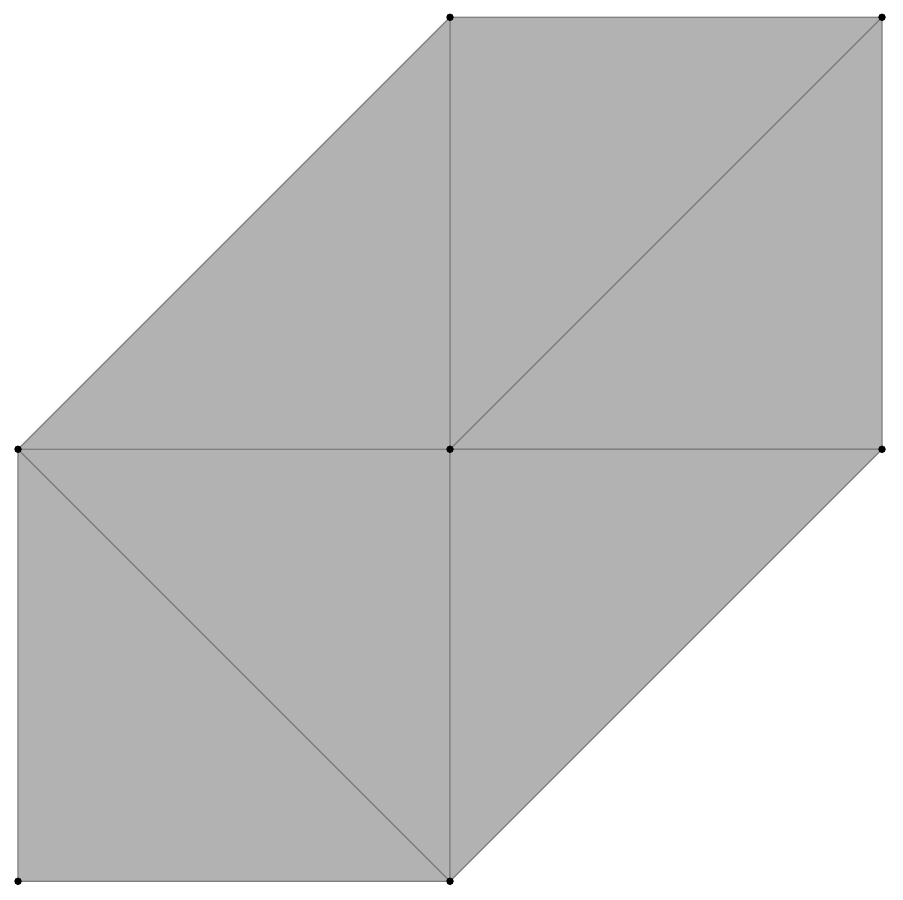}} &
\multicolumn{3}{c}{\includegraphics[width=2.4cm]{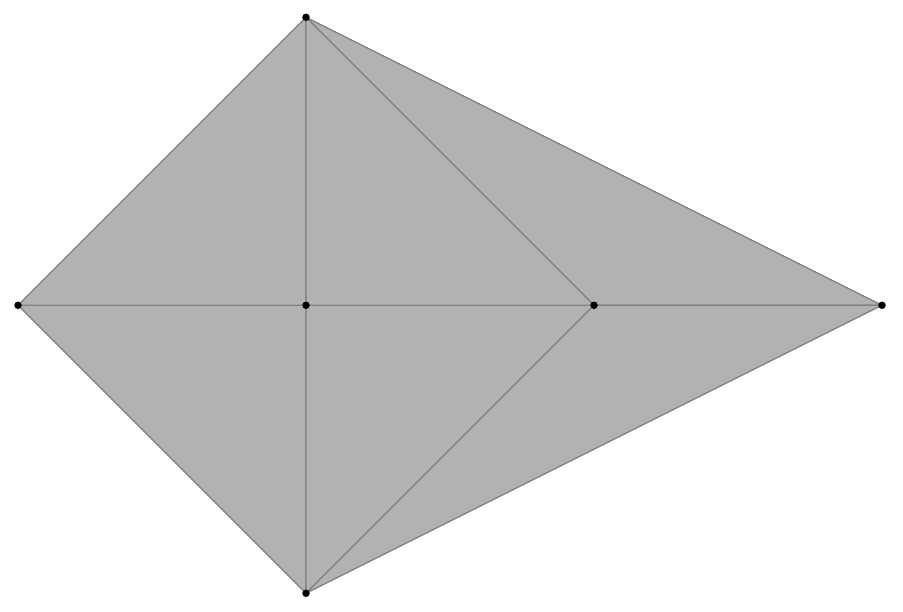}} \\
\multicolumn{3}{c}{6} &
\multicolumn{3}{c}{7} &
\multicolumn{3}{c}{8} &
\multicolumn{3}{c}{9} \\ \hline
\multicolumn{3}{c}{\includegraphics[width=2.4cm]{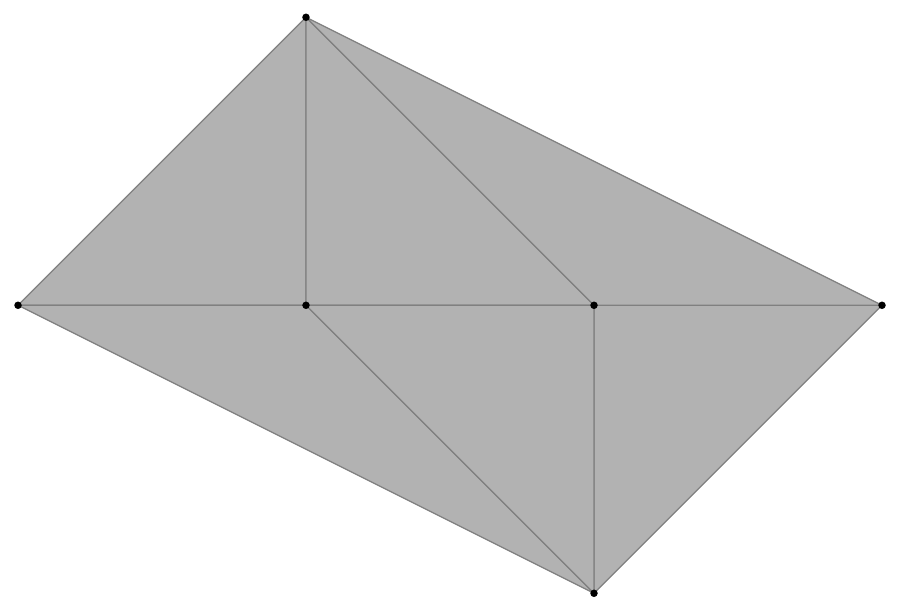}} &
\multicolumn{3}{c}{ \includegraphics[width=2.4cm]{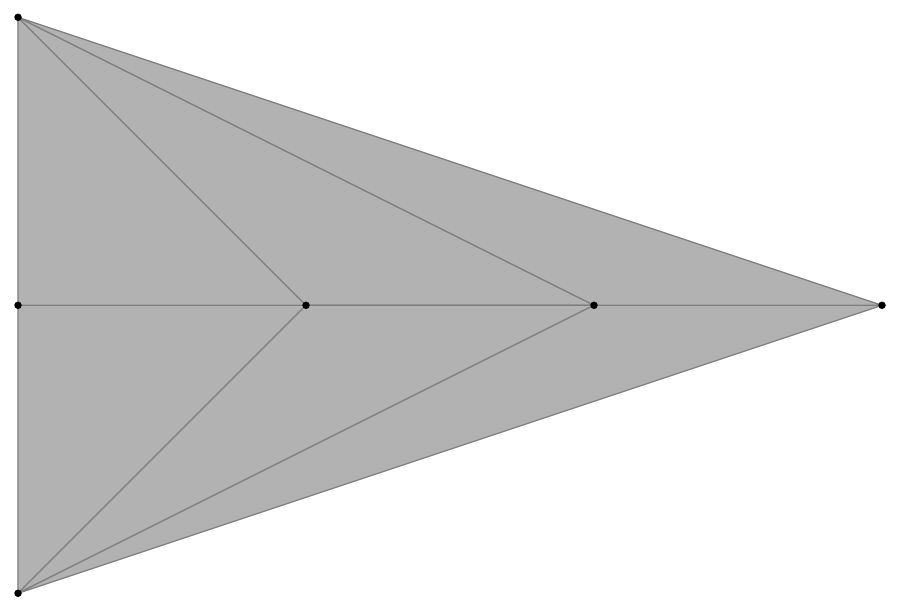}} &
\multicolumn{3}{c}{\includegraphics[width=2.4cm]{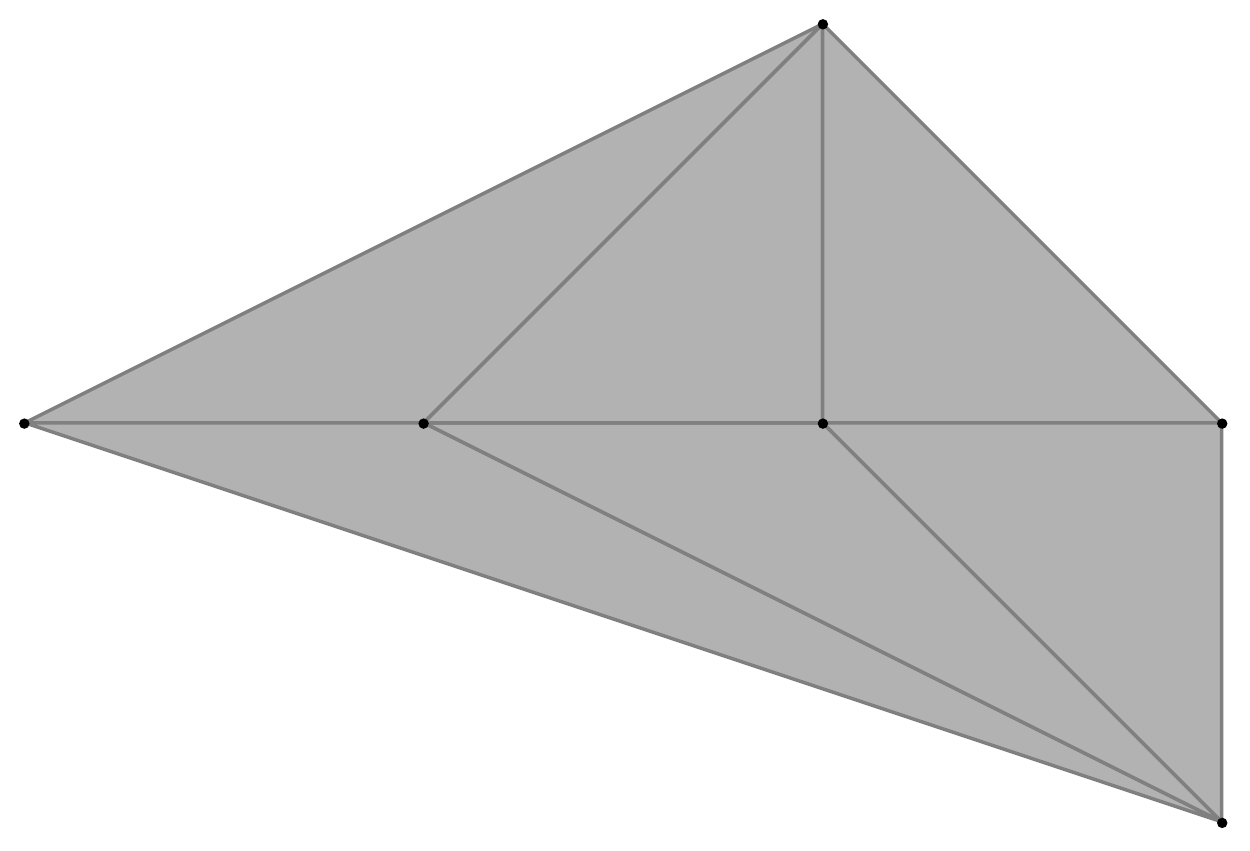}} &
\multicolumn{3}{c}{\includegraphics[width=2.4cm]{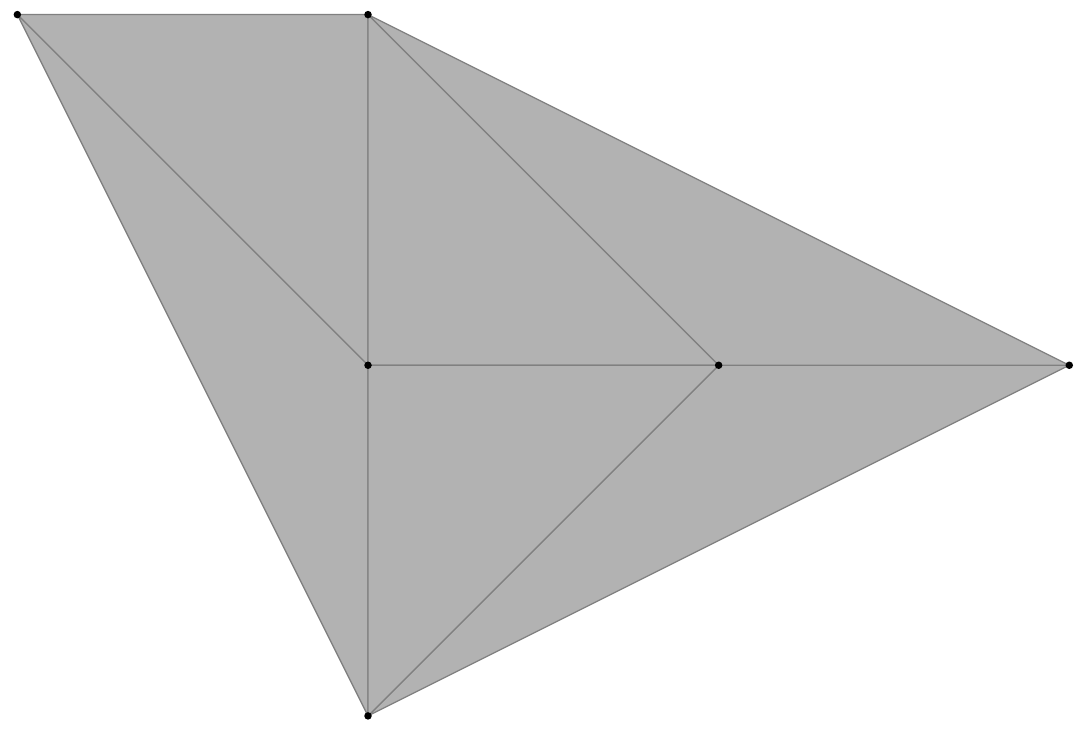}} \\
\multicolumn{3}{c}{10} &
\multicolumn{3}{c}{11} &
\multicolumn{3}{c}{12} &
\multicolumn{3}{c}{13} \\ \hline \hline
\\ \caption{Toric diagrams of area 6.}
\label{table:convpoly6}
\end{longtable}
\normalsize
\end{center}

\vspace{-1.7cm}
\begin{center}
\small
\begin{longtable}{cccccccccccc}
\hline \hline 
\multicolumn{6}{c}{ \includegraphics[width=5.6cm]{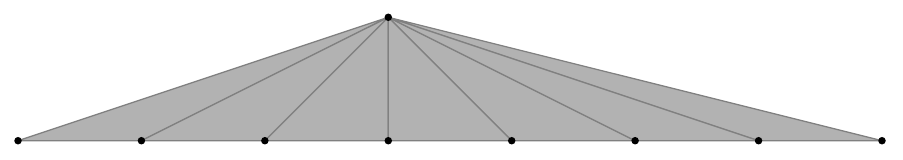}} &
\multicolumn{6}{c}{\includegraphics[width=4.8cm]{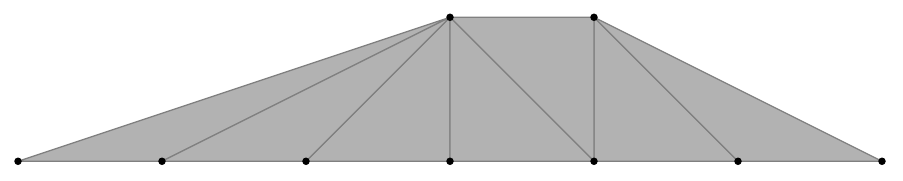}} \\
\multicolumn{6}{c}{1} &
\multicolumn{6}{c}{2} \\ \hline
\multicolumn{6}{c}{\includegraphics[width=4cm]{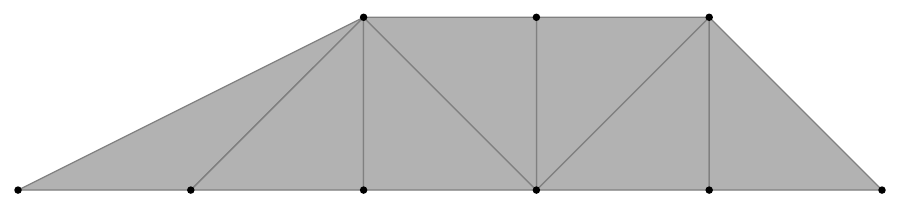}} &
\multicolumn{6}{c}{\includegraphics[width=3.2cm]{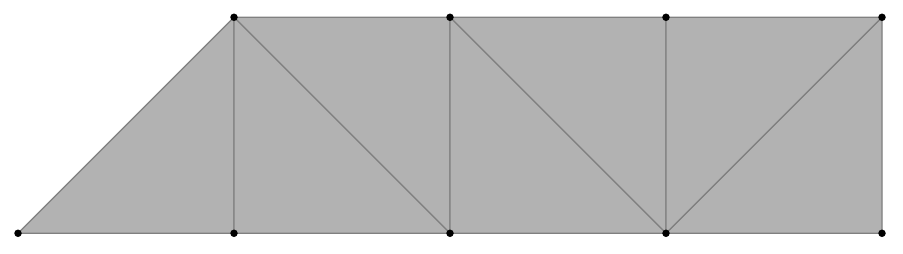}} \\
\multicolumn{6}{c}{3} &
\multicolumn{6}{c}{4} \\ \hline
\multicolumn{3}{c}{\includegraphics[width=2.4cm]{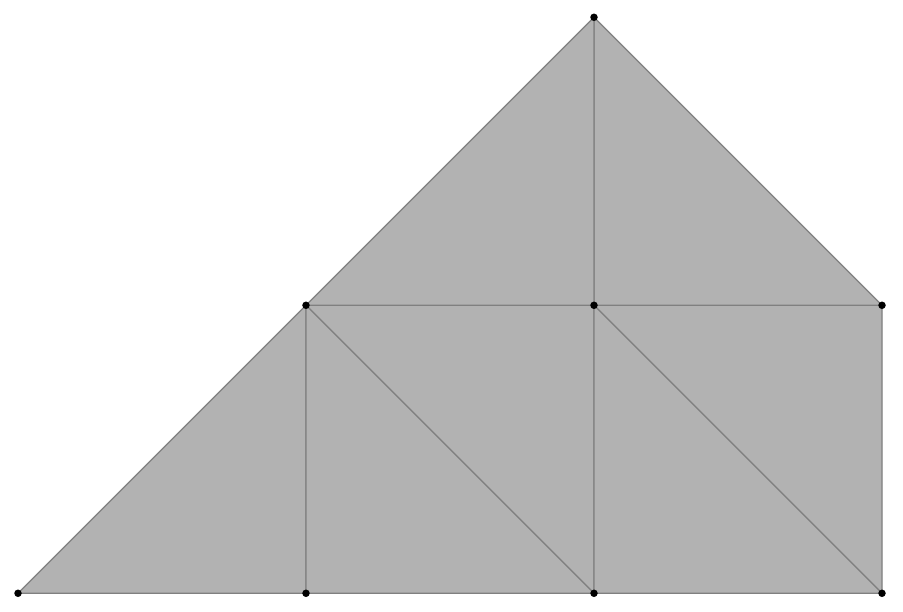}} &
\multicolumn{6}{c}{\includegraphics[width=1.6cm]{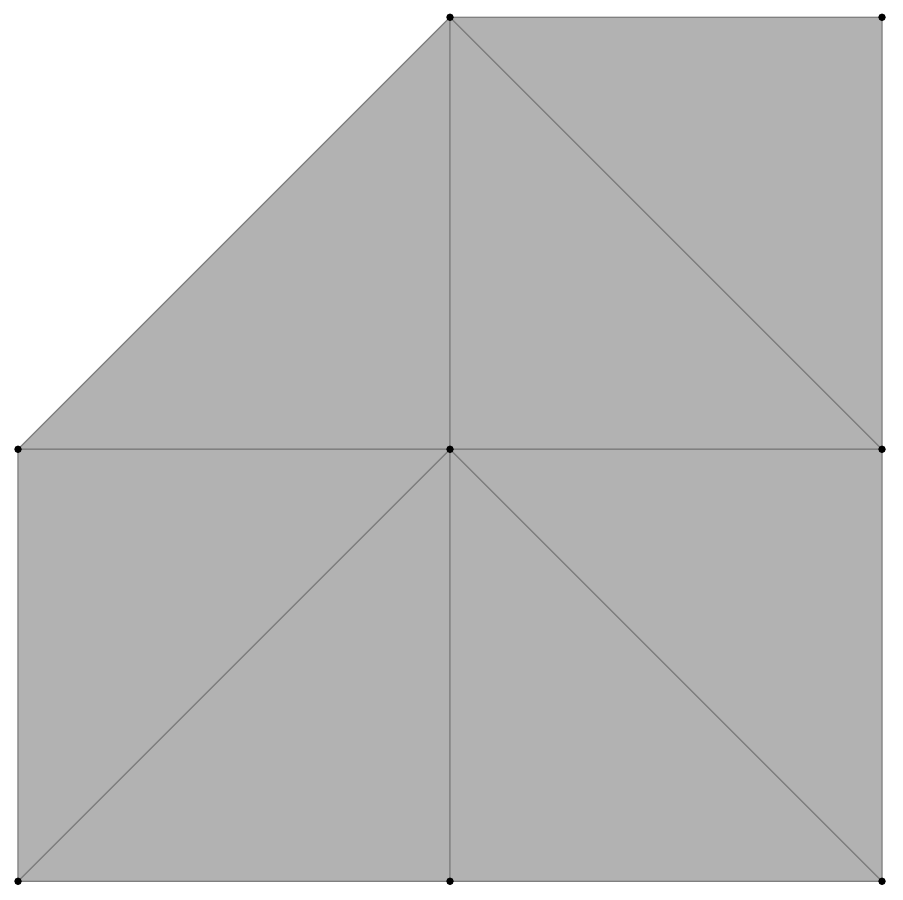}} &
\multicolumn{3}{c}{\includegraphics[width=2.4cm]{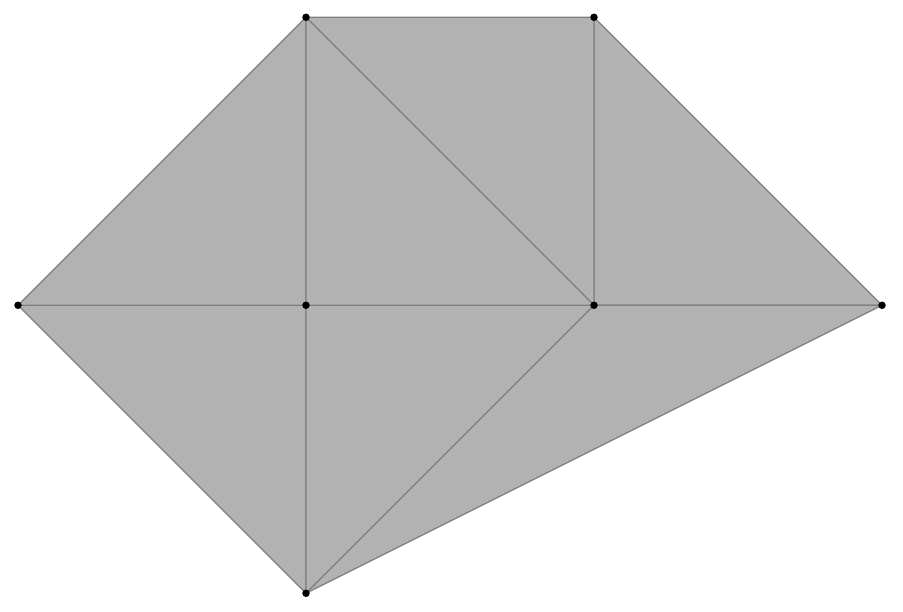}} \\
\multicolumn{3}{c}{5} &
\multicolumn{6}{c}{6} &
\multicolumn{3}{c}{7} \\ \hline
\multicolumn{3}{c}{ \includegraphics[width=2.4cm]{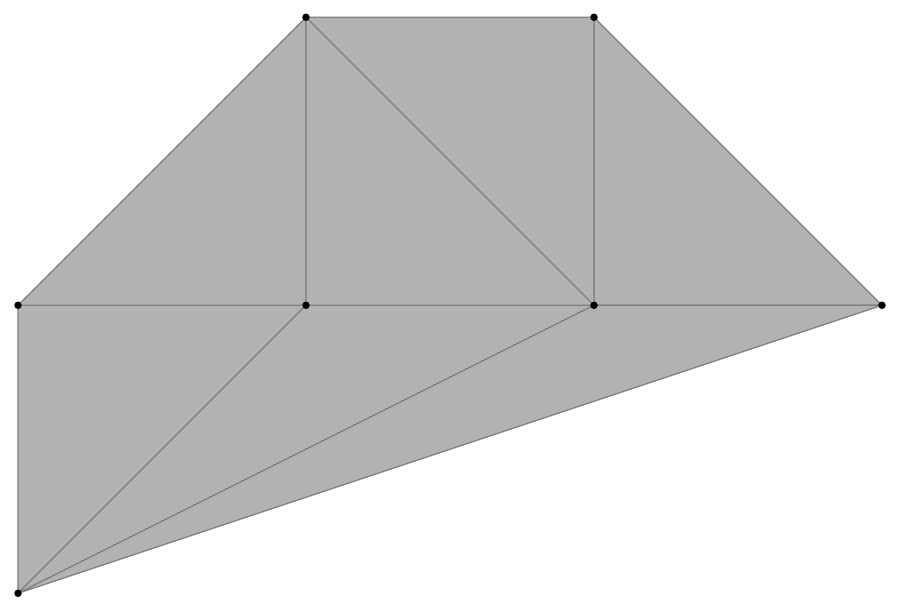}} &
\multicolumn{3}{c}{\includegraphics[width=1.6cm]{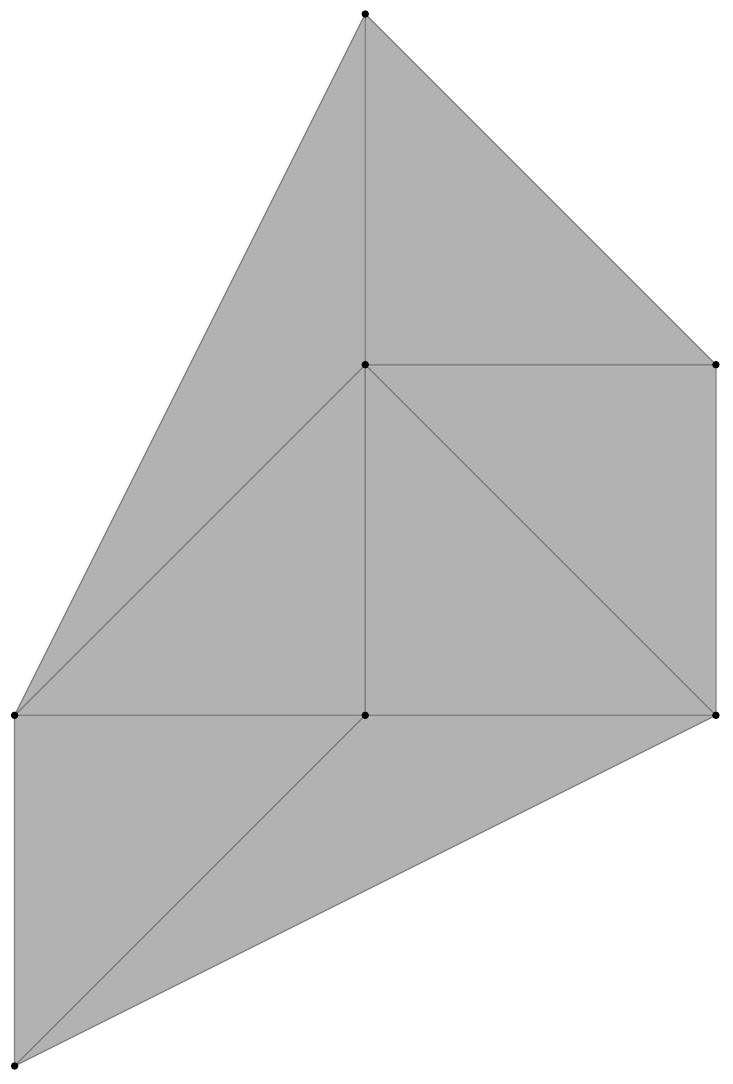}} &
\multicolumn{3}{c}{\includegraphics[width=1.6cm]{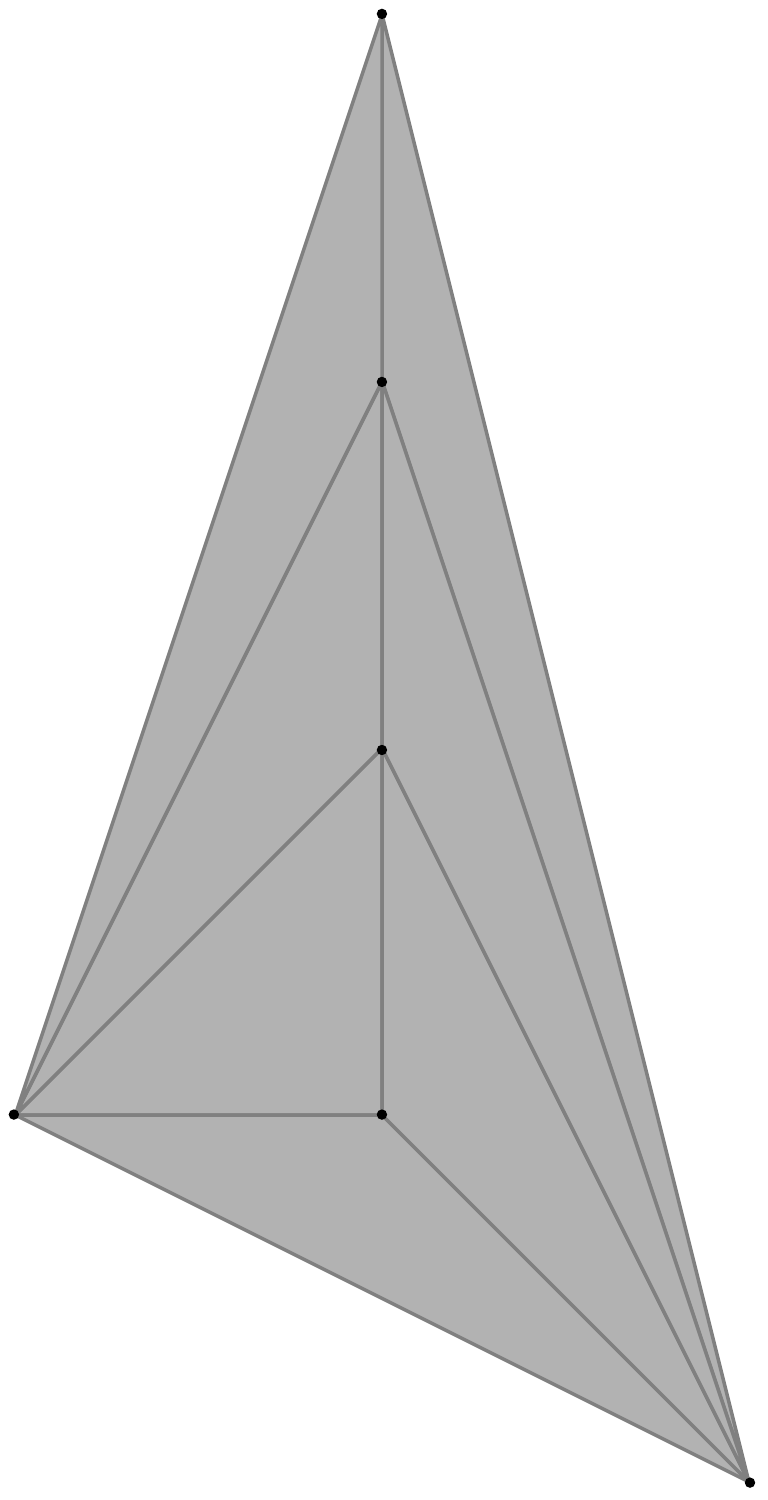}} &
\multicolumn{3}{c}{ \includegraphics[width=2.4cm]{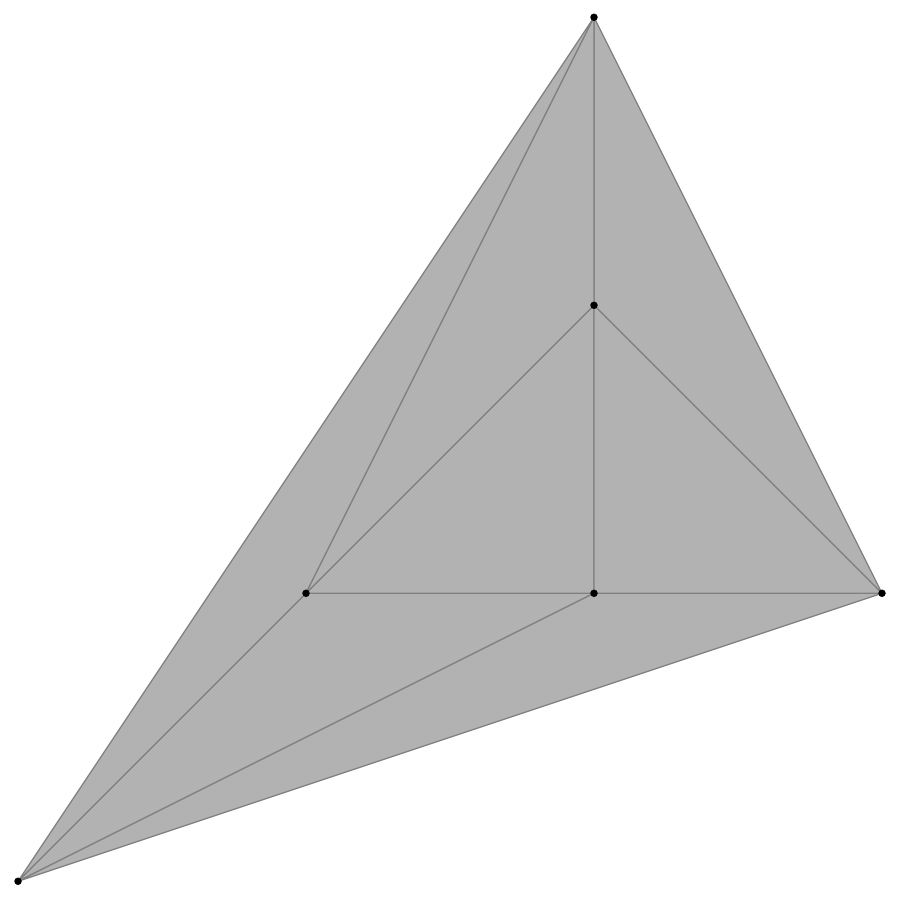}} \\
\multicolumn{3}{c}{8} &
\multicolumn{3}{c}{9} &
\multicolumn{3}{c}{10} &
\multicolumn{3}{c}{11} \\ \hline \hline
\\ \caption{Toric diagrams of area 7.}
\label{table:convpoly7}
\end{longtable}
\normalsize
\end{center}

\begin{center}
\small
\begin{longtable}{cccccccccccc}
\hline \hline 
\multicolumn{3}{c}{\includegraphics[width=2.1cm]{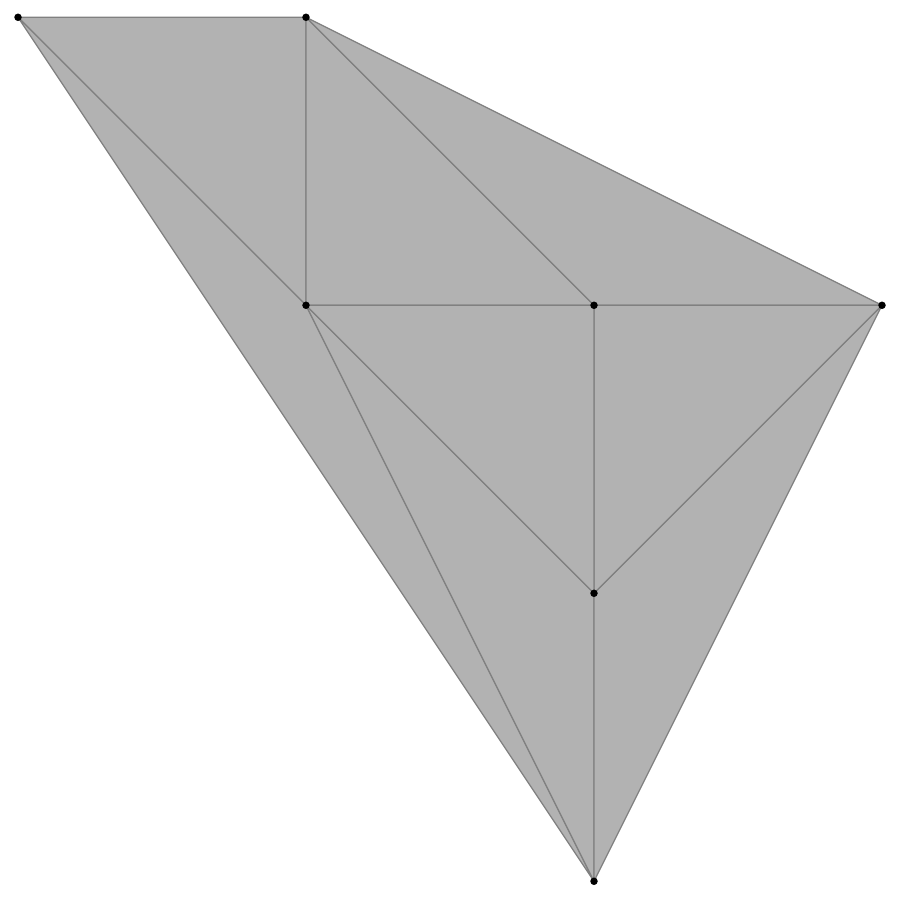}} &
\multicolumn{3}{c}{\includegraphics[width=2.1cm]{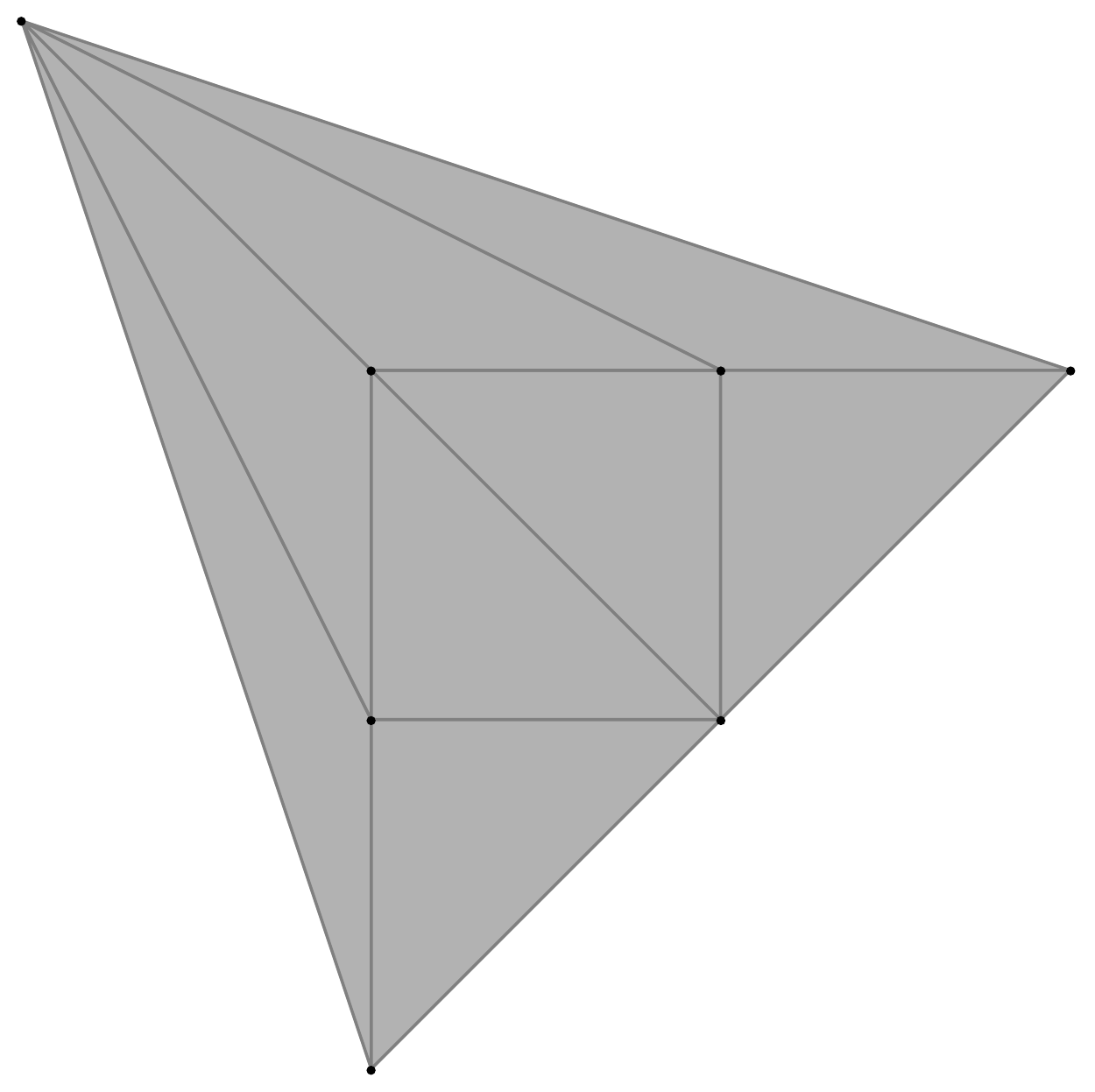}} &
\multicolumn{3}{c}{\includegraphics[width=2.8cm]{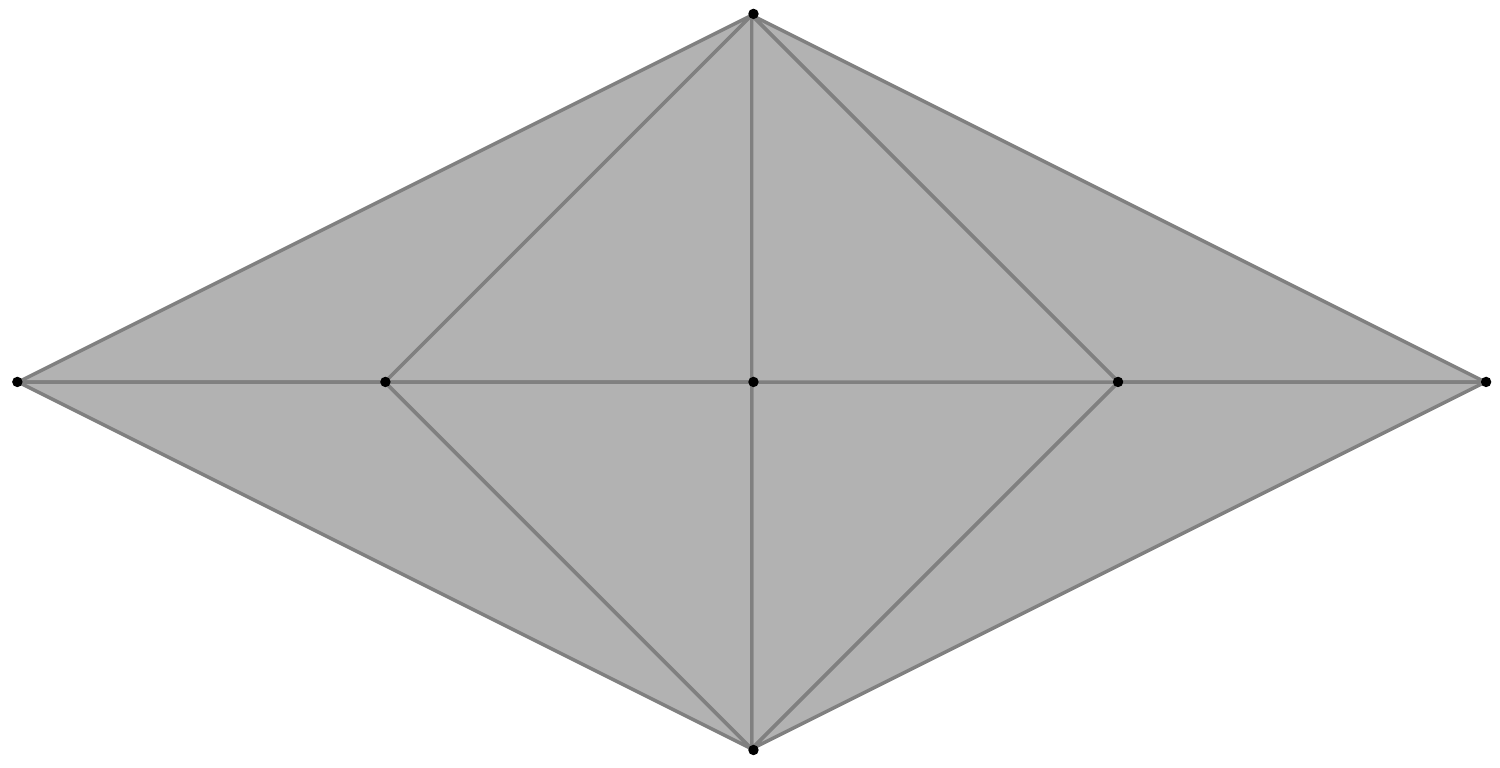}} &
\multicolumn{3}{c}{\includegraphics[width=2.8cm]{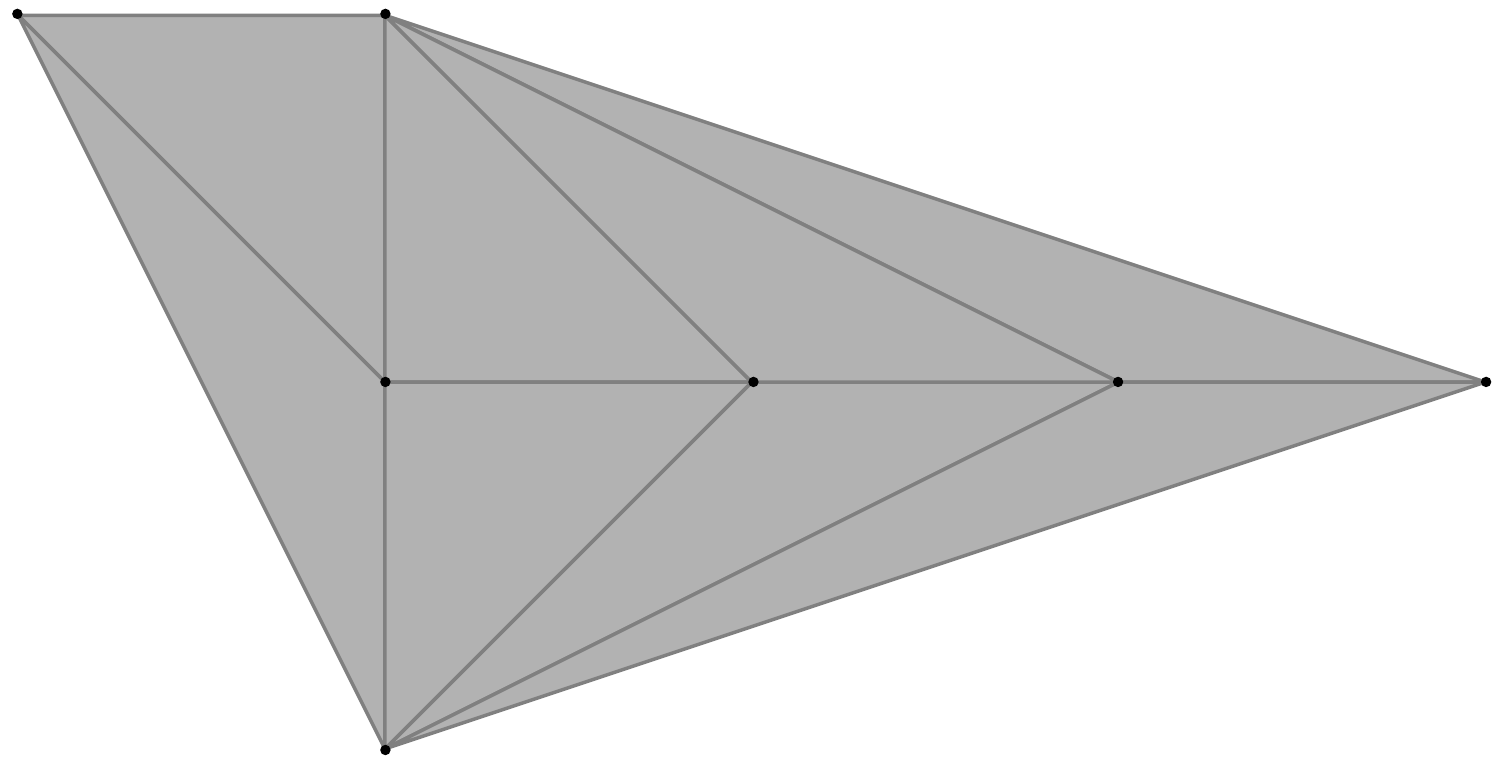}} \\
\multicolumn{3}{c}{1} &
\multicolumn{3}{c}{2} &
\multicolumn{3}{c}{3} &
\multicolumn{3}{c}{4} \\ \hline
\multicolumn{3}{c}{\includegraphics[width=2.8cm]{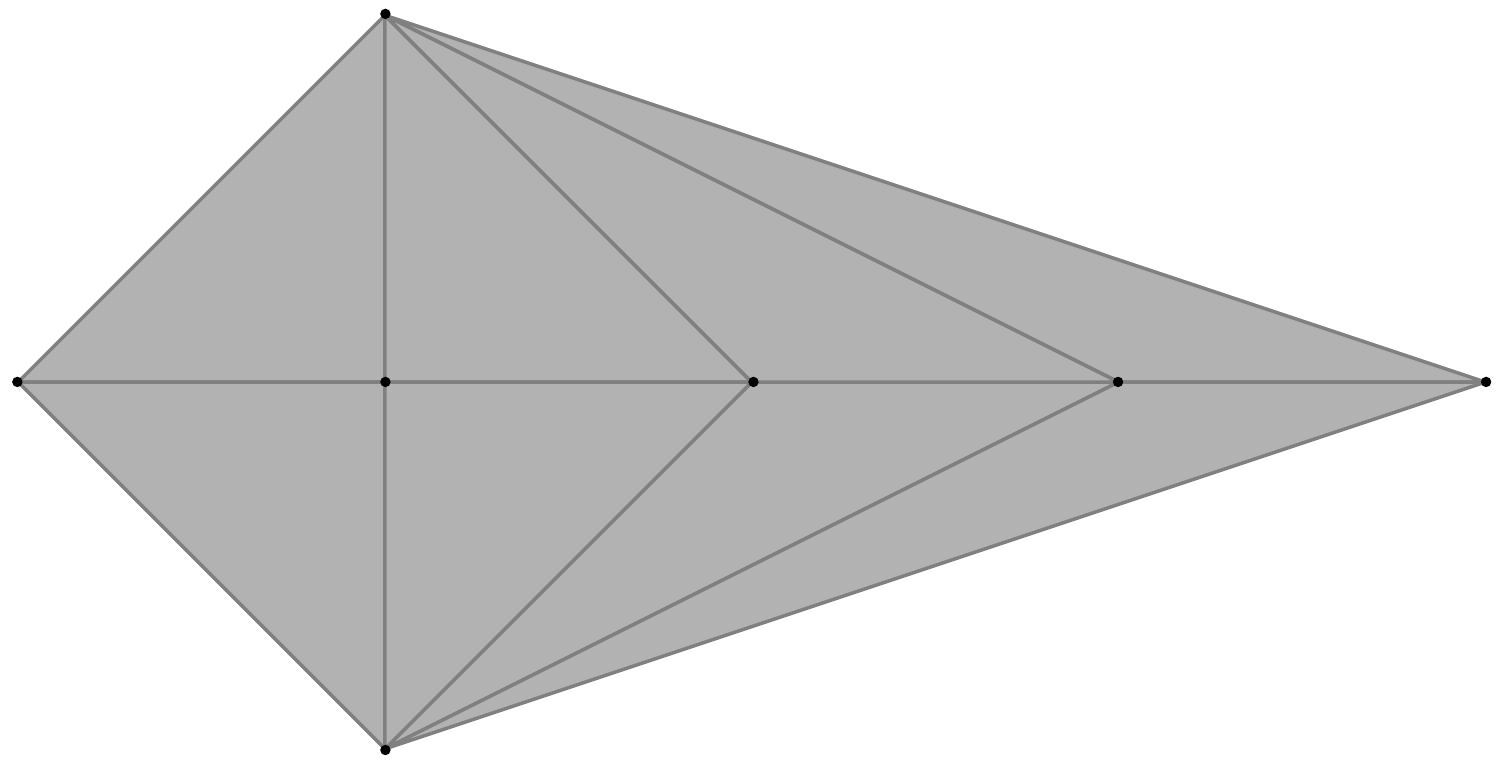}} &
\multicolumn{3}{c}{\includegraphics[width=2.8cm]{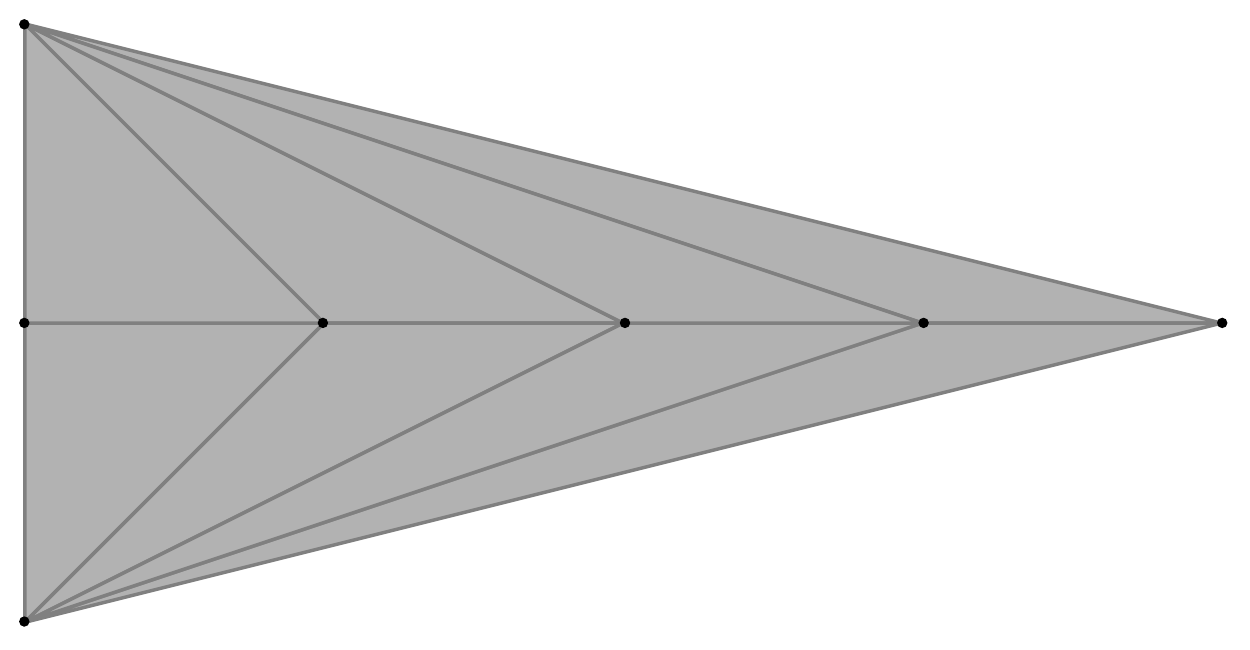}} &
\multicolumn{3}{c}{\includegraphics[width=2.8cm]{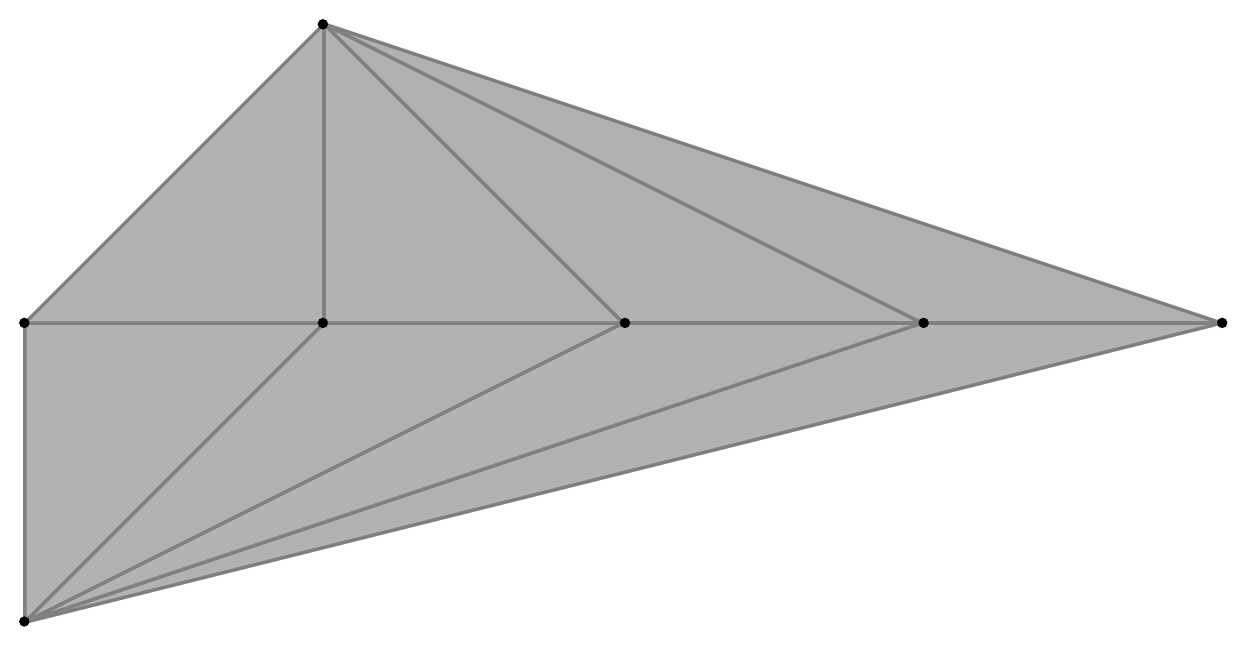}} &
\multicolumn{3}{c}{\includegraphics[width=2.8cm]{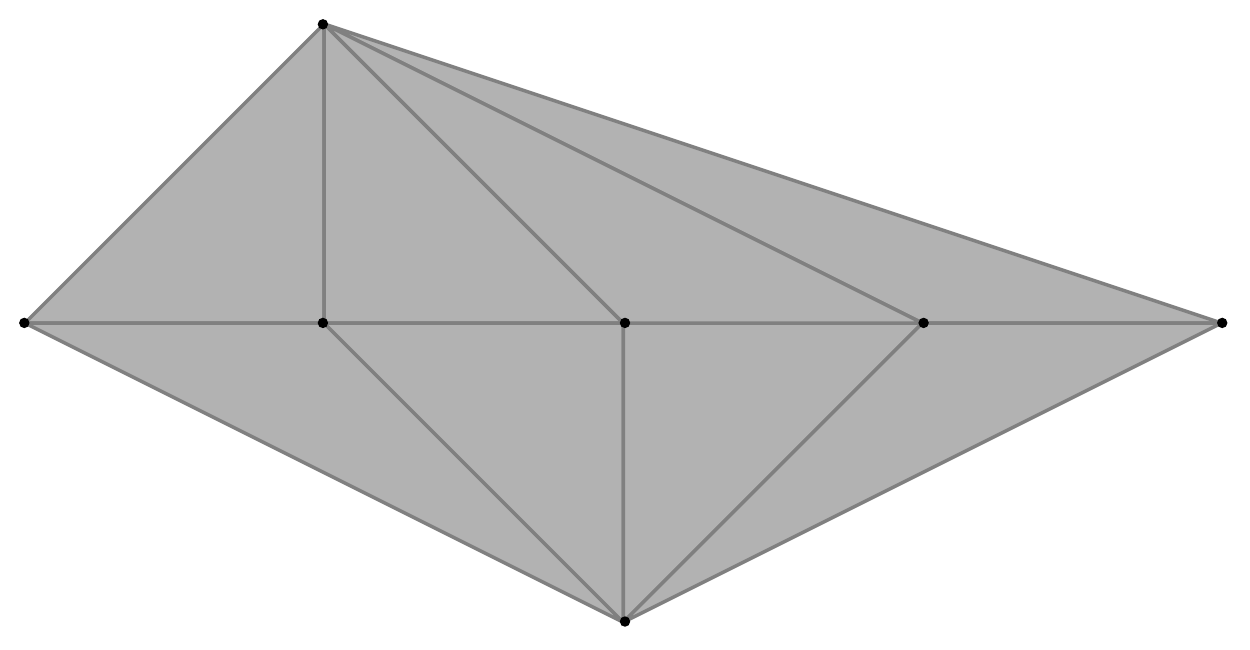}} \\
\multicolumn{3}{c}{5} &
\multicolumn{3}{c}{6} &
\multicolumn{3}{c}{7} &
\multicolumn{3}{c}{8} \\ \hline
\multicolumn{3}{c}{\includegraphics[width=2.1cm]{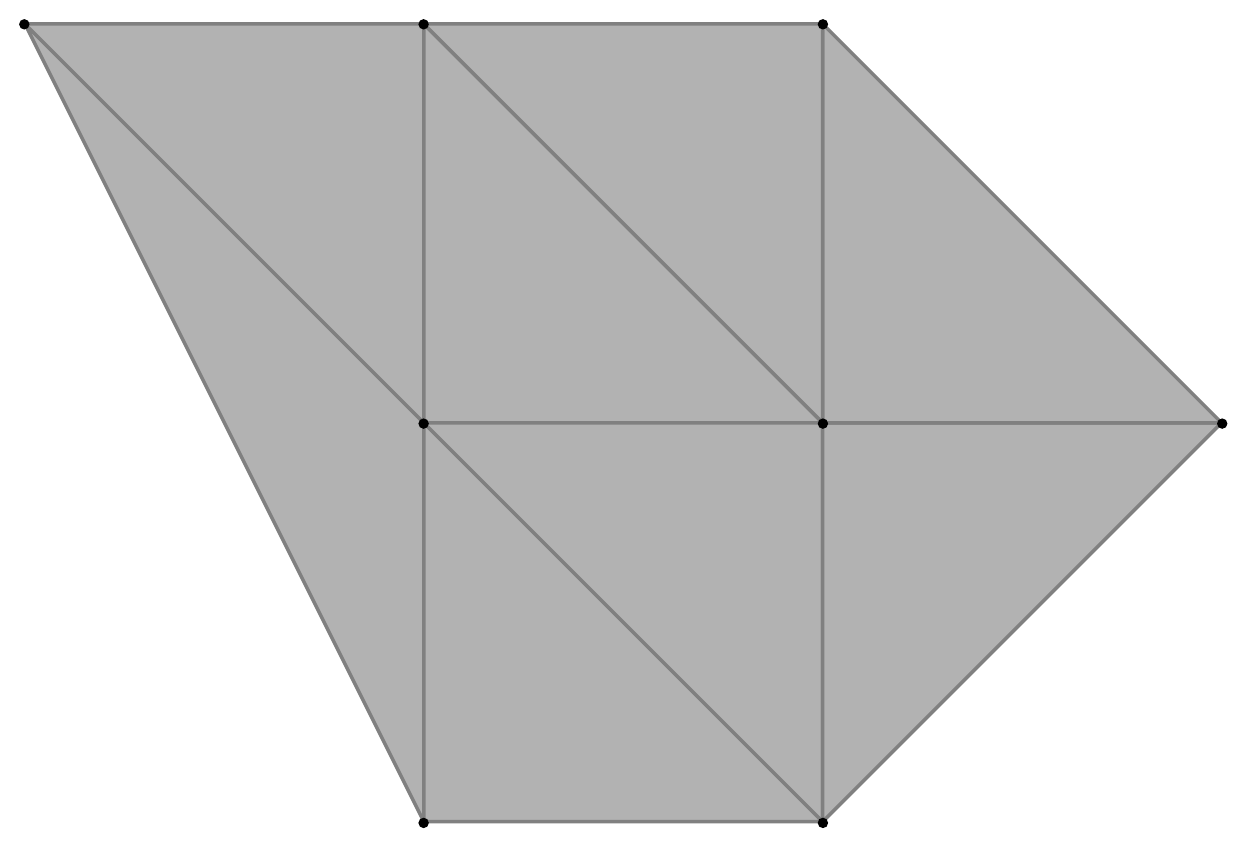}} &
\multicolumn{3}{c}{\includegraphics[width=2.1cm]{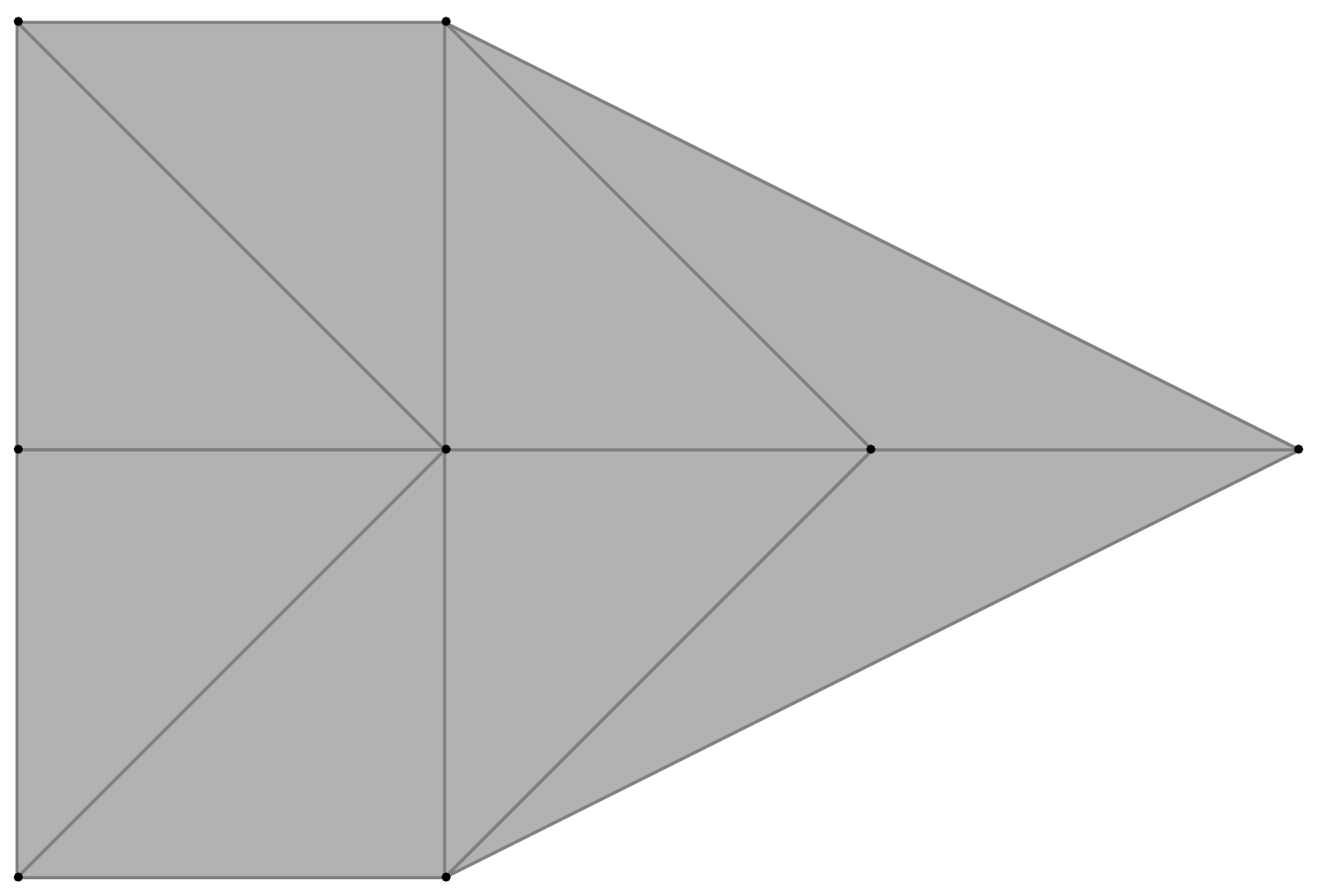}} &
\multicolumn{3}{c}{\includegraphics[width=2.1cm]{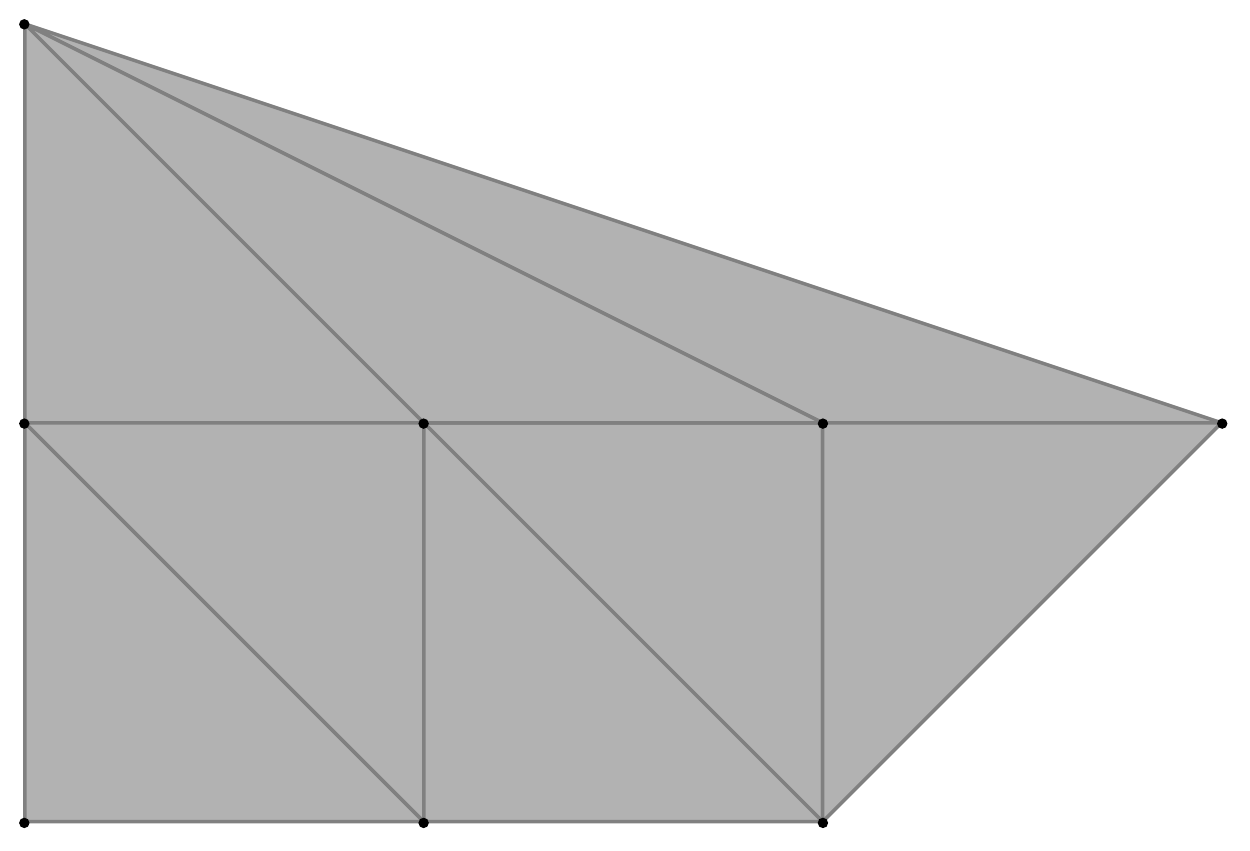}} &
\multicolumn{3}{c}{\includegraphics[width=2.1cm]{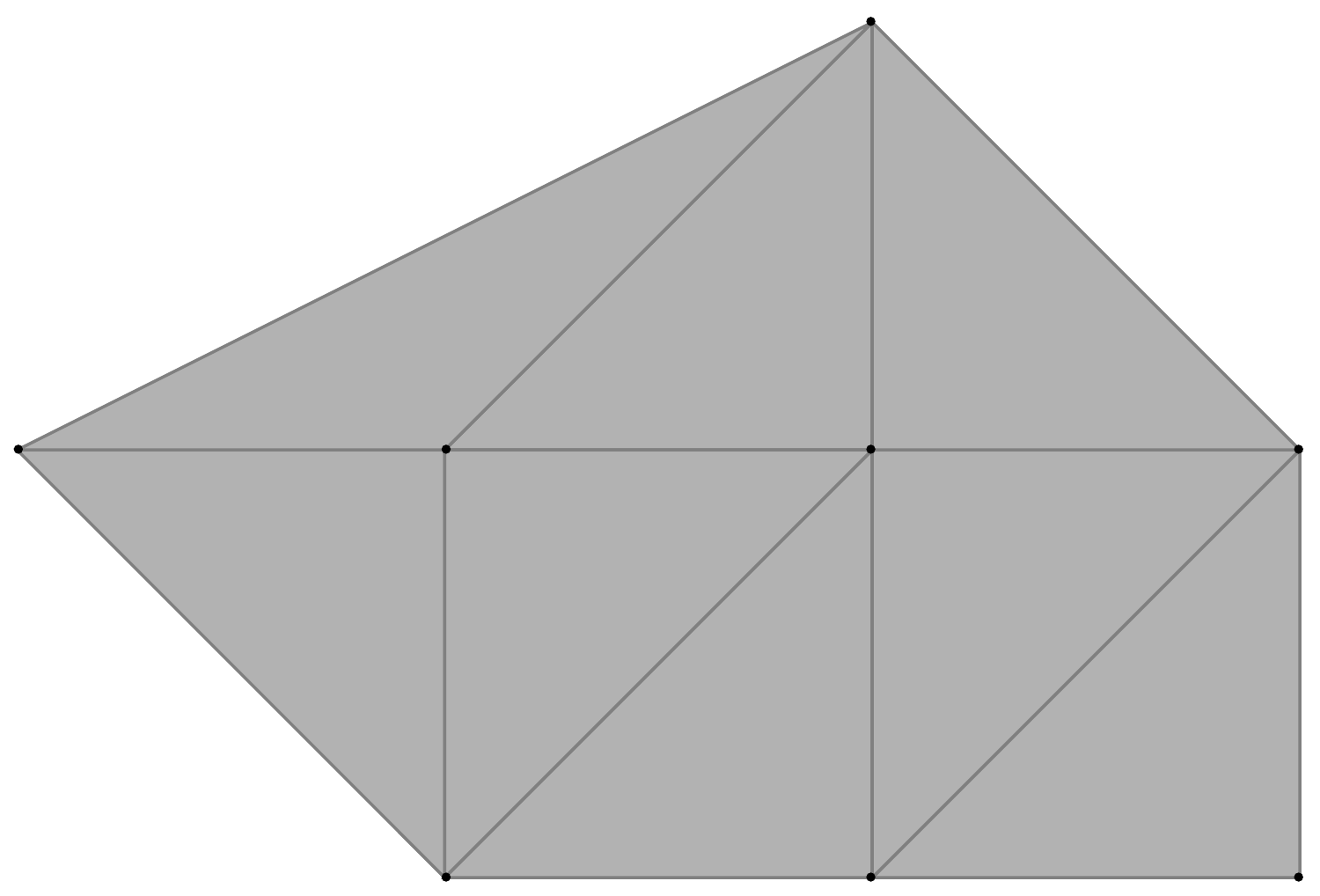}} \\
\multicolumn{3}{c}{9} &
\multicolumn{3}{c}{10} &
\multicolumn{3}{c}{11} &
\multicolumn{3}{c}{12} \\ \hline
\multicolumn{3}{c}{\includegraphics[width=2.1cm]{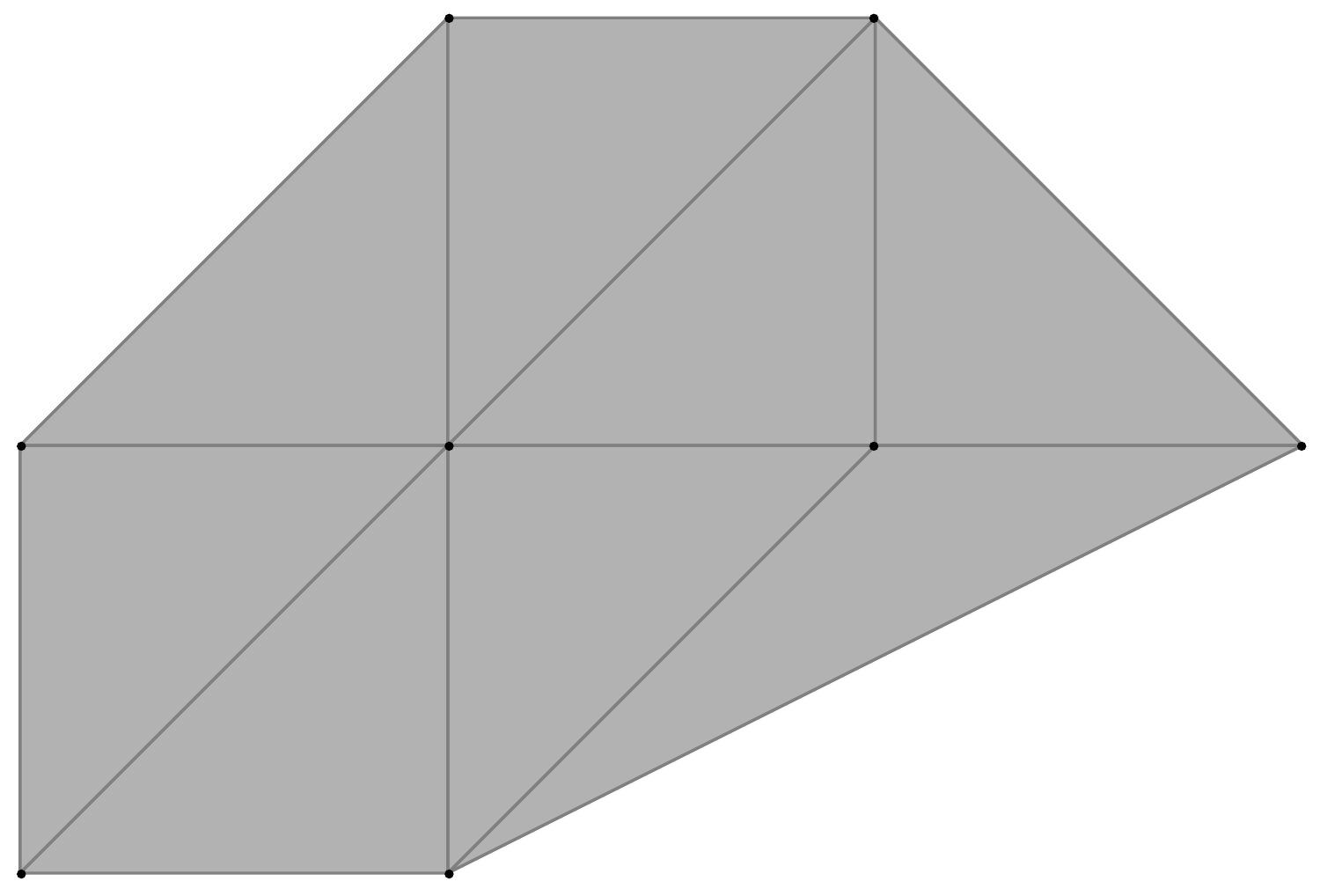}} &
\multicolumn{3}{c}{\includegraphics[width=2.1cm]{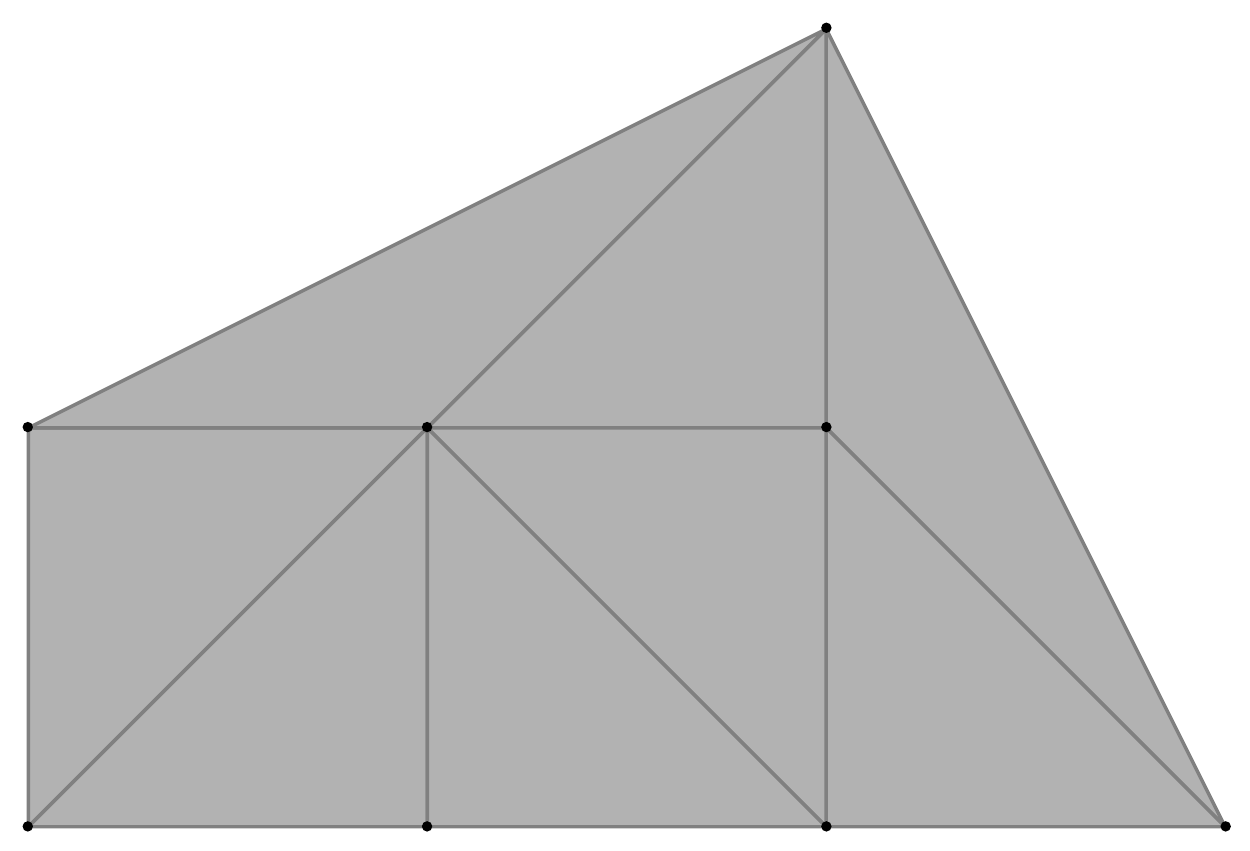}} &
\multicolumn{3}{c}{\includegraphics[width=2.1cm]{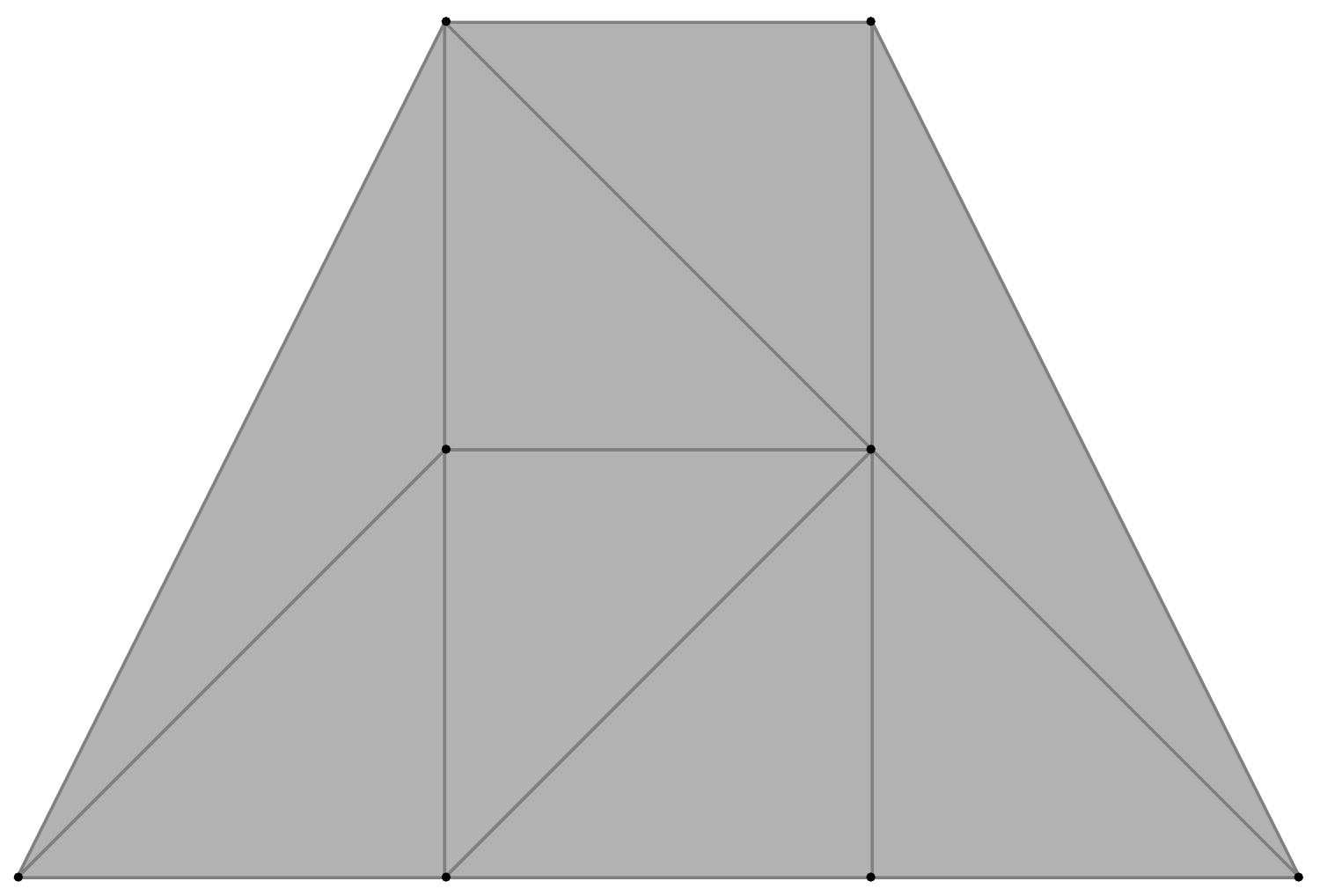}} &
\multicolumn{3}{c}{\includegraphics[width=2.1cm]{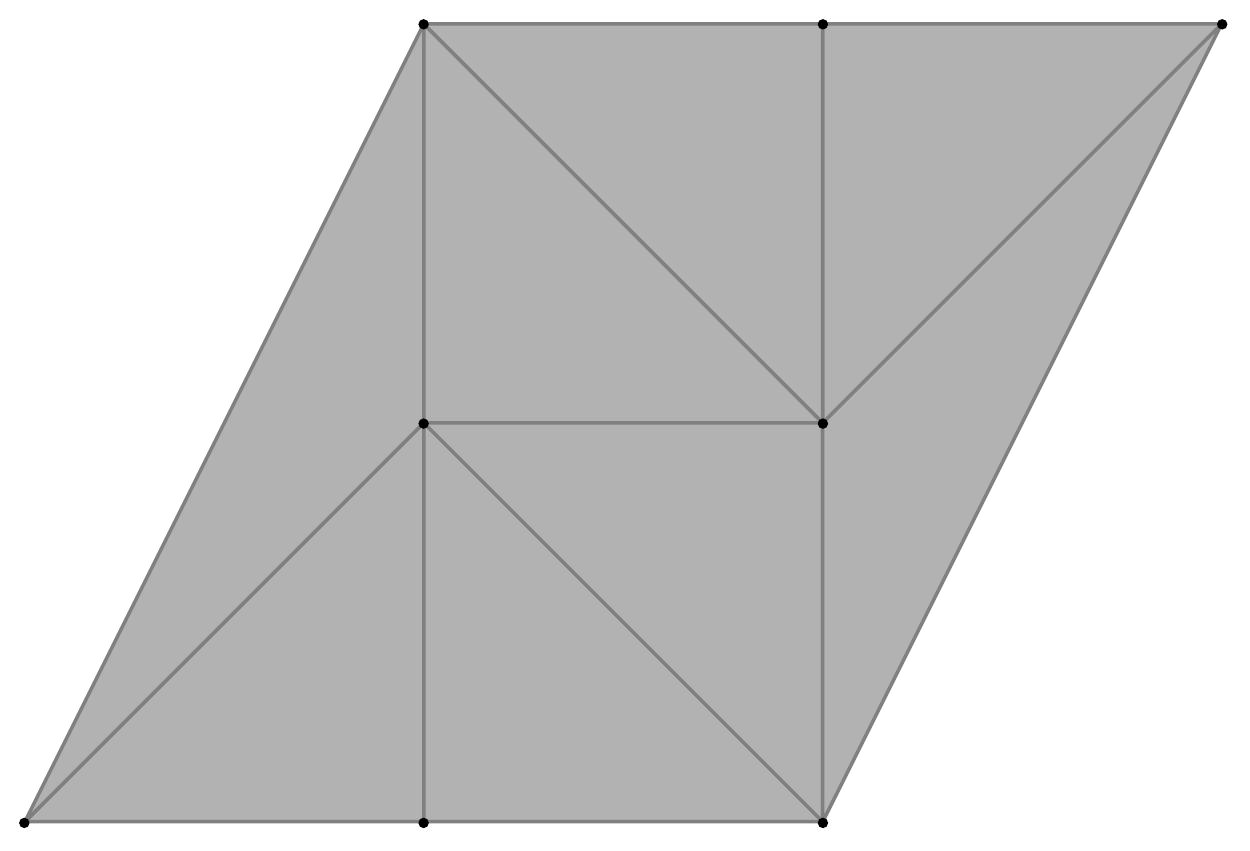}} \\
\multicolumn{3}{c}{13} &
\multicolumn{3}{c}{14} &
\multicolumn{3}{c}{15} &
\multicolumn{3}{c}{16} \\ \hline
\multicolumn{3}{c}{\includegraphics[width=2.1cm]{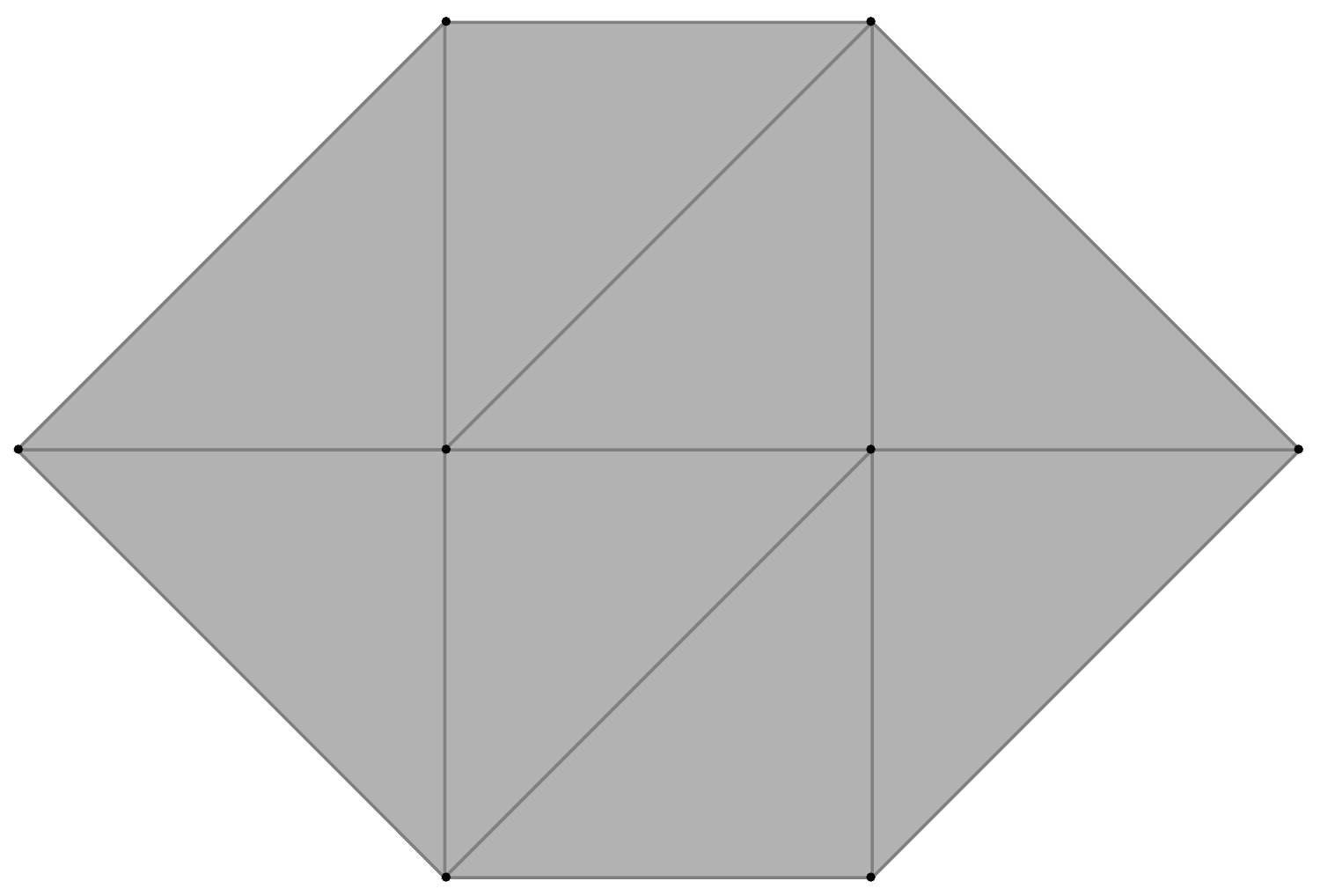}} &
\multicolumn{3}{c}{\includegraphics[width=2.1cm]{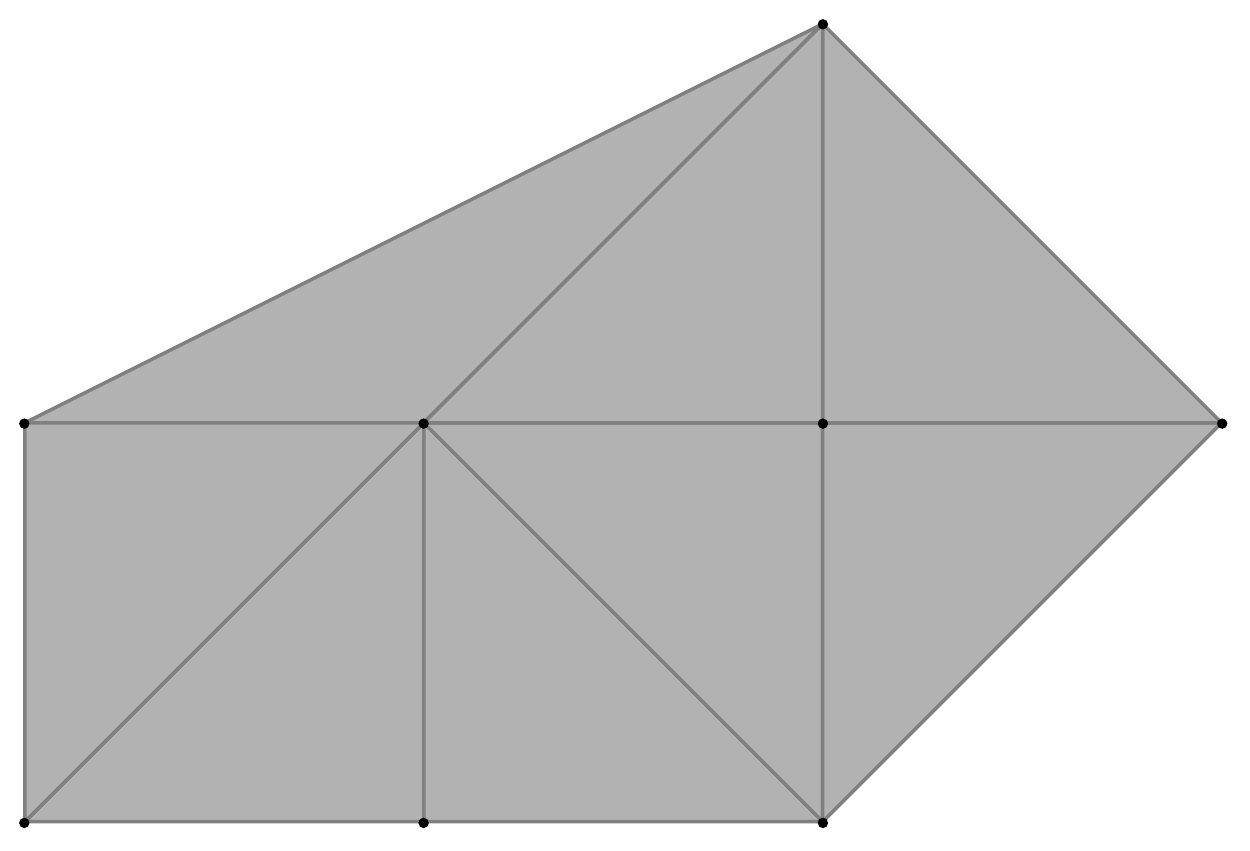}} &
\multicolumn{3}{c}{\includegraphics[width=2.8cm]{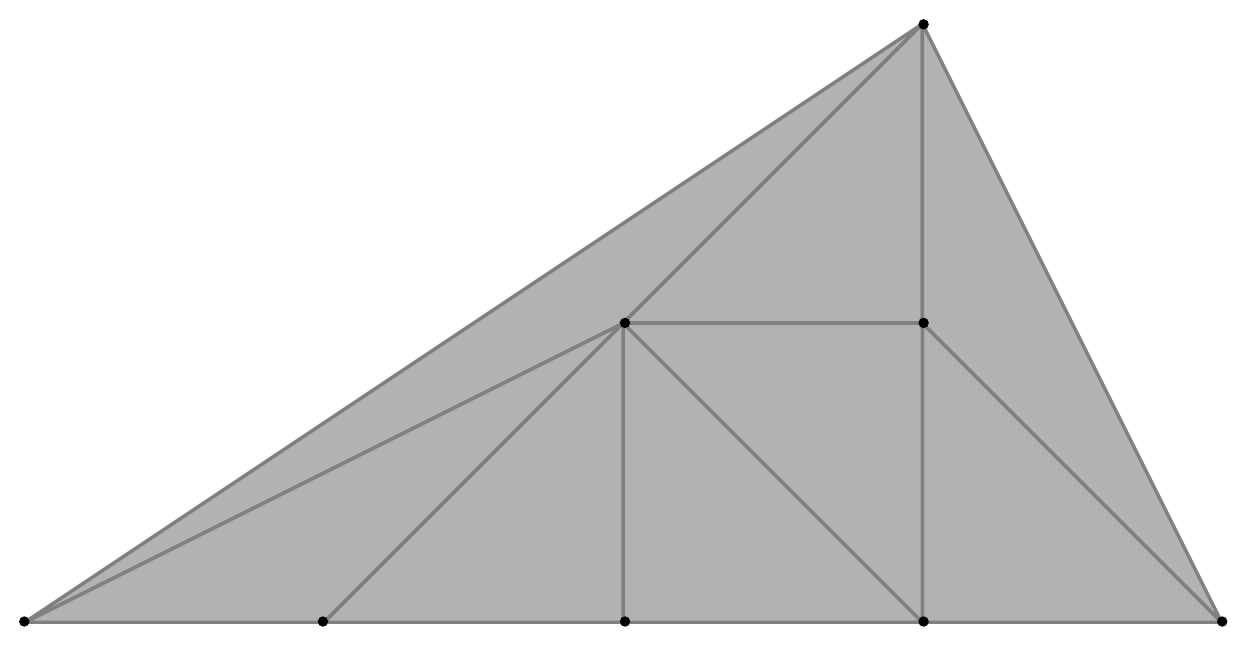}} &
\multicolumn{3}{c}{\includegraphics[width=1.4cm]{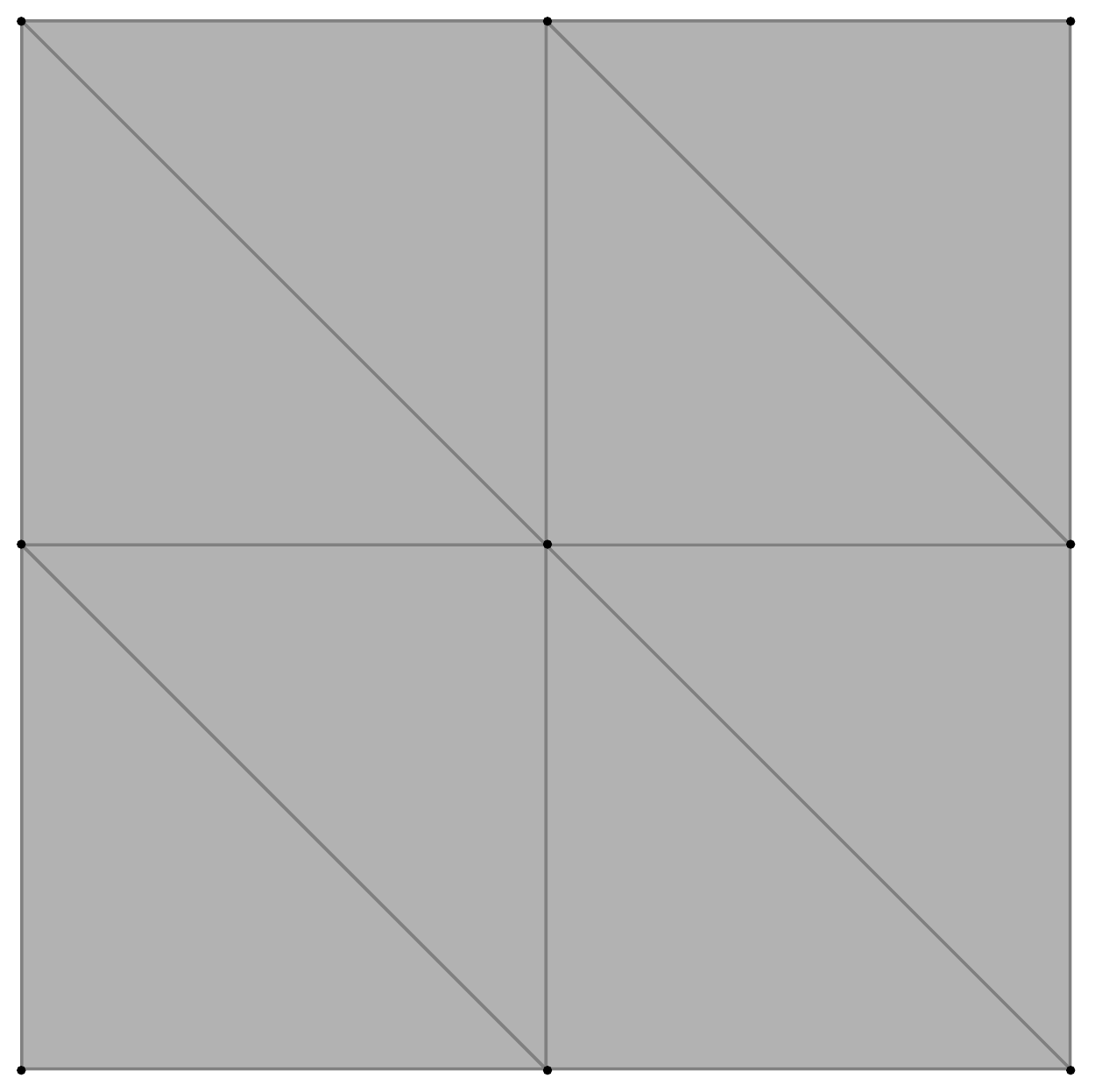}} \\
\multicolumn{3}{c}{17} &
\multicolumn{3}{c}{18} &
\multicolumn{3}{c}{19} &
\multicolumn{3}{c}{20} \\ \hline
\multicolumn{3}{c}{ \includegraphics[width=2.1cm]{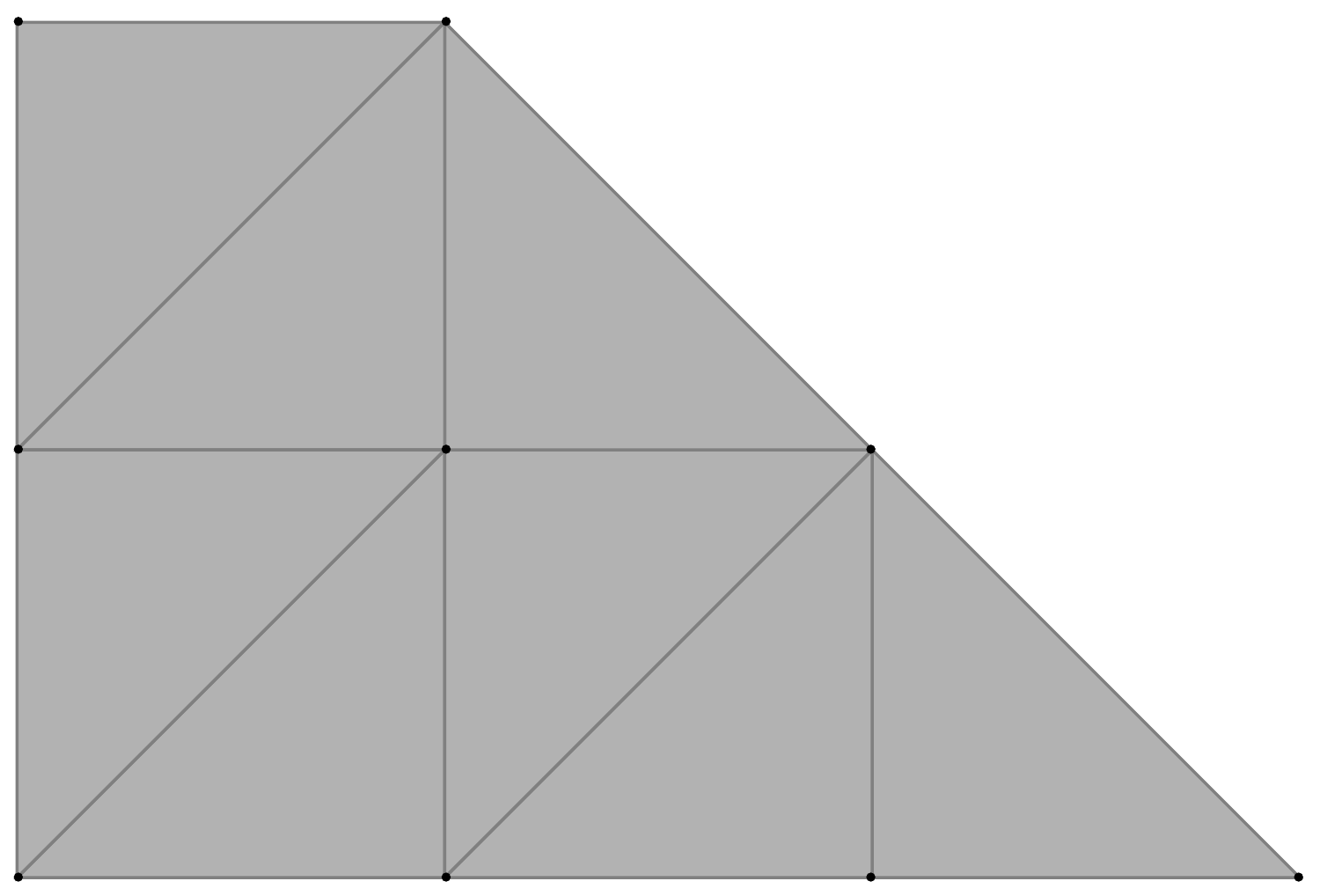}} &
\multicolumn{6}{c}{\includegraphics[width=2.8cm]{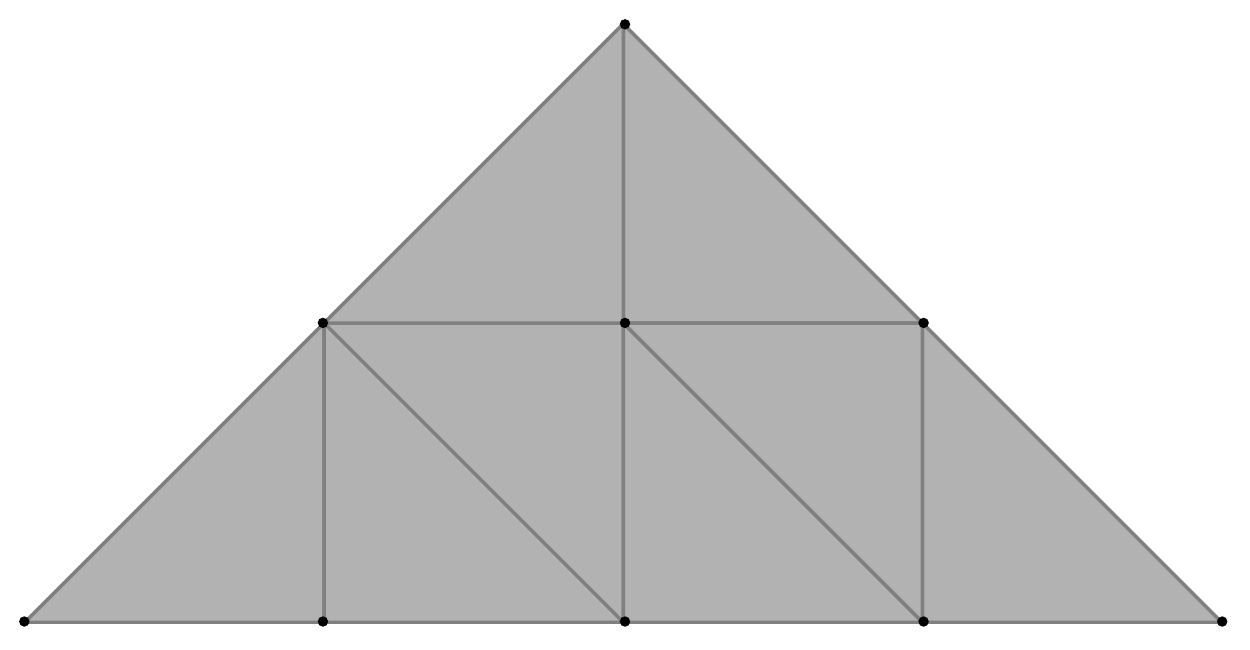}} &
\multicolumn{3}{c}{\includegraphics[width=2.8cm]{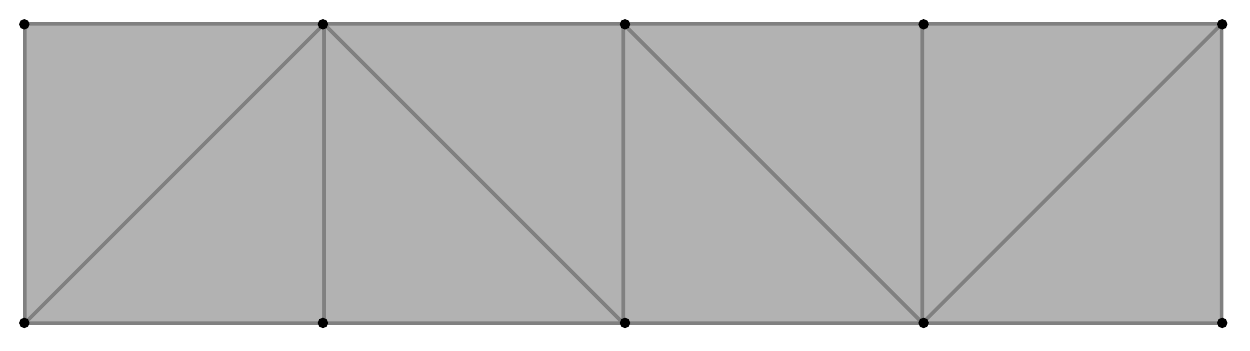}} \\
\multicolumn{3}{c}{21} &
\multicolumn{6}{c}{22} &
\multicolumn{3}{c}{23} \\ \hline 
\multicolumn{6}{c}{ \includegraphics[width=3.5cm]{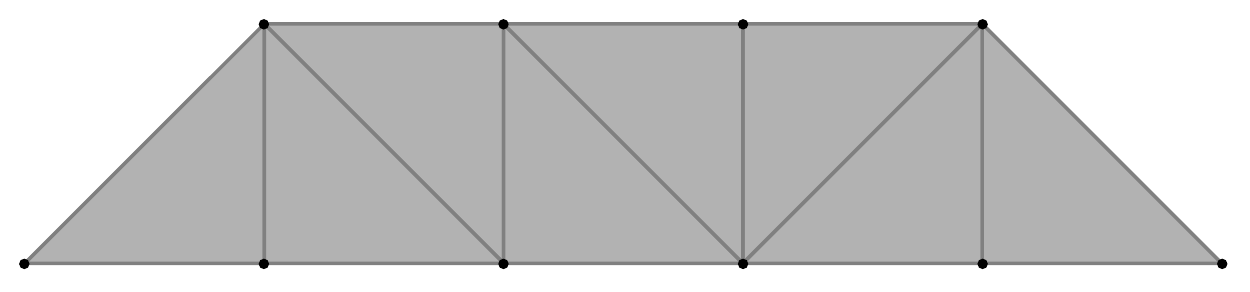}} &
\multicolumn{6}{c}{\includegraphics[width=4.2cm]{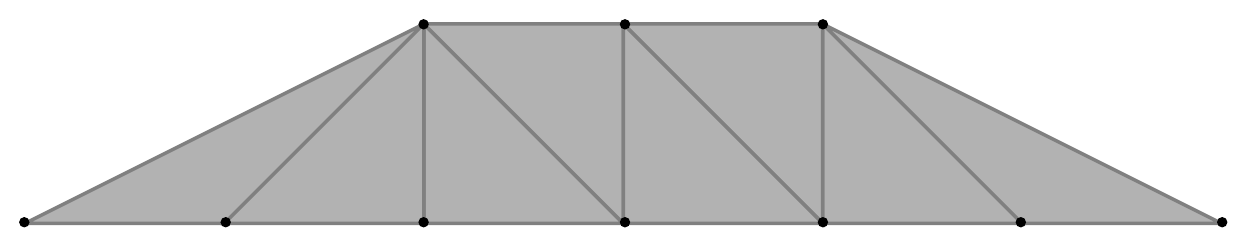}} \\
\multicolumn{6}{c}{24} &
\multicolumn{6}{c}{25} \\ \hline
\multicolumn{6}{c}{ \includegraphics[width=4.9cm]{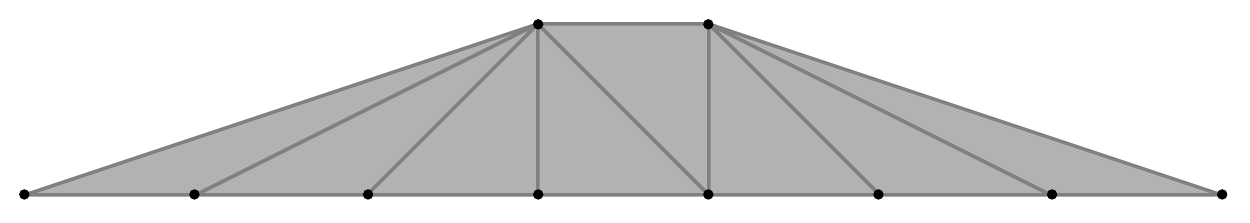}} &
\multicolumn{6}{c}{\includegraphics[width=5.6cm]{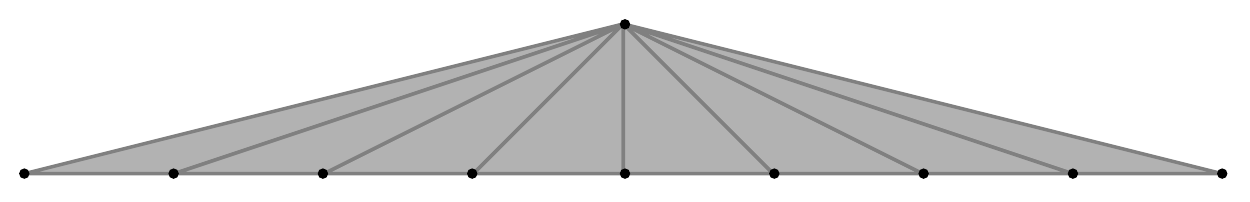}} \\
\multicolumn{6}{c}{26} &
\multicolumn{6}{c}{27} \\ \hline \hline
\\ \caption{Toric diagrams of area 8.}
\label{table:convpoly8}
\end{longtable}
\normalsize
\end{center}

\section{Results}

\label{section_results}

We now present the classification of brane tilings obtained when implementing the ideas outlined in section \sref{section_brane_tiling_technology} to the geometries presented in section \sref{section_geometries}. We provide one brane tiling per geometry for toric diagrams with areas 6 to 8. Below, the order of toric diagrams is as given in Tables 4, 5 and 6. While some of these theories have previously appeared in the literature, ours is the first exhaustive classification. Generically, there can be multiple brane tilings for a given CY$_3$. It is straightforward to generate all of them by systematically acting with Seiberg duality on the brane tilings that we present.

The geometries associated to toric diagrams without internal points give rise to non-chiral gauge theories, which are not so interesting from a model building point of view. For areas 6 to 8, they correspond to cones over $L^{a,b,a}$ manifolds \cite{Franco:2005sm}. 

\subsection{Area 6}

\begin{table}[!htbp] 
\centering 
\renewcommand{\arraystretch}{1.2}
\resizebox{\linewidth}{!}{%
\tiny
\begin{tabular}{|c|c|c|}
 \hline 
Toric Diagram & Brane Tiling & Quiver  \\
\hline 
\adjustimage{width=3cm,valign=m}{area6td1} & 
\adjustimage{height=4.1cm,valign=m}{area6s1d1x} &  
\adjustimage{height=3cm,valign=m}{area6s1d2x} \\  
\hline
\multicolumn{3}{|c|}{{\tiny $\begin{array}{cl}
W = &  -X_{11}X_{12}X_{21}-X_{22}X_{23}X_{32}-X_{33}X_{34}X_{43}-X_{44}X_{45}X_{54}-X_{55}X_{56}X_{65}-X_{66}X_{61}X_{16} \\
& + X_{11}X_{16}X_{61}+X_{22}X_{21}X_{12}+X_{33}X_{32}X_{23}+X_{44}X_{43}X_{34}+X_{55}X_{54}X_{45}+X_{66}X_{65}X_{56}
\end{array}$}} \\ \hline
 \hline 
\adjustimage{width=2.5cm,valign=m}{area6td2} & 
\adjustimage{height=3.6cm,valign=m}{area6s2d1} &
\adjustimage{height=3.5cm,valign=m}{area6s2d2} \\
\hline
\multicolumn{3}{|c|}{{\tiny $\begin{array}{cl}
W = &  -X_{34} X_{44} X_{43} -X_{16} X_{66} X_{61}  -X_{12} X_{24} X_{42} X_{21}  -X_{33} X_{35} X_{53} -X_{55} X_{56} X_{65} \\
& + X_{33} X_{34} X_{43} + X_{35} X_{53} X_{55} + X_{56} X_{66} X_{65}  + X_{12} X_{21} X_{16}  X_{61} + X_{24} X_{44} X_{42} 
\end{array}$}} \\ \hline
\end{tabular}}
\end{table}


\newpage

\begin{table}[!htbp] 
\centering 
\renewcommand{\arraystretch}{1.2}
\resizebox{\linewidth}{!}{
\tiny
\begin{tabular}{|c|c|c|}
 \hline 
Toric Diagram & Brane Tiling & Quiver  \\
\hline 
\adjustimage{height=.8cm,valign=m}{area6td3} & 
\adjustimage{height=4cm,valign=m}{area6s3d1} &
\adjustimage{height=3cm,valign=m}{area6s3d2} \\
\hline
\multicolumn{3}{|c|}{{\tiny $\begin{array}{cl}
W = &  -X_{34} X_{44} X_{43}  -X_{13} X_{33} X_{31}  -X_{15} X_{52} X_{25} X_{51}  -X_{26} X_{64} X_{46} X_{62}  \\
& + X_{33} X_{34} X_{43} + X_{13} X_{31} X_{15}  X_{51} + X_{25} X_{52} X_{26}  X_{62} + X_{44} X_{46} X_{64} 
\end{array}$}} \\ \hline
 \hline 
\adjustimage{height=.8cm,valign=m}{area6td4} & 
\adjustimage{height=4cm,valign=m}{area6s4d1} &
\adjustimage{height=3cm,valign=m}{area6s4d2} \\
\hline
\multicolumn{3}{|c|}{{\tiny $\begin{array}{cl}
W = & -X_{15} X_{52} X_{25} X_{51}  -X_{14}  X_{43} X_{34} X_{41} -X_{26} X_{63} X_{36} X_{62}  \\
& + X_{14} X_{41} X_{15}  X_{51} + X_{34} X_{43} X_{36} X_{63} + X_{25} X_{52} X_{26}  X_{62} 
\end{array}$}} \\ \hline
 \hline 
\adjustimage{height=1.6cm,valign=m}{area6td5} & 
\adjustimage{height=4cm,valign=m}{area6s5d1} &
\adjustimage{height=3cm,valign=m}{area6s5d2} \\
\hline
\multicolumn{3}{|c|}{{\tiny $\begin{array}{cl}
W = &  -X_{12} X_{23} X_{31} -X_{14} X_{45} X_{51} -X_{24} X_{46} X_{62} -X_{25} X_{53} X_{32} -X_{15} X_{56} X_{61} -X_{36} X_{64} X_{43}  \\
& + X_{12} X_{25} X_{51} + X_{45} X_{56} X_{64} + X_{24} X_{43}X_{32} + X_{15} X_{53} X_{31} + X_{14} X_{46} X_{61} + X_{23} X_{36} X_{62} 
\end{array}$}} \\ \hline
\end{tabular}}
\end{table}


\newpage

\begin{table}[!htbp] 
\centering 
\renewcommand{\arraystretch}{1.2}
\resizebox{\linewidth}{!}{
\tiny
\begin{tabular}{|c|c|c|}
 \hline 
Toric Diagram & Brane Tiling & Quiver  \\
\hline 
\adjustimage{height=1.6cm,valign=m}{area6td6} & 
\adjustimage{height=3.5cm,valign=m}{area6s6d1} &
\adjustimage{height=3cm,valign=m}{area6s6d2} \\
\hline
\multicolumn{3}{|c|}{{\tiny $\begin{array}{cl}
W = &  -X_{13} X_{32} X_{21} -X_{12} X_{24} X_{41} -X_{26} X_{63} X_{35} X_{52} -X_{15} X_{54} X_{46}  X_{61} \\
& + X_{13} X_{35} X_{54} X_{41} + X_{24}  X_{46} X_{63} X_{32} + X_{15} X_{52} X_{21} + X_{12} X_{26} X_{61}
\end{array}$}} \\ \hline
 \hline 
\adjustimage{height=1.6cm,valign=m}{area6td7} & 
\adjustimage{height=3.5cm,valign=m}{area6s7d1} &
\adjustimage{height=3cm,valign=m}{area6s7d2} \\
\hline
\multicolumn{3}{|c|}{{\tiny $\begin{array}{cl}
W = & -X_{12} X_{23} X_{31} -X_{12} X_{24} X_{41} -X_{25} X_{53} X_{36}  X_{62} -X_{15} X_{54} X_{46}  X_{61} \\
& + X_{12} X_{25} X_{54} X_{41} + X_{24} X_{46} X_{62} + X_{15} X_{53} X_{31} + X_{12} X_{23} X_{36} X_{61}
\end{array}$}} \\ \hline
 \hline 
\adjustimage{height=1.6cm,valign=m}{area6td8} & 
\adjustimage{height=3.5cm,valign=m}{area6s8d1} &
\adjustimage{height=3cm,valign=m}{area6s8d2} \\
\hline
\multicolumn{3}{|c|}{{\tiny $\begin{array}{cl}
W = &  -X_{16}  X_{65} X_{53} X_{32} X_{24}  X_{41}  -X_{15} X_{52} X_{21} -X_{34} X_{46} X_{63} \\
& + X_{16} X_{63} X_{32} X_{21} + X_{15} X_{53} X_{34} X_{41} + X_{24} X_{46} X_{65} X_{52} 
\end{array}$}} \\ \hline
\end{tabular}}
\end{table}


\newpage

\begin{table}[!htbp] 
\centering 
\renewcommand{\arraystretch}{1.2}
\resizebox{\linewidth}{!}{%
\tiny
\begin{tabular}{|c|c|c|}
 \hline 
Toric Diagram & Brane Tiling & Quiver  \\
\hline 
\adjustimage{height=1.6cm,valign=m}{area6td9} & 
\adjustimage{height=4.5cm,valign=m}{area6s9d1} &
\adjustimage{height=3.2cm,valign=m}{area6s9d2} \\
\hline
\multicolumn{3}{|c|}{{\tiny $\begin{array}{cl}
W = & -X_{24} X_{43} X_{32}  -X_{26} X_{65} X_{52} -X_{36} X_{65} X_{53} -X_{12} X_{24} X_{41} -X_{12} X_{26} X_{61} -X_{36} X_{64} X_{43}  \\
& +  X_{12} X_{24} X_{41} + X_{26} X_{65} X_{52} + X_{36} X_{64} X_{43} + X_{12} X_{26} X_{61} + X_{36} X_{65} X_{53} + X_{24} X_{43} X_{32}
\end{array}$}} \\ \hline
 \hline 
\adjustimage{height=1.6cm,valign=m}{area6td10} & 
\adjustimage{height=4.5cm,valign=m}{area6s10d1} &
\adjustimage{height=3.5cm,valign=m}{area6s10d2} \\
\hline
\multicolumn{3}{|c|}{{\tiny $\begin{array}{cl}
W = & -X_{13} X_{34} X_{41} -X_{16}  X_{65} X_{54} X_{41}-X_{36}  X_{65} X_{53}-X_{23} X_{34} X_{42} -X_{23} X_{36} X_{62} \\
& + X_{13} X_{34} X_{41} + X_{16}  X_{65} X_{54} X_{41} + X_{23} X_{36} X_{62} + X_{36} X_{65} X_{53} + X_{23} X_{34} X_{42} 
\end{array}$}} \\ \hline
 \hline 
\adjustimage{height=1.6cm,valign=m}{area6td11} & 
\adjustimage{height=4.5cm,valign=m}{area6s11d1} &
\adjustimage{height=3.5cm,valign=m}{area6s11d2} \\
\hline
\multicolumn{3}{|c|}{{\tiny $\begin{array}{cl}
W = &   -X_{13} X_{32} X_{21}  -X_{14} X_{45} X_{51} -X_{26} X_{64} X_{42} -X_{26} X_{63} X_{32} -X_{13} X_{35} X_{51} -X_{45} X_{56} X_{64} \\
& + X_{13} X_{35} X_{51} + X_{45} X_{56} X_{64} + X_{26} X_{63} X_{32} + X_{13} X_{32} X_{21} + X_{14} X_{45} X_{51} + X_{26} X_{64} X_{42} 
\end{array}$}} \\ \hline
\end{tabular}}
\end{table}


\newpage

\begin{table}[H] 
\centering 
\vspace{3.2cm}\renewcommand{\arraystretch}{1.2}
\resizebox{\linewidth}{!}{
\tiny
\begin{tabular}{|c|c|c|}
 \hline 
Toric Diagram & Brane Tiling & Quiver  \\
\hline 
\adjustimage{height=1.6cm,valign=m}{area6td12} & 
\adjustimage{height=2.8cm,valign=m}{area6s12d1} &
\adjustimage{height=3cm,valign=m}{area6s12d2} \\
\hline
\multicolumn{3}{|c|}{{\tiny $\begin{array}{cl}
W = &  -X_{34} X_{45} X_{53} -X_{45} X_{56} X_{64} -X_{15} X_{56} X_{61} -X_{13} X_{32} X_{26}  X_{61} -X_{12} X_{24} X_{41} \\
& + X_{24} X_{45} X_{53} X_{32} + X_{45} X_{56} X_{64} + X_{15} X_{56} X_{61} + X_{12} X_{26} X_{61} + X_{13} X_{34} X_{41}
\end{array}$}} \\ \hline
 \hline 
\adjustimage{height=1.6cm,valign=m}{area6td13} & 
\adjustimage{height=4.5cm,valign=m}{area6s13d1} &
\adjustimage{height=3.5cm,valign=m}{area6s13d2} \\
\hline
\multicolumn{3}{|c|}{{\tiny $\begin{array}{cl}
W = &  -X_{16} X_{65} X_{51} -X_{25}  X_{54} X_{43}  X_{32} -X_{12} X_{25} X_{51} -X_{16} X_{64} X_{41}  -X_{36} X_{64} X_{43} \\
& + X_{12} X_{25} X_{51} + X_{16} X_{65} X_{51} + X_{36} X_{64} X_{43} + X_{25} X_{54} X_{43} X_{32} + X_{16} X_{64} X_{41}  
\end{array}$}} \\ \hline
\end{tabular}}
\end{table}


\newpage

\subsection{Area 7}

\begin{table}[!htbp] 
\centering 
\renewcommand{\arraystretch}{1.2}
\resizebox{\linewidth}{!}{
\tiny
\begin{tabular}{|c|c|c|}
 \hline 
Toric Diagram & Brane Tiling & Quiver  \\
\hline 
\adjustimage{height=.7cm,valign=m}{area7td1} & 
\adjustimage{height=6.5cm,valign=m}{area7s1d1x} &  
\adjustimage{height=4cm,valign=m}{area7s1d2x} \\  
\hline
\multicolumn{3}{|c|}{{\tiny $\begin{array}{cl}
W = &  -X_{11}X_{12}X_{21}-X_{22}X_{23}X_{32}-X_{33}X_{34}X_{43}-X_{44}X_{45}X_{54} \\
& -X_{55}X_{56}X_{65}-X_{66}X_{67}X_{76} - X_{77}X_{71}X_{17} \\
& + X_{11}X_{17}X_{71}+X_{22}X_{21}X_{12}+X_{33}X_{32}X_{23}+X_{44}X_{43}X_{34}\\
& +X_{55}X_{54}X_{45}+X_{66}X_{65}X_{56} + X_{77}X_{76}X_{67}
\end{array}$}} \\ \hline
\hline 
\adjustimage{height=.7cm,valign=m}{area7td2} & 
\adjustimage{height=5.7cm,valign=m}{area7s2d1x} &  
\adjustimage{height=4cm,valign=m}{area7s2d2x} \\  
\hline
\multicolumn{3}{|c|}{{\tiny $\begin{array}{cl}
W = &  -X_{11}X_{12}X_{21}-X_{22}X_{23}X_{32}-X_{33}X_{34}X_{43}-X_{44}X_{45}X_{54} -X_{55}X_{56}X_{65}-X_{67}X_{71}X_{17}X_{76} \\
& + X_{11}X_{17}X_{71}+X_{22}X_{21}X_{12}+X_{33}X_{32}X_{23}+X_{44}X_{43}X_{34}+X_{55}X_{54}X_{45}+X_{67}X_{76}X_{65}X_{56}
\end{array}$}} \\ \hline
\end{tabular}}
\end{table}


\newpage

\begin{table}[!htbp] 
\centering 
\vspace{.45cm}\renewcommand{\arraystretch}{1.2}
\resizebox{\linewidth}{!}{
\tiny
\begin{tabular}{|c|c|c|}
 \hline 
Toric Diagram & Brane Tiling & Quiver  \\
\hline 
\adjustimage{height=.7cm,valign=m}{area7td3} & 
\adjustimage{width=2.8cm,valign=m}{area7s3d1x} &  
\adjustimage{height=4cm,valign=m}{area7s3d2x} \\  
\hline
\multicolumn{3}{|c|}{{\tiny $\begin{array}{cl}
W = &  -X_{11}X_{12}X_{21}-X_{22}X_{23}X_{32}-X_{33}X_{34}X_{43}-X_{45}X_{56}X_{65}X_{54}-X_{67}X_{71}X_{17}X_{76}\\
& + X_{11}X_{17}X_{71}+X_{22}X_{21}X_{12}+X_{33}X_{32}X_{23}+X_{34}X_{45}X_{54}X_{43}+X_{56}X_{67}X_{76}X_{65}
\end{array}$}} \\ \hline
 \hline 
\adjustimage{height=.7cm,valign=m}{area7td4} & 
\adjustimage{width=2.8cm,valign=m}{area7s4d1x} &  
\adjustimage{height=4cm,valign=m}{area7s4d2x} \\  
\hline
\multicolumn{3}{|c|}{{\tiny $\begin{array}{cl}
W = &  -X_{11}X_{12}X_{21}-X_{23}X_{34}X_{43}X_{32}-X_{45}X_{56}X_{65}X_{54}-X_{67}X_{71}X_{17}X_{76} \\
& + X_{11}X_{17}X_{71}+X_{12}X_{23}X_{32}X_{21}+X_{34}X_{45}X_{54}X_{43}+X_{56}X_{67}X_{76}X_{65}
\end{array}$}} \\ \hline
\end{tabular}}
\end{table}


\newpage

\begin{table}[!htbp] 
\centering 
\renewcommand{\arraystretch}{1.2}
\resizebox{\linewidth}{!}{
\tiny
\begin{tabular}{|c|c|c|}
 \hline 
Toric Diagram & Brane Tiling & Quiver  \\
\hline 
\adjustimage{height=1.6cm,valign=m}{area7td5} & 
\adjustimage{height=4.2cm,valign=m}{area7s5d1} &
\adjustimage{height=3.5cm,valign=m}{area7s5d2} \\
\hline
\multicolumn{3}{|c|}{{\tiny $\begin{array}{cl}
W = & -X_{23} X_{34} X_{42} -X_{14} X_{43} X_{31} -X_{25} X_{56} X_{62}-X_{16} X_{67} X_{72} X_{21}-X_{35} X_{57} X_{73}-X_{47}  X_{75} X_{54} \\
& + X_{14} X_{42} X_{21} + X_{35} X_{54} X_{43} + X_{16} X_{62} X_{23} X_{31} + X_{25} X_{57} X_{72} + X_{34} X_{47} X_{73} + X_{56} X_{67} X_{75}
\end{array}$}} \\ \hline
 \hline 
\adjustimage{height=1.6cm,valign=m}{area7td6} & 
\adjustimage{height=4.2cm,valign=m}{area7s6d1} &
\adjustimage{height=3.5cm,valign=m}{area7s6d2} \\
\hline
\multicolumn{3}{|c|}{{\tiny $\begin{array}{cl}
W = & -X_{36}  X_{67} X_{75} X_{53}-X_{16}  X_{62} X_{25} X_{51}-X_{37}  X_{74} X_{43}-X_{13}  X_{32} X_{24} X_{41} \\
& + X_{25}  X_{53} X_{32} + X_{13} X_{37}  X_{75} X_{51} + X_{16}  X_{67} X_{74} X_{41} + X_{24}  X_{43} X_{36} X_{62}
\end{array}$}} \\ \hline
 \hline 
\adjustimage{height=1.6cm,valign=m}{area7td7} & 
\adjustimage{height=4.2cm,valign=m}{area7s7d1} &
\adjustimage{height=3.2cm,valign=m}{area7s7d2} \\
\hline
\multicolumn{3}{|c|}{{\tiny $\begin{array}{cl}
W = &   -X_{24} X_{43} X_{32}  -X_{25} X_{56} X_{62} -X_{35} X_{57} X_{73} -X_{27}  X_{74} X_{46} X_{62} -X_{15} X_{56} X_{61} -X_{15} X_{57} X_{71} \\
 & + X_{24} X_{46} X_{62} + X_{15} X_{56} X_{61} + X_{35}  X_{57} X_{74} X_{43} + X_{25} X_{56} X_{62} + X_{15} X_{57} X_{71} + X_{27}  X_{73} X_{32}
\end{array}$}} \\ \hline
\end{tabular}}
\end{table}


\newpage

\begin{table}[!htbp] 
\centering 
\renewcommand{\arraystretch}{1.2}
\resizebox{\linewidth}{!}{
\tiny
\begin{tabular}{|c|c|c|}
 \hline 
Toric Diagram & Brane Tiling & Quiver  \\
\hline 
\adjustimage{height=1.6cm,valign=m}{area7td8} & 
\adjustimage{height=4.7cm,valign=m}{area7s8d1} &
\adjustimage{height=3.4cm,valign=m}{area7s8d2} \\
\hline
\multicolumn{3}{|c|}{{\tiny $\begin{array}{cl}
W = & -X_{24} X_{43}  X_{32} -X_{25} X_{56} X_{62} -X_{16} X_{63} X_{37} X_{75} X_{51} -X_{12} X_{24} X_{41} -X_{37} X_{74} X_{43} \\
& + X_{16}  X_{62} X_{24} X_{41} + X_{37}  X_{75} X_{56} X_{63} + X_{37}  X_{74} X_{43} + X_{12} X_{25} X_{51} + X_{24}  X_{43} X_{32}
\end{array}$}} \\ \hline
 \hline 
\adjustimage{height=2.4cm,valign=m}{area7td9} & 
\adjustimage{height=3.1cm,valign=m}{area7s9d1} &
\adjustimage{height=4.2cm,valign=m}{area7s9d2} \\
\hline
\multicolumn{3}{|c|}{{\tiny $\begin{array}{cl}
W = &  -X_{13}  X_{32} X_{24} X_{41} -X_{16}  X_{62} X_{25} X_{51}-X_{37}  X_{75} X_{53} -X_{47}  X_{76} X_{64} -X_{37}  X_{76} X_{63} \\
& + X_{13} X_{37}  X_{75} X_{51} + X_{25}  X_{53} X_{32} + X_{37}  X_{76} X_{63} + X_{16}  X_{64} X_{41} + X_{24} X_{47}  X_{76} X_{62}
\end{array}$}} \\ \hline
 \hline 
\adjustimage{height=3.6cm,valign=m}{area7td10} & 
\adjustimage{height=5.5cm,valign=m}{area7s10d1} &
\adjustimage{height=4.2cm,valign=m}{area7s10d2} \\
\hline
\multicolumn{3}{|c|}{{\tiny $\begin{array}{cl}
W = & -X_{13}  X_{32} X_{21} -X_{14} X_{45} X_{51} -X_{46} X_{67} X_{74} -X_{26} X_{67} X_{72} -X_{26}  X_{63} X_{32} -X_{13} X_{35} X_{51} -X_{45} X_{57} X_{74} \\
& + X_{13} X_{35} X_{51} + X_{45} X_{57} X_{74} + X_{26} X_{67} X_{72} + X_{26}  X_{63} X_{32} + X_{13} X_{32} X_{21} + X_{14} X_{45} X_{51} + X_{46} X_{67} X_{74} 
\end{array}$}} \\ \hline
\end{tabular}}
\end{table}


\newpage

\begin{table}[!htbp] 
\centering 
\renewcommand{\arraystretch}{1.2}
\resizebox{\linewidth}{!}{
\tiny
\begin{tabular}{|c|c|c|}
 \hline 
Toric Diagram & Brane Tiling & Quiver  \\
\hline 
\adjustimage{height=2.4cm,valign=m}{area7td11} & 
\adjustimage{height=4.3cm,valign=m}{area7s11d1} & 
\adjustimage{height=3.5cm,valign=m}{area7s11d2} \\ 
\hline
\multicolumn{3}{|c|}{{\tiny $\begin{array}{cl}
W = &  -X_{23} X_{34} X_{42} -X_{25} X_{56} X_{62} -X_{35} X_{57} X_{73} -X_{13} X_{36} X_{61} -X_{14} X_{45} X_{51} -X_{17}  X_{72} X_{21} -X_{47}  X_{76} X_{64} \\
& + X_{23} X_{36} X_{62} + X_{45} X_{56} X_{64} + X_{13} X_{35} X_{51} + X_{17} X_{76} X_{61} + X_{14} X_{42} X_{21} + X_{25} X_{57} X_{72} + X_{34} X_{47} X_{73}
\end{array}$}} \\ \hline
\end{tabular}}
\end{table}

\subsection{Area 8}

\begin{table}[!htbp] 
\centering 
\renewcommand{\arraystretch}{1.2}
\resizebox{\linewidth}{!}{
\tiny
\begin{tabular}{|c|c|c|}
 \hline 
Toric Diagram & Brane Tiling & Quiver  \\
\hline 
\adjustimage{height=2.4cm,valign=m}{area8s1d3} & 
\adjustimage{height=5cm,valign=m}{area8s1d1} &
\adjustimage{height=5cm,valign=m}{area8s1d2} \\
\hline
\multicolumn{3}{|c|}{{\tiny $\begin{array}{cl}
W = &  -X_{35} X_{54} X_{43} -X_{37} X_{76} X_{63} -X_{46} X_{68} X_{84} -X_{15} X_{58} X_{87} X_{71} -X_{25} X_{56} X_{62} -X_{12} X_{24} X_{41} -X_{23} X_{38} X_{82} \\
& + X_{35} X_{56} X_{63} + X_{24} X_{46} X_{62} + X_{15}  X_{54} X_{41} + X_{25} X_{58} X_{82} + X_{12} X_{23} X_{37} X_{71} + X_{68}  X_{87} X_{76} + X_{38} X_{84} X_{43} 
\end{array}$}} \\ \hline
 \hline 
\adjustimage{height=2.4cm,valign=m}{area8s2d3} & 
\adjustimage{height=5cm,valign=m}{area8s2d1} &
\adjustimage{height=4.7cm,valign=m}{area8s2d2} \\
\hline
\multicolumn{3}{|c|}{{\tiny $\begin{array}{cl}
W = &   -X_{26} X_{67} X_{72} -X_{58} X_{86} X_{65} -X_{18} X_{82} X_{21} -X_{17} X_{73}  X_{31}-X_{47} X_{78} X_{84} -X_{15} X_{54}  X_{41}-X_{25}  X_{53} X_{32} -X_{34} X_{46} X_{63} \\
& + X_{17}  X_{72} X_{21} + X_{67} X_{78} X_{86} + X_{25} X_{58} X_{82} + X_{15} X_{53} X_{31} + X_{34} X_{47} X_{73} + X_{18} X_{84} X_{41} + X_{26} X_{63} X_{32} + X_{46} X_{65} X_{54} 
\end{array}$}} \\ \hline
\end{tabular}}
\end{table}


\newpage

\begin{table}[!htbp] 
\centering 
\renewcommand{\arraystretch}{1.2}
\resizebox{\linewidth}{!}{
\tiny
\begin{tabular}{|c|c|c|}
 \hline 
Toric Diagram & Brane Tiling & Quiver  \\
\hline 
\adjustimage{height=1.6cm,valign=m}{area8s3d3} & 
\adjustimage{height=3cm,valign=m}{area8s3d1} &
\adjustimage{height=3.5cm,valign=m}{area8s3d2} \\
\hline
\multicolumn{3}{|c|}{{\tiny $\begin{array}{cl}
W = & -X_{35} X_{54} X_{43} -X_{36} X_{64} X_{43} -X_{15} X_{57} X_{78} X_{81} -X_{35} X_{57} X_{73}-X_{16} X_{62} X_{28} X_{81} -X_{23} X_{36} X_{62}\\
& + X_{35} X_{57} X_{73} + X_{35} X_{54} X_{43} + X_{15} X_{57} X_{78} X_{81} + X_{16} X_{62} X_{28} X_{81} + X_{23} X_{36} X_{62} + X_{36} X_{64} X_{43}
\end{array}$}} \\ \hline
 \hline 
\adjustimage{height=1.6cm,valign=m}{area8s4d3} & 
\adjustimage{height=3cm,valign=m}{area8s4d1} &
\adjustimage{height=3.5cm,valign=m}{area8s4d2} \\
\hline
\multicolumn{3}{|c|}{{\tiny $\begin{array}{cl}
W = & -X_{45} X_{56} X_{64} -X_{56} X_{68} X_{87}  X_{75}-X_{16} X_{67} X_{71} -X_{12} X_{28} X_{81}-X_{12} X_{23} X_{31} -X_{23} X_{34} X_{42}-X_{34} X_{45} X_{53} \\
& +  X_{45} X_{56} X_{64} + X_{56} X_{67} X_{75} + X_{16} X_{68} X_{81} + X_{12} X_{28} X_{87} X_{71} + X_{12} X_{23} X_{31} + X_{23} X_{34} X_{42} + X_{34} X_{45} X_{53}
\end{array}$}} \\ \hline
 \hline 
\adjustimage{height=1.6cm,valign=m}{area8s5d3} & 
\adjustimage{height=3cm,valign=m}{area8s5d1} &
\adjustimage{height=3.5cm,valign=m}{area8s5d2} \\
\hline
\multicolumn{3}{|c|}{{\tiny $\begin{array}{cl}
W = & -X_{15}  X_{54} X_{46} X_{61} -X_{15} X_{57} X_{71}-X_{57} X_{78} X_{85} -X_{27} X_{78} X_{82}-X_{23} X_{38} X_{82} -X_{23} X_{34} X_{46} X_{62}\\
& + X_{15} X_{57} X_{71} + X_{57} X_{78} X_{85} + X_{27} X_{78} X_{82} + X_{23} X_{38} X_{82} + X_{23} X_{34} X_{46} X_{62} + X_{15} X_{54} X_{46} X_{61}
\end{array}$}} \\ \hline
 \hline 
Toric Diagram & Brane Tiling & Quiver  \\
\hline 
\adjustimage{height=1.6cm,valign=m}{area8s6d3} & 
\adjustimage{height=5.7cm,valign=m}{area8s6d1} &
\adjustimage{height=4cm,valign=m}{area8s6d2} \\
\hline
\multicolumn{3}{|c|}{{\tiny $\begin{array}{cl}
W = & -X_{24} X_{43} X_{32} -X_{25} X_{56} X_{62}-X_{57} X_{78} X_{85} -X_{17} X_{73} X_{31} -X_{14} X_{43} X_{31} -X_{24} X_{46} X_{62}-X_{56} X_{68} X_{85} -X_{17} X_{78} X_{81} \\
& + X_{24} X_{46} X_{62} + X_{56} X_{68} X_{85} + X_{17} X_{78} X_{81} + X_{14} X_{43}  X_{31} + X_{24} X_{43}  X_{32} + X_{25} X_{56} X_{62} + X_{57} X_{78} X_{85} + X_{17} X_{73}  X_{31}
\end{array}$}} \\ \hline 
\end{tabular}}
\end{table}


\newpage

\begin{table}[!htbp] 
\centering 
\renewcommand{\arraystretch}{1.2}
\resizebox{\linewidth}{!}{
\tiny
\begin{tabular}{|c|c|c|}
 \hline 
Toric Diagram & Brane Tiling & Quiver  \\
 \hline 
\adjustimage{height=1.6cm,valign=m}{area8s7d3} & 
\adjustimage{height=5.2cm,valign=m}{area8s7d1} &
\adjustimage{height=4.4cm,valign=m}{area8s7d2} \\
\hline
\multicolumn{3}{|c|}{{\tiny $\begin{array}{cl}
W = & -X_{13}  X_{32} X_{24} X_{41} -X_{15} X_{56} X_{61} -X_{57} X_{78} X_{85} -X_{24} X_{47} X_{72} -X_{13} X_{36} X_{61} -X_{56} X_{68} X_{85}-X_{47} X_{78} X_{84} \\
& +  X_{13} X_{36} X_{61} + X_{56} X_{68} X_{85} + X_{47} X_{78} X_{84} + X_{13} X_{32} X_{24} X_{41} + X_{15} X_{56} X_{61}+ X_{57} X_{78} X_{85} + X_{24} X_{47} X_{72}
\end{array}$}} \\ \hline
 \hline 
\adjustimage{height=1.6cm,valign=m}{area8s8d3} & 
\adjustimage{height=5.2cm,valign=m}{area8s8d1} &
\adjustimage{height=4.2cm,valign=m}{area8s8d2} \\
\hline
\multicolumn{3}{|c|}{{\tiny $\begin{array}{cl}
W = & -X_{14}  X_{43} X_{31}-X_{17} X_{75}  X_{56} X_{61}-X_{58} X_{87} X_{75} -X_{38} X_{84} X_{43} -X_{14} X_{46} X_{61} -X_{28} X_{87} X_{72} -X_{28} X_{84} X_{42} \\
 & + X_{14} X_{46} X_{61} + X_{17} X_{75} X_{56} X_{61} + X_{28} X_{87} X_{72} + X_{38} X_{84} X_{43} + X_{58} X_{87} X_{75} + X_{28} X_{84} X_{42} + X_{14} X_{43} X_{31} 
\end{array}$}} \\ \hline
 \hline 
Toric Diagram & Brane Tiling & Quiver  \\
\hline 
\adjustimage{height=1.6cm,valign=m}{area8s9d3} & 
\adjustimage{height=4.7cm,valign=m}{area8s9d1} &
\adjustimage{height=3.3cm,valign=m}{area8s9d2} \\
\hline
\multicolumn{3}{|c|}{{\tiny $\begin{array}{cl}
W = & -X_{24} X_{43} X_{32} -X_{26} X_{65} X_{52} -X_{37} X_{75} X_{58} X_{83}-X_{14} X_{45} X_{52} X_{21} -X_{18} X_{86} X_{67} X_{71} \\
& + X_{24} X_{45} X_{52} + X_{26} X_{67} X_{75} X_{52} + X_{14} X_{43} X_{37} X_{71} + X_{58} X_{86}X_{65} + X_{18} X_{83} X_{32} X_{21} 
\end{array}$}} \\ \hline 
\end{tabular}}
\end{table}


\newpage

\begin{table}[!htbp] 
\centering 
\renewcommand{\arraystretch}{1.2}
\resizebox{\linewidth}{!}{
\tiny
\begin{tabular}{|c|c|c|}
 \hline 
Toric Diagram & Brane Tiling & Quiver  \\
 \hline 
\adjustimage{height=1.6cm,valign=m}{area8s10d3} & 
\adjustimage{height=3cm,valign=m}{area8s10d1} &
\adjustimage{height=3cm,valign=m}{area8s10d2} \\
\hline
\multicolumn{3}{|c|}{{\tiny $\begin{array}{cl}
W = & -X_{34} X_{45} X_{53} -X_{37} X_{74} X_{46} X_{63} -X_{38} X_{86} X_{63} -X_{18} X_{86} X_{61} -X_{18} X_{82} X_{25} X_{51} -X_{17} X_{72} X_{21} \\
& + X_{38} X_{86} X_{63} + X_{34} X_{46} X_{63} +X_{18} X_{82} X_{21} + X_{18} X_{86} X_{61} + X_{25} X_{53} X_{37} X_{72} + X_{17} X_{74} X_{45} X_{51}
\end{array}$}} \\ \hline
 \hline 
\adjustimage{height=1.6cm,valign=m}{area8s11d3} & 
\adjustimage{height=4.1cm,valign=m}{area8s11d1} &
\adjustimage{height=3cm,valign=m}{area8s11d2} \\
\hline
\multicolumn{3}{|c|}{{\tiny $\begin{array}{cl}
W = & -X_{35}  X_{54} X_{43} -X_{36} X_{67} X_{73}-X_{48} X_{86} X_{64} -X_{15} X_{54} X_{41} -X_{23} X_{35} X_{57} X_{72} -X_{16} X_{62} X_{28} X_{81}\\
& + X_{35} X_{57} X_{73} + X_{28} X_{86} X_{67} X_{72} + X_{15} X_{54} X_{48} X_{81} + X_{35}X_{54}  X_{43} + X_{23} X_{36} X_{62} + X_{16} X_{64} X_{41}
\end{array}$}} \\ \hline
\hline 
\adjustimage{height=1.6cm,valign=m}{area8s12d3} & 
\adjustimage{height=3cm,valign=m}{area8s12d1} &
\adjustimage{height=3cm,valign=m}{area8s12d2} \\
\hline
\multicolumn{3}{|c|}{{\tiny $\begin{array}{cl}
W = & -X_{35}  X_{54} X_{43} -X_{37} X_{74} X_{46} X_{63} -X_{56} X_{68} X_{85} -X_{16} X_{68} X_{81}-X_{13} X_{32} X_{28} X_{81} -X_{12} X_{27} X_{71}\\
& + X_{35} X_{56} X_{63} + X_{46} X_{68} X_{85} X_{54} + X_{16} X_{68} X_{81} + X_{12} X_{28} X_{81}+ X_{13} X_{37} X_{71} + X_{27} X_{74} X_{43} X_{32} 
\end{array}$}} \\ \hline 
 \hline 
\adjustimage{height=1.6cm,valign=m}{area8s13d3} & 
\adjustimage{height=3cm,valign=m}{area8s13d1} &
\adjustimage{height=3cm,valign=m}{area8s13d2} \\
\hline
\multicolumn{3}{|c|}{{\tiny $\begin{array}{cl}
W = & -X_{24} X_{43} X_{35} X_{52} -X_{26} X_{63} X_{32} -X_{14} X_{47} X_{78} X_{86} X_{61} -X_{24} X_{47} X_{72}-X_{15} X_{58} X_{81} \\
& +  X_{24} X_{47} X_{72} + X_{24} X_{43} X_{32} + X_{14} X_{47} X_{78} X_{81} + X_{15} X_{52} X_{26} X_{61} + X_{35} X_{58} X_{86} X_{63}
\end{array}$}} \\ \hline
\end{tabular}}
\end{table}


\newpage

\begin{table}[!htbp] 
\centering 
\renewcommand{\arraystretch}{1.2}
\resizebox{\linewidth}{!}{
\tiny
\begin{tabular}{|c|c|c|}
 \hline 
Toric Diagram & Brane Tiling & Quiver  \\
 \hline 
\adjustimage{height=2cm,valign=m}{area8s14d3} & 
\adjustimage{height=4.5cm,valign=m}{area8s14d1} &
\adjustimage{height=3.8cm,valign=m}{area8s14d2} \\
\hline
\multicolumn{3}{|c|}{{\tiny $\begin{array}{cl}
W = & -X_{12} X_{23} X_{31} -X_{16} X_{64} X_{45} X_{51} -X_{27} X_{74} X_{48} X_{82} -X_{13} X_{37} X_{71}-X_{35} X_{56} X_{63} -X_{57} X_{78} X_{85}\\
& + X_{13} X_{35} X_{51} + X_{45} X_{57} X_{74} + X_{12} X_{27} X_{71} + X_{16} X_{63} X_{31} + X_{48} X_{85} X_{56} X_{64} + X_{23} X_{37} X_{78} X_{82}
\end{array}$}} \\ \hline
\hline 
\adjustimage{height=1.6cm,valign=m}{area8s15d3} & 
\adjustimage{height=3cm,valign=m}{area8s15d1} &
\adjustimage{height=3.7cm,valign=m}{area8s15d2} \\
\hline
\multicolumn{3}{|c|}{{\tiny $\begin{array}{cl}
W = & -X_{46} X_{65} X_{54} -X_{45} X_{57} X_{74}-X_{68} X_{87} X_{76} -X_{16} X_{67} X_{71}-X_{25} X_{53} X_{38} X_{82} -X_{13} X_{34} X_{42} X_{21} \\
& + X_{46} X_{67} X_{74} + X_{57} X_{76} X_{65} + X_{16} X_{68} X_{82} X_{21} + X_{13} X_{38} X_{87} X_{71} + X_{25} X_{54} X_{42} + X_{34} X_{45} X_{53}
\end{array}$}} \\ \hline 
 \hline 
\adjustimage{height=1.6cm,valign=m}{area8s16d3} & 
\adjustimage{height=4.5cm,valign=m}{area8s16d1} &
\adjustimage{height=3.9cm,valign=m}{area8s16d2} \\
\hline
\multicolumn{3}{|c|}{{\tiny $\begin{array}{cl}
W = & -X_{26} X_{64} X_{45} X_{52} -X_{37} X_{74} X_{48} X_{83}-X_{16} X_{63} X_{35} X_{51} -X_{17} X_{72} X_{28} X_{81}\\
& + X_{28} X_{83} X_{35} X_{52} + X_{17} X_{74} X_{45} X_{51} + X_{26} X_{63} X_{37} X_{72} + X_{16} X_{64} X_{48} X_{81}
\end{array}$}} \\ \hline
\end{tabular}}
\end{table}


\newpage

\begin{table}[!htbp] 
\centering 
\renewcommand{\arraystretch}{1.2}
\resizebox{\linewidth}{!}{%
\tiny
\begin{tabular}{|c|c|c|}
 \hline 
Toric Diagram & Brane Tiling & Quiver  \\
\hline 
\adjustimage{height=1.6cm,valign=m}{area8s17d3} & 
\adjustimage{height=3.7cm,valign=m}{area8s17d1} &
\adjustimage{height=4.7cm,valign=m}{area8s17d2} \\
\hline
\multicolumn{3}{|c|}{{\tiny $\begin{array}{cl}
W = & -X_{15} X_{54} X_{46} X_{61} -X_{14} X_{47} X_{71}-X_{28} X_{87} X_{75} X_{52} -X_{38} X_{84}X_{46} X_{63} -X_{23} X_{34} X_{42} \\
& +  X_{15} X_{52} X_{23} X_{38} X_{87} X_{71} + X_{47} X_{75} X_{54} + X_{28} X_{84} X_{42} + X_{14} X_{46} X_{61} + X_{34} X_{46} X_{63}
\end{array}$}} \\ \hline
\hline 
\adjustimage{height=1.6cm,valign=m}{area8s18d3} & 
\adjustimage{height=5.5cm,valign=m}{area8s18d1} &
\adjustimage{height=4.5cm,valign=m}{area8s18d2} \\
\hline
\multicolumn{3}{|c|}{{\tiny $\begin{array}{cl}
W = & -X_{13} X_{32} X_{24} X_{41}-X_{15} X_{56} X_{61}-X_{27} X_{75} X_{56} X_{68} X_{82} -X_{36} X_{64}X_{43}-X_{48} X_{87} X_{74} \\
& + X_{13} X_{36} X_{61} + X_{56} X_{68} X_{87} X_{75} + X_{27} X_{74} X_{43} X_{32} + X_{15} X_{56} X_{64} X_{41} + X_{24} X_{48} X_{82}
\end{array}$}} \\ \hline 
 \hline 
\adjustimage{height=1.6cm,valign=m}{area8s19d3} & 
\adjustimage{height=6.3cm,valign=m}{area8s19d1} &
\adjustimage{height=5.5cm,valign=m}{area8s19d2} \\
\hline
\multicolumn{3}{|c|}{{\tiny $\begin{array}{cl}
W = & -X_{34} X_{45} X_{53} -X_{36} X_{67} X_{73}-X_{18} X_{86} X_{61} -X_{24} X_{48} X_{82}-X_{47} X_{75} X_{54} -X_{13} X_{37} X_{71}-X_{16} X_{62} X_{21} -X_{28} X_{85} X_{52} \\
& + X_{34} X_{47} X_{73} + X_{16} X_{67} X_{71} + X_{28} X_{86} X_{62} + X_{48} X_{85} X_{54} + X_{37} X_{75} X_{53} + X_{13} X_{36} X_{61} + X_{18} X_{82} X_{21} + X_{24} X_{45} X_{52}
\end{array}$}} \\ \hline
\end{tabular}}
\end{table}


\newpage

\begin{table}[!htbp] 
\centering 
\renewcommand{\arraystretch}{1.2}
\resizebox{\linewidth}{!}{
\tiny
\begin{tabular}{|c|c|c|}
 \hline 
Toric Diagram & Brane Tiling & Quiver  \\
\hline 
\adjustimage{height=1.6cm,valign=m}{area8s20d3} & 
\adjustimage{height=4.5cm,valign=m}{area8s20d1} &
\adjustimage{height=4.5cm,valign=m}{area8s20d2} \\
\hline
\multicolumn{3}{|c|}{{\tiny $\begin{array}{cl}
W = &  -X_{13} X_{32} X_{24} X_{41} -X_{16} X_{62} X_{25} X_{51}-X_{37} X_{74} X_{48} X_{83} -X_{58} X_{86} X_{67} X_{75} \\
& + X_{13} X_{37} X_{75} X_{51} + X_{25} X_{58} X_{83} X_{32} + X_{16} X_{67} X_{74} X_{41} + X_{24} X_{48} X_{86} X_{62}
\end{array}$}} \\ \hline
\hline 
\adjustimage{height=1.6cm,valign=m}{area8s21d3} & 
\adjustimage{height=4.5cm,valign=m}{area8s21d1} &
\adjustimage{height=4.5cm,valign=m}{area8s21d2} \\
\hline
\multicolumn{3}{|c|}{{\tiny $\begin{array}{cl}
W = & -X_{46} X_{65} X_{54} -X_{45} X_{57} X_{74}-X_{68} X_{87} X_{76} -X_{16} X_{67} X_{71}-X_{24} X_{43} X_{38} X_{82} -X_{13} X_{35} X_{52} X_{21} \\
& + X_{46} X_{67} X_{74} + X_{57} X_{76} X_{65} + X_{16} X_{68} X_{82} X_{21} + X_{13} X_{38} X_{87} X_{71} + X_{24} X_{45} X_{52} + X_{35} X_{54} X_{43}
\end{array}$}} \\ \hline 
 \hline 
\adjustimage{height=1.6cm,valign=m}{area8s22d3} & 
\adjustimage{height=6.8cm,valign=m}{area8s22d1} &
\adjustimage{height=4.8cm,valign=m}{area8s22d2} \\
\hline
\multicolumn{3}{|c|}{{\tiny $\begin{array}{cl}
W = & -X_{34} X_{45} X_{53} -X_{36} X_{67} X_{73}-X_{18} X_{86} X_{61} -X_{28} X_{84} X_{42} -X_{47} X_{75} X_{54} -X_{13} X_{37} X_{71}-X_{16} X_{62} X_{21} -X_{25} X_{58} X_{82}\\
& + X_{34} X_{47} X_{73} + X_{16} X_{67} X_{71} + X_{28} X_{86} X_{62} + X_{25} X_{54} X_{42} + X_{37} X_{75} X_{53} + X_{13} X_{36} X_{61} + X_{18} X_{82} X_{21} + X_{45} X_{58} X_{84}
\end{array}$}} \\ \hline

\end{tabular}}
\end{table}


\newpage

\begin{table}[!htbp] 
\centering 
\renewcommand{\arraystretch}{1.2}
\resizebox{\linewidth}{!}{
\tiny
\begin{tabular}{|c|c|c|}
 \hline 
Toric Diagram & Brane Tiling & Quiver  \\
 \hline 
\adjustimage{height=.7cm,valign=m}{area8s23d3} & 
\adjustimage{height=4cm,valign=m}{area8s23d1x} &  
\adjustimage{height=3.3cm,valign=m}{area8s23d2x} \\   
\hline
\multicolumn{3}{|c|}{{\tiny $\begin{array}{cl}
W = &  -X_{12}X_{23}X_{32}X_{21} -X_{34}X_{45}X_{54}X_{43} -X_{56}X_{67}X_{76}X_{65} -X_{78}X_{81}X_{18}X_{87} \\
& + X_{23}X_{34}X_{43}X_{32}+ X_{45}X_{56}X_{65}X_{54}+ X_{67}X_{78}X_{87}X_{76}+ X_{81}X_{12}X_{21}X_{18}
\end{array}$}} \\ \hline
\hline 
\adjustimage{height=.7cm,valign=m}{area8s24d3} & 
\adjustimage{height=5cm,valign=m}{area8s24d1x} &  
\adjustimage{width=3.3cm,valign=m}{area8s24d2x} \\   
\hline
\multicolumn{3}{|c|}{{\tiny $\begin{array}{cl}
W = & - X_{11}X_{12}X_{21}-X_{22}X_{23}X_{32}-X_{34}X_{45}X_{54}X_{43}-X_{56}X_{67}X_{76}X_{65}-X_{78}X_{81}X_{18}X_{87} \\
& + X_{11}X_{18}X_{81}+X_{22}X_{21}X_{12}+X_{23}X_{34}X_{43}X_{32}+X_{45}X_{56}X_{65}X_{54}+X_{67}X_{78}X_{87}X_{76} 
\end{array}$}} \\ \hline 
 \hline 
\adjustimage{height=.7cm,valign=m}{area8s25d3} & 
\adjustimage{height=5cm,valign=m}{area8s25d1x} &  
\adjustimage{height=3.5cm,valign=m}{area8s25d2x} \\   
\hline
\multicolumn{3}{|c|}{{\tiny $\begin{array}{cl}
W = & -X_{11}X_{12}X_{21}-X_{22}X_{23}X_{32}-X_{33}X_{34}X_{43}-X_{44}X_{45}X_{54}-X_{56}X_{67}X_{76}X_{65}-X_{78}X_{81}X_{18}X_{87}\\
& + X_{11}X_{18}X_{81}+X_{22}X_{21}X_{12}+X_{33}X_{32}X_{23}+X_{44}X_{43}X_{34}+X_{45}X_{56}X_{65}X_{54}+X_{67}X_{78}X_{87}X_{76}
\end{array}$}} \\ \hline
\end{tabular}}
\end{table}


\newpage

\begin{table}[!htbp] 
\centering 
\renewcommand{\arraystretch}{1.2}
\resizebox{\linewidth}{!}{
\tiny
\begin{tabular}{|c|c|c|}
 \hline 
Toric Diagram & Brane Tiling & Quiver  \\
 \hline 
\adjustimage{height=.7cm,valign=m}{area8s26d3} & 
\adjustimage{height=6.5cm,valign=m}{area8s26d1x} &  
\adjustimage{height=4.7cm,valign=m}{area8s26d2x} \\   
\hline
\multicolumn{3}{|c|}{{\tiny $\begin{array}{cl}
W = &  -X_{11}X_{12}X_{21}-X_{22}X_{23}X_{32}-X_{33}X_{34}X_{43}-X_{44}X_{45}X_{54}-X_{55}X_{56}X_{65}-X_{66}X_{67}X_{76}-X_{67}X_{78}X_{87}X_{76} \\
& +  X_{11}X_{18}X_{81}+X_{22}X_{21}X_{12}+X_{33}X_{32}X_{23}+X_{44}X_{43}X_{34}+X_{55}X_{54}X_{45}+X_{66}X_{65}X_{56}+X_{78}X_{81}X_{18}X_{87}
\end{array}$}} \\ \hline
\hline 
\adjustimage{height=.7cm,valign=m}{area8s27d3} & 
\adjustimage{height=6.5cm,valign=m}{area8s27d1x} &  
\adjustimage{height=4.7cm,valign=m}{area8s27d2x} \\   
\hline
\multicolumn{3}{|c|}{{\tiny $\begin{array}{cl}
W = &  -X_{11}X_{12}X_{21}-X_{22}X_{23}X_{32}-X_{33}X_{34}X_{43}-X_{44}X_{45}X_{54} \\
& -X_{55}X_{56}X_{65}-X_{66}X_{67}X_{76} - X_{77}X_{78}X_{87}-X_{88}X_{81}X_{18} \\
& + X_{11}X_{18}X_{81}+X_{22}X_{21}X_{12}+X_{33}X_{32}X_{23}+X_{44}X_{43}X_{34}\\
& +X_{55}X_{54}X_{45}+X_{66}X_{65}X_{56} + X_{77}X_{76}X_{67}+X_{88}X_{87}X_{78}
\end{array}$}} \\ \hline
\end{tabular}}
\end{table}

\section{Conclusions}

\label{section_conclusions}

Since their introduction, brane tilings have hugely simplified the connection between gauge theories on D3-branes and the toric CY 3-folds they probe. While given an arbitrary toric singularity there are well-defined methods for obtaining the corresponding brane tiling, it is of great interest to work out catalogues of explicit examples. Such databases are useful, for example, for uncovering general properties of these theories and for identifying the best models for specific applications. 

In this paper, we classified all toric CY 3-folds with toric diagrams up to area 8 and constructed a brane tiling for each of them. To do so, we developed implementations of dimer model techniques specifically tailored for partial resolution. We also created computational modules for a wide range of manipulations and computations involving brane tilings. They can be accessed at \cite{DimerSystem}.

There are various directions for future investigation. First, additional information can be added to our catalogue.  We found one brane tiling for every toric CY$_3$ but, generically, each geometry is associated to more than one brane tiling. These so-called {\it toric phases} are related to each other by Seiberg duality and it would be interesting to provide a complete classification of them for the geometries in our list. Ideally, we would also like to determine extra data such as $R$-charges, $j$-invariants for the dessins, etc.

In future work, we plan to use our classification of brane tilings as the starting point for local model building of Standard Model (SM)-like theories with realistic spectra, hierarchies of masses, flavor mixings, etc. The main idea of this kind of construction is to consider a singularity that gives rise to a reasonable spectrum, such as the cone over $dP_0$ and embed it into a slightly larger one, such as the cone over $dP_3$. This particular example was studied in great detail in \cite{Krippendorf:2010hj, Dolan:2011qu}, with encouraging results. The finite size of the resolved cycles map to non-vanishing vevs for the scalar components of some bifundamental chiral multiplets. By construction, the resulting low energy theory is the desired SM quiver, but with the vevs appearing as new parameters that can be tuned to control the flavor structure. These vevs appear in very specific ways in the superpotential, leading to a constrained and predictive scenario. We will undertake a systematic large scale investigation of local model building using the entire set of area 6 to 8 toric CY 3-folds as parent geometries. We will identify those that are phenomenologically promising and work out the detailed features of the low energy theories.

\acknowledgments

The work of S. F. is supported by the U.S. National Science Foundation grant PHY-1518967 and by a PSC-CUNY award. Y.-H. H. would like to thank the Science and Technology Facilities Council, UK, for grant ST/J00037X/1, the Chinese Ministry of Education, for a Chang-Jiang Chair Professorship at NanKai University as well as the City of Tian-Jin for a Qian-Ren Scholarship, and Merton College, Oxford, for her enduring support.

\appendix

\section{Computational Modules}

\label{appendix_modules}

We created various {\it Mathematica} modules that implement the ideas presented in this paper. Their applicability goes well beyond the classification of brane tilings we presented and should be useful for a wide community. They are publicly available at \cite{DimerSystem}. Here we summarize some of the basic commands.

So far the package is for a standard \code{Unix} environment, where the default directory for storing the intermediate output is the user's home directory \code{\$HOME}.  

The $m\times n$ rectangular brane tilings for $\mathcal{C}/\mathbb{Z}_m \times \mathbb{Z}_n$ play a central role in our studies, since we use them as simple starting points for partial resolution. For this reason, we created a module called \code{RecDimerModels[$m$,$n$]}, which generates the brane tiling for $\mathcal{C}/(\mathbb{Z}_m \times \mathbb{Z}_n)$ with all its elements properly labeled and generates its Kasteleyn matrix. The intermediate data is stored in the file \code{\$HOME.dimer.model.tmp.txt}. 

Next, the \code{ToricInfo[$KM$]} module takes a Kasteleyn matrix as input and produces the corresponding perfect matchings and toric diagram.

For triangulating toric diagrams, we provide \code{TriangDimer[$ToricPts$]}, which is a modified version of the \code{DelaunayMesh[]} command in {\it Mathematica}.

The module \code{RemovePoints[$KM$,$Ptsremove$]} generates all possible collections of vevs, or equivalently edges to be removed, that give rise to a desired partial resolution defined by a starting toric diagram and the points we want to delete from it ({\it Ptsremove}). The data is loaded in \code{\$HOME.dimer.model.tmp.txt}. This is the most computationally intensive module, even though we use parallel computing and an optimized algorithm to enumerate all collections of removed edges. The output is in the form of a list containing all the possible higgsings ({\it PossibleHiggsings}). With this information, it is straightforward to determine the brane tiling resulting from any of these higgsings using \code{HiggsingDimerSU[$Kmatrix$,$possiblehiggsing$]}. This module also produces the quiver and superpotential for the brane tiling.

Algorithm \ref{alg:the_alg} provides a brief summary of how these modules were exploited for the classification of brane tilings carried out in this paper.

\begin{algorithm}[H]
  \caption{Classification of dimer models for all toric diagrams with a given area}
  \label{alg:the_alg}
  \begin{algorithmic}
  \State Initialise $Models$ as empty set. \Comment used as storing physical models.
  \State Load $PSets$ as all the inequivalent toric diagrams with a given area.
  \For {$toric$ in $PSets$}
  \State Define $K matrix$ by using \code{RecDimerModels[$m$,$n$]}. The integers $m$ and $n$ must define a rectangular toric diagram in which $toric$ can be embedded.
  \State Define $ptsremove$ as the set containing points to be removed from the rectangular toric diagram.
  \State Determine $PossibleHiggsings$, the collections of vevs that produce a given partial resolution, using \code{RemovePoints[$Kmatrix$,$ptsremove$]}.
  \For {$possiblehiggsing$ in $PossibleHiggsings$}
  \State 
  Use \code{HiggsingDimer[$Kmatrix$,$higgsantz$]} to compute the brane tiling, quiver and superpotential for every $possiblehiggsing$.
  \State Save this information into $Models$.
  \EndFor
  \EndFor 
  \end{algorithmic}
\end{algorithm}

\bibliographystyle{JHEP}
\bibliography{mybib}

\end{document}